\documentclass[prd,amsmath,aps,floats,amssymb, floatfix,
  superscriptaddress,nofootinbib,twocolumn,preprintnumbers]{revtex4-1}
\usepackage[utf8]{inputenc}
\usepackage[english]{babel}
\usepackage{natbib}
\def\apj{ApJ}%
\def\apjl{ApJ}%
\def\apjs{ApJS}%
\def\ao{Appl.~Opt.}%
\def\aap{A\&A}%
\def\mnras{MNRAS}%
\def\prd{Phys.~Rev.~D}%
\def\pasp{PASP}%
\def\jcap{JCAP}%
\usepackage{newtxtext,newtxmath}
\usepackage[T1]{fontenc}
\usepackage{ae,aecompl}
\usepackage{xcolor}

\usepackage{graphicx}	% Including figure files
\usepackage{amsmath}	% Advanced maths commands
\usepackage{amssymb}	% Extra maths symbols
\usepackage{braket}
\usepackage{siunitx} %v. useful units package 
\usepackage{multirow}
\usepackage{mathtools}
\usepackage{makecell}

\usepackage[caption = false]{subfig}
\usepackage{multirow}
\usepackage{ulem}
\usepackage{float}
\restylefloat{table}

\newcommand{\gammat}{\ensuremath{\gamma_{\rm t}(\theta)}}
\newcommand{\wtheta}{\ensuremath{w(\theta)}}

\newcommand{\pgg}{\ensuremath{P_{\mathrm{gg}}}}
\newcommand{\pgm}{\ensuremath{P_{\mathrm{gm}}}}
\newcommand{\xigg}{\ensuremath{\xi_{\mathrm{gg}}} }
\newcommand{\xigm}{\ensuremath{\xi_{\mathrm{gm}}} }

\DeclareMathAlphabet\mathbfcal{OMS}{cmsy}{b}{n}

\DeclareSIUnit \megaparsec {Mpc}
\DeclareSIUnit \h {\mbox{$h$}}

\newcommand{\om}{\ensuremath{\Omega_{\mathrm m}}}

\newcommand{\lcdm}{$\Lambda$CDM}
\newcommand{\wcdm}{$w$CDM}

\newcommand\fig[1]{Figure~\ref{#1}}

\newcommand{\redmagic}{\texttt{redMaGiC} }
\newcommand{\mice}{\texttt{MICE} }
\newcommand{\maglim}{\texttt{MagLim} }
\newcommand{\gold}{\texttt{GOLD} }
\newcommand{\buzzard}{\texttt{Buzzard} }

\newcommand{\beq}{\begin{equation}}
\newcommand{\eeq}{\end{equation}}

\newcommand{\balrog}{{\textsc{Balrog}}}

\newcommand{\isd}{\textsc{ISD}\xspace}
\newcommand{\enet}{\textsc{ENet}\xspace}

\newcommand{\metacal}{{\textsc{Metacalibration}~}}

\usepackage{xspace} %This allows to add extra space when needed in new commands

\newcommand{\Planck}{{\slshape Planck~}}

\newcommand{\Yv}[1]{{\color{blue}[YYY]}}

\begin{document}
\preprint{DES-2019-0481}
\preprint{FERMILAB-PUB-21-249-SCD-T}
\title[Short title, max. 45 characters]{Dark Energy Survey Year 3 Results: Constraints on cosmological parameters and galaxy bias models from galaxy clustering and galaxy-galaxy lensing using the \redmagic sample}

\label{firstpage}
% \pagerange{\pageref{firstpage}--\pageref{lastpage}}

% Abstract of the paper
\begin{abstract}
We constrain cosmological parameters and galaxy-bias parameters using the combination of galaxy clustering and galaxy-galaxy lensing measurements from the Dark Energy Survey (DES) Year-3 data. We describe our modeling framework and choice of scales analyzed, validating their robustness to theoretical uncertainties in small-scale clustering by analyzing simulated data. Using a linear galaxy bias model and \redmagic galaxy sample, we obtain 10\% constraints on the matter density of the universe. 
% to be $\Omega_{\rm m} = 0.325^{+0.033}_{-0.034}$. 
We also implement a non-linear galaxy bias model to probe smaller scales that includes parameterization based on hybrid perturbation theory, and find that it leads to a 17\% gain in cosmological constraining power. We perform robustness tests of our methodology pipeline and demonstrate stability of the constraints to changes in the theory model. Using the \redmagic galaxy sample as foreground lens galaxies, and adopting the best-fitting cosmological parameters from DES Year-1 data, we find the galaxy clustering and galaxy-galaxy lensing measurements to exhibit significant signals akin to de-correlation between galaxies and mass on large scales, which is not expected in any current models. This likely systematic measurement error biases our constraints on galaxy bias and the $S_8$ parameter. We find that a scale-, redshift- and sky-area-independent phenomenological de-correlation parameter can effectively capture this inconsistency between the galaxy clustering and galaxy-galaxy lensing. We trace the source of this correlation to a color-dependent photometric issue and minimize its impact on our result by changing the selection criteria of \redmagic galaxies. Using this new sample, our constraints on the $S_8$ parameter are consistent with previous studies and we find a small shift in the $\Omega_{\rm m}$ constraints compared to the fiducial \redmagic sample. We infer the constraints on the mean host halo mass of the \redmagic galaxies in this new sample from the large-scale bias constraints, finding the galaxies occupy halos of mass approximately $1.6 \times 10^{13} M_{\odot}/h$. 
\end{abstract}

% Select between one and six entries from the list of approved keywords.
% Don't make up new ones.
% \begin{keywords}
% keyword1 -- keyword2 -- keyword3
% \end{keywords}

\author{S.~Pandey}\email{shivamp@sas.upenn.edu}
\affiliation{Department of Physics and Astronomy, University of Pennsylvania, Philadelphia, PA 19104, USA}
\author{E.~Krause}
\affiliation{Department of Astronomy/Steward Observatory, University of Arizona, 933 North Cherry Avenue, Tucson, AZ 85721-0065, USA}
\author{J.~DeRose}
\affiliation{Lawrence Berkeley National Laboratory, 1 Cyclotron Road, Berkeley, CA 94720, USA}
\author{N.~MacCrann}
\affiliation{Department of Applied Mathematics and Theoretical Physics, University of Cambridge, Cambridge CB3 0WA, UK}
\author{B.~Jain}
\affiliation{Department of Physics and Astronomy, University of Pennsylvania, Philadelphia, PA 19104, USA}
\author{M.~Crocce}
\affiliation{Institut d'Estudis Espacials de Catalunya (IEEC), 08034 Barcelona, Spain}
\affiliation{Institute of Space Sciences (ICE, CSIC),  Campus UAB, Carrer de Can Magrans, s/n,  08193 Barcelona, Spain}
\author{J.~Blazek}
\affiliation{Department of Physics, Northeastern University, Boston, MA 02115, USA}
\affiliation{Laboratory of Astrophysics, \'Ecole Polytechnique F\'ed\'erale de Lausanne (EPFL), Observatoire de Sauverny, 1290 Versoix, Switzerland}
\author{A.~Choi}
\affiliation{Center for Cosmology and Astro-Particle Physics, The Ohio State University, Columbus, OH 43210, USA}
\author{H.~Huang}
\affiliation{Department of Physics, University of Arizona, Tucson, AZ 85721, USA}
\author{C.~To}
\affiliation{Department of Physics, Stanford University, 382 Via Pueblo Mall, Stanford, CA 94305, USA}
\affiliation{Kavli Institute for Particle Astrophysics \& Cosmology, P. O. Box 2450, Stanford University, Stanford, CA 94305, USA}
\affiliation{SLAC National Accelerator Laboratory, Menlo Park, CA 94025, USA}
% ...

\collaboration{DES Collaboration}

\maketitle
%%%%%%%%%%%%%%%%%%%%%%%%%%%%%%%%%%%%%%%%%%%%%%%%%%

%%%%%%%%%%%%%%%%% BODY OF PAPER %%%%%%%%%%%%%%%%%%

\section{Introduction}
%\IR{Mostly a placeholder introduction, will fill more details soon}
\label{sec:intro}

Wide-area imaging surveys of galaxies provide cosmological information through measurements of galaxy clustering and weak gravitational lensing. Galaxies are useful tracers of the full matter distribution, and their spatial clustering is used to infer the matter power spectrum. The shapes of distant galaxies are lensed by the intervening matter, providing a second way to probe the mass distribution. With wide-area galaxy surveys, these two probes of the late time universe have provided  information on both the geometry and growth of structure in the universe. 
In recent years, the combination of two-point correlations--- galaxy-galaxy lensing (the cross-correlation of lens galaxy positions with background source galaxy shear) and the angular auto-correlation of the lens galaxy positions---have been developed in a theoretical framework \citep{Cacciato_2009,Baldauf_2010,Cacciato_2012,van_den_Bosch_2013, Wibking_2018} and used to constrain cosmological parameters  \citep{Cacciato_2013,Mandelbaum_2013,Kwan_2016,More_2015,Dvornik_2018,Coupon_2015, Singh_2019, Wibking_2019}. In practice, two galaxy samples are used:  {\it lens} galaxies tracing the foreground large scale structure, and background {\it source} galaxies whose shapes are used to infer the lensing shear and this combination of galaxy-galaxy lensing and galaxy clustering is refereed to as ``2$\times$2pt'' datavector. This is generally complemented with the two-point of cosmic shear (the lensing shear auto-correlation, referred to as $1 \times 2$pt). 
The Dark Energy Survey (DES) presented cosmological constraints from their Year 1 (Y1) data set from cosmic shear \citep{Troxel_2018} and a joint analysis of all three two-point correlations (henceforth called the ``$3\times2$pt'' datavector) \citep{Abbott_2018}. 

This paper is part of a series describing the methodology and results of DES Year 3 (Y3) $3\times2$pt analysis. The cosmological constraints are presented for cosmic shear \citep{y3-cosmicshear1,y3-cosmicshear2}, the combination of galaxy clustering and galaxy-galaxy lensing using two different lens galaxy samples \citep[this paper; ][]{y3-2x2ptaltlensresults,y3-2x2ptmagnification}, as well as the $3\times2$pt analysis \citep{y3-3x2ptkp}. These cosmological results are enabled by extensive methodology developments at all stages of the analysis from pixels to cosmology, which are referenced throughout. This paper presents the modeling methodology and cosmology inference from DES Y3 galaxy clustering \citep{y3-galaxyclustering} and galaxy-galaxy lensing \citep{y3-gglensing} measurements. 
We focus on the \redmagic \citep{Rozo_2016} galaxy sample, described further below. A parallel analysis using a different galaxy sample, the \maglim sample \citep{y3-2x2maglimforecast}, is presented in a separate paper \citep{y3-2x2ptaltlensresults}.

Incomplete theoretical understanding of the relationship of galaxies to the mass distribution, called galaxy bias, has been a limiting factor in interpreting the lens galaxy auto-correlation function (denoted $w(\theta)$) and galaxy-galaxy lensing (and denoted $\gamma_{\rm t}(\theta)$). At large scales, galaxy bias can be described by a single number, the linear bias $b_1$. On smaller scales, bias is non-local and non-linear, and its description is complicated \citep{Fry_93,Scherrer_98}. Perturbation theory (PT) approaches have been developed for quasi-linear scales $\sim 10$ Mpc, though the precise range of scales of its validity is a subtle question that depends on the galaxy population, the theoretical model, and the statistical power of the survey. 

With a model for galaxy bias, $w(\theta)$ and $\gamma_{\rm t}$ measurements, together called the ``$2\times2$pt'' datavector, can probe the underlying matter power spectrum. They are also sensitive to the distance-redshift relation over the redshift range of the lens and source galaxy distributions.  These two datavectors constitute a useful subset of the full $3\times 2$pt datavector, since bias and cosmological parameters can both be constrained (though the uncertainty in galaxy bias would limit either $w(\theta)$ or $\gamma_{\rm t}(\theta)$ individually).

A major part of the modeling and validation involves PT models of galaxy bias and tests using mock catalogs based on N-body simulations with various schemes of populating galaxies. 
Approaches based on the halo occupation distribution (HOD) have been widely developed and are used for the DES galaxy samples. For the Year 3 (Y3) dataset of DES, two independent sets of mock catalogs have been developed, based on the $\buzzard$\citep{DeRose2019} and \mice simulations \citep{Fosalba_2014, Crocce_2015, Fosalba_2015}.

An interesting recent development in cosmology is a possible disagreement between the inference of the expansion rate and the amplitude of mass fluctuations (denoted $\sigma_8$) and direct measurements or the inference of these quantities in the late-time universe. The predictions are anchored via measurements of the cosmic microwave background (CMB) and use general relativity and a cosmological model of the universe to extrapolate to late times. This cosmological model, denoted by $\Lambda$CDM, relies on two ingredients in the energy budget of the universe that have yet to be directly detected: cold dark matter (CDM) and dark energy in the form of a cosmological constant denoted as $\Lambda$. The experiments that infer the cosmological constraints using the lensing of source galaxies, particularly using the cosmic-shear 2pt correlation are unable to generally break the degeneracy between $\om$ and $\sigma_8$. A derived parameter, $S_8 = \sigma_8 (\Omega_{\rm m}/0.3)^{0.5}$, is well constrained as it approximately controls the amplitude of the cosmic shear correlation function. The value of $S_8$ or $\sigma_8$ inferred from measurements of cosmic shear and the $3\times 2$pt datavector \citep{Abbott_2018,Troxel_2018, Heymans_2013, Heymans_2021, Hikage_2019, y3-3x2ptkp, y3-cosmicshear1, y3-cosmicshear2}, from galaxy clusters \citep{Abbott_2020_clusters,To_2021} and the redshift-space power spectrum \citep{Philcox_2020} tends to be lower than the CMB prediction. The significance of this tension is a work in progress and crucial to the viability of $\Lambda$CDM. The Hubble tension refers to the measured expansion rate being higher than predicted by the CMB. The resolution of the two tensions, and their possible relationship, is an active area of research in cosmology and provides additional context for the analysis presented here. 

\begin{figure}
\includegraphics[width=\columnwidth]{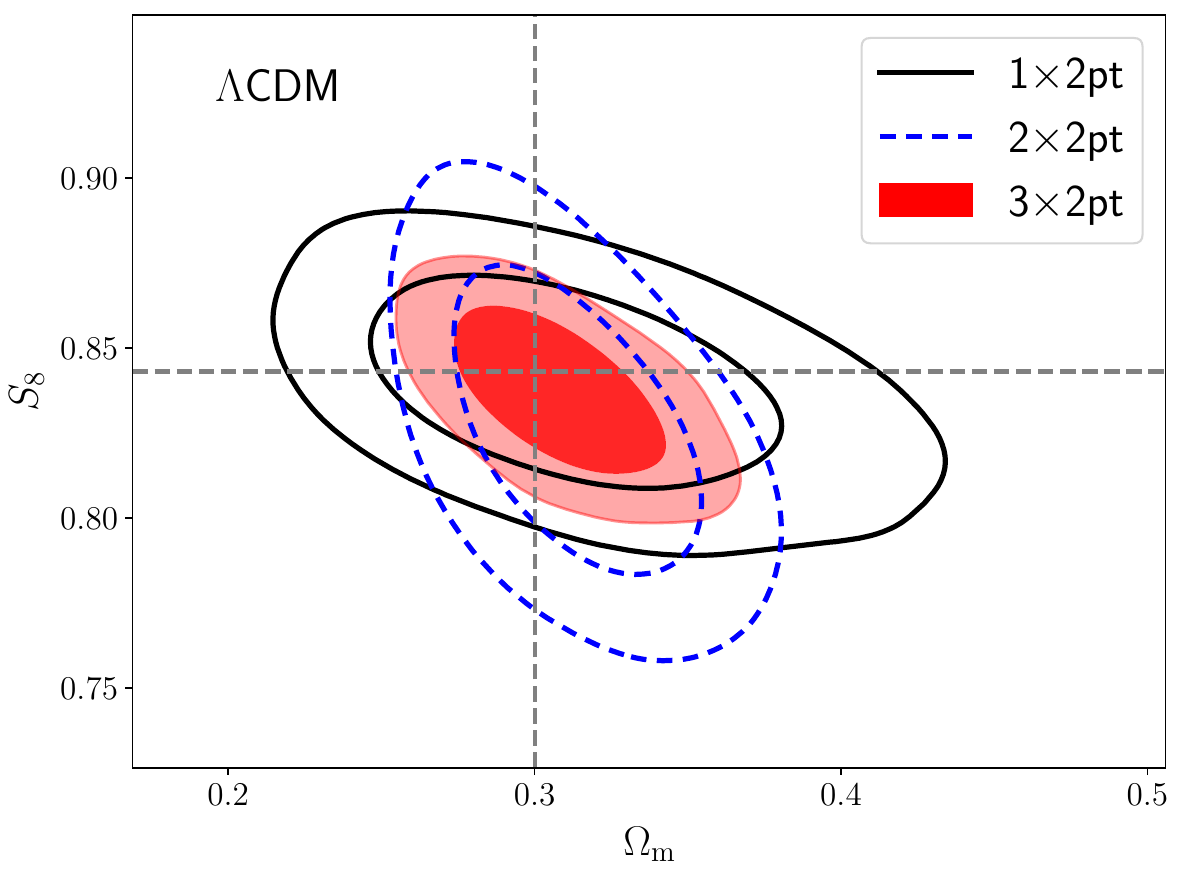}
\includegraphics[width=\columnwidth]{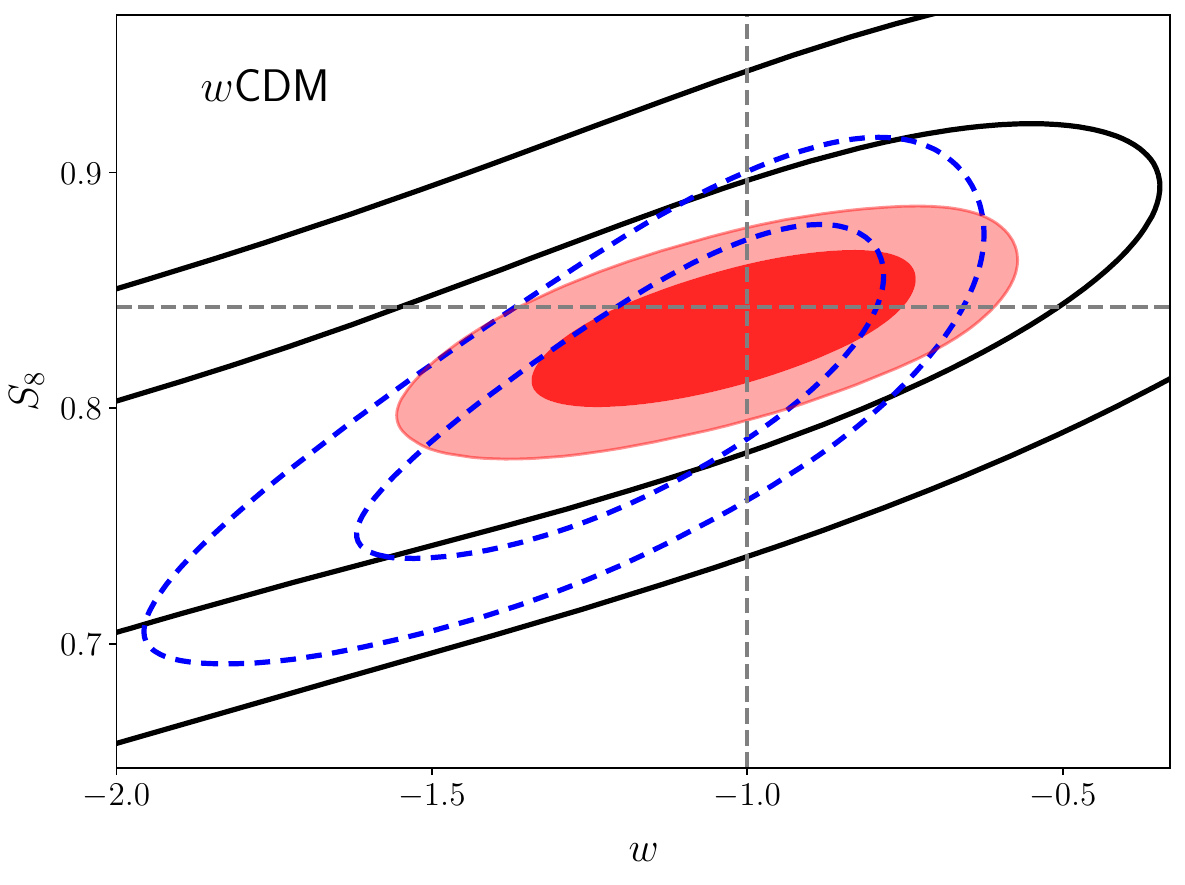}
\caption[]{Comparison of \textit{simulated} constraints on cosmological parameters $\Omega_{\mathrm{m}}$ and $S_8$ from cosmic shear alone ($1\times2$pt), galaxy clustering + galaxy-galaxy lensing ($2\times2$pt) and including all three probes ($3\times2$pt). This plot uses a \textit{simulated noise-less baseline datavector} (see \S\ref{sec:simlike_analysis}) and shows that $2\times2$pt adds complementary information to cosmic shear constraints, particularly, providing stronger constraints on $\Omega_{\mathrm{m}}$ and $w$.}
\label{fig:all2pt_comp}
\end{figure}

Figure~\ref{fig:all2pt_comp}, based on simulated data, shows the expected constraints on $S_8$ and $\Omega_{\rm m}$ from the $2\times2$pt datavector and cosmic shear ($1\times 2$pt). It is evident that the two have some complementarity, which enables the breaking of degeneracies in both $\Lambda$CDM and $w$CDM cosmological models (where $w$ is the dark energy equation of state parameter and $w \neq -1$ points towards the departure from standard $\Lambda$CDM model). Particularly noteworthy are the significantly better constraints compared to $1\times 2$pt on the parameter $w$ and $\Omega_{\rm m}$ using $2\times2$pt in the $w$CDM and $\Lambda$CDM models respectively. Note that unlike in $1\times2$pt, where all the matter in front of source galaxy contributes to its signal, $2\times2$pt receives contribution only from galaxies within the narrow lens redshift bins. Therefore, we attribute better constraints on these cosmological parameters from $2\times2$pt to significantly more precise redshifts of the lens galaxy sample. This allows for precise tomographic measurements of $2\times2$pt datavector which constrains the background geometric parameters like $w$ and $\Omega_{\rm m}$. With data, these somewhat independent avenues to cosmology provide a valuable cross-check, as the leading sources of systematics are largely different.

The formalism used to compute the $2\times2$pt datavector is presented in \S\ref{sec:stat_theory}. The description of the lens and source galaxy samples, their redshift distributions and measurement methodology of our datavector and its covariance estimation are presented in \S\ref{sec:data}. In  \S\ref{sec:param_inf} we validate our methodology using N-body simulations and determine the scale cuts for our analysis. Note that in this paper we focus on validation of analysis when using the \redmagic lens galaxy sample and we refer the reader to \citet{y3-2x2ptaltlensresults} for validation of analysis choices for the \maglim sample. The results on data are presented in \S\ref{sec:results}, and we conclude in \S\ref{sec:conclusions}. 

%\section{Summary Statistics and theory}
\section{Theoretical model}
\label{sec:stat_theory}
\subsection{Two-point correlations}
Here we describe the hybrid perturbation theory (PT) model used to make theoretical predictions for the two-point statistics $w(\theta)$ and $\gamma_t(\theta)$.

\subsubsection{Power spectrum}
\label{sec:Pk_pred}

To compute the two-point projected statistics $\wtheta$ and $\gammat$, we first describe our methodology of predicting galaxy-galaxy and galaxy-matter power spectra ($\pgg$ and $\pgm$ respectively). PT provides a framework to describe the distribution of biased tracers of the underlying dark matter field in quasi-linear and linear scales. This framework allows for an order-by-order controlled expansion of the overdensity of biased tracer (here galaxies) in terms of the overdensity of the dark matter field where successively higher-order non-linearities dominate only in successively smaller-scale modes. We will analyze two PT models in this analysis, an hybrid linear bias model (that is complete only at first order) and an hybrid one--loop PT model (that is complete up to third order). 

For the linear bias model, we can write the galaxy-matter cross spectrum as $P_{\mathrm{gm}}(k) = b_1 P_{\mathrm{mm}}$ and auto-power spectrum of the galaxies as $P_{\mathrm{gg}}(k) = b_1^2 P_{\mathrm{mm}}(k)$. Here $b_1$ is the linear bias parameter and $P_{\mathrm{mm}}(k)$ is the \textit{non-linear} power spectrum of the matter field. We use the non-linear matter power spectrum prediction from \citet{Takahashi:2012em} to model $P_{\mathrm{mm}}(k)$ (referred to as \textsc{Halofit} hereafter). We use the  \citet{Bird_halofit} prescription to model the impact of massive neutrinos in this \textsc{Halofit} fitting formula. We refer the reader to \citet{y3-generalmethods} for robustness of our results despite the limitations of these modeling choices (c.f. \cite{Mead_21} for an alternative modeling scheme).

In the hybrid one--loop PT model used here, $P_{\mathrm{gm}}$ and $P_{\mathrm{gg}}$ can be expressed as:
% % \begin{linenomath*}
% \begin{align}\label{eq:P_gg_gm}
%     P_{\mathrm{gm}}(k, z) &= b_1 P_{\mathrm{mm}}(k, z) +  \frac{1}{2} b_2 P_{\rm b_1 b_2}(k, z) + \frac{1}{2} b_{\mathrm{s}} P_{\rm b_1 s^2}(k, z) \nonumber  \\
%     & + \frac{1}{2} b_{\rm 3nl}P_{\rm b_1 b_{\rm 3nl}}(k, z)+ b_{\rm k} k^2 P_{\mathrm{mm}}(k, z)  \\
%     P_{\mathrm{gg}}(k, z) &= b_1^2 P_{\mathrm{mm}}(k, z) + b_1 b_2 P_{\rm b_1 b_2}(k, z) + b_1 b_{\mathrm{s}}P_{\rm b_1 s^2}(k, z) \nonumber \\ 
%     & + b_1b_{\rm 3nl} P_{\rm b_1 b_{\rm 3nl} }(k, z) + \frac{1}{4}b_2^2 P_{\rm b_2 b_2}(k, z)  \nonumber \\
%     &  + \frac{1}{2} b_2 b_{\mathrm{s}}P_{\rm b_2 s^2}(k, z) + \frac{1}{4} b^2_{\mathrm{s}} P_{\rm s^2 s^2}(k, z) + 2 b_1 b_{\rm k} k^2 P_{\mathrm{mm}}(k, z).  
% \end{align}
% % \end{linenomath*}

% \begin{linenomath*}
\begin{widetext}
\begin{align}\label{eq:P_gg_gm}
    P_{\mathrm{gm}}(k, z) &= b_1 P_{\mathrm{mm}}(k, z) +  \frac{1}{2} b_2 P_{\rm b_1 b_2}(k, z) + \frac{1}{2} b_{\mathrm{s}} P_{\rm b_1 s^2}(k, z) + \frac{1}{2} b_{\rm 3nl}P_{\rm b_1 b_{\rm 3nl}}(k, z)+ b_{\rm k} k^2 P_{\mathrm{mm}}(k, z)  \\
    P_{\mathrm{gg}}(k, z) &= b_1^2 P_{\mathrm{mm}}(k, z) + b_1 b_2 P_{\rm b_1 b_2}(k, z) + b_1 b_{\mathrm{s}}P_{\rm b_1 s^2}(k, z) + b_1b_{\rm 3nl} P_{\rm b_1 b_{\rm 3nl} }(k, z) + \frac{1}{4}b_2^2 P_{\rm b_2 b_2}(k, z)  \nonumber \\
    &  + \frac{1}{2} b_2 b_{\mathrm{s}}P_{\rm b_2 s^2}(k, z) + \frac{1}{4} b^2_{\mathrm{s}} P_{\rm s^2 s^2}(k, z) + 2 b_1 b_{\rm k} k^2 P_{\mathrm{mm}}(k, z).  
\end{align}
% \end{linenomath*}
\end{widetext}

Here the parameters $ b_1 $, $ b_2 $, $ b_{\mathrm{s}}, b_{\rm 3nl} $ and $ b_{\rm k} $ are the renormalized bias parameters \citep{McDonald2009}. The kernels $P_{\rm b_1 b_2}$, $P_{\rm b_1 s^2}$, $P_{\rm b_1 b_{\rm 3nl}}$, $P_{\rm b_2 b_2}$, $P_{\rm b_2 s^2}$ and $P_{\rm s^2 s^2}$ are described in \citet{Saito2014a} and are calculable from the linear matter power spectrum. 
We validated this model in \citet{p2020perturbation} using 3D correlation functions, $\xigg$ and $\xigm$, of \redmagic galaxies measured in DES-like \mice simulations \citep{Fosalba_2014, Crocce_2015, Fosalba_2015}. These configuration space statistics are the Fourier transforms of the power spectra mentioned above. We found this model to describe the high signal-to-noise 3D measurements on the simulations above scales of 4 Mpc/$h$ and redshift $z < 1$ with a reduced $\chi^2$ consistent with one. Our tests also showed that at the projected precision of this analysis, two of the nonlinear bias parameters ($ b_{\mathrm{s}} $ and $ b_{\rm 3nl} $) can be fixed to their co-evolution values given by $b_{\rm s} = (-4/7) (b_1 - 1)$ and $b_{\rm 3nl} = (b_1 - 1)$; while $b_{\rm k}$ can be fixed to zero. We will use this result as our \textit{fiducial} modeling choice for the one--loop PT model. 

We note that there are alternative ways of modeling the scale-dependent non-linear biasing of galaxies. For example, \citet{Simon:2018} used a template fitting procedure using the Millennium simulation suite \citep{Springel_2005}, using a semi-analytic galaxy model \citep{Henriques:2015} and obtaining a significantly higher resolution of inter-halo physics. However, in order to validate the bias model on the scales above 4Mpc/$h$ on mock catalogs, the simulation volume (to minimize cosmic variance) and realistic galaxy selection are more important than resolution within individual halos. Due to the larger volume of the $\buzzard$ flock compared to the Millennium simulation, and detailed DES galaxy selection modeling in $\buzzard$ and $\mice$, we believe that the bias modeling validation performed in \citet{p2020perturbation} (and in \S~\ref{sec:sims}) is more direct (not affected by model imperfections of an intermediate fitting function), more stringent (due to lower cosmic variance) and more specific (due to DES-specific galaxy selection). Finally, as described in \citet{Goldstein:2021}, our two-parameter model also fits the 3D correlations measurements between matter and galaxy catalogs at various limiting magnitudes at 2\% level, in both configuration and Fourier spaces in a high-resolution simulation suite.

\subsubsection{Angular correlations} \label{sec:proj_2pt}
In order to calculate our observables $\wtheta$ and $\gammat$, we project the 3D power spectra described above to angular space. 
The projected galaxy clustering and galaxy-galaxy lensing angular power spectra of tomography bins $i,j$ are given by:
% \begin{linenomath*}
\begin{align}\label{eq:Cl_exact}
    C^{ij}_{AB}(\ell) &= \frac{2}{\pi} \int d\chi_1 W^{\rm i}_{A}(\chi_1) \int d\chi_2 W^{\rm j}_{B}(\chi_2) \nonumber \\
    &\mathrel{\phantom{=}} \int dk \ k^2 \ P_{ AB}[k,z(\chi_1),z(\chi_2)] j_{\ell}(k \chi_1) j_{\ell}(k \chi_2)\,,
\end{align}
% \end{linenomath*}
where, $AB = \rm{gg}$ models galaxy clustering and $AB={\rm g\kappa}$, where $\kappa$ denotes the convergence field, models galaxy-galaxy lensing. Here $W^{i}_{\rm g}(\chi) = n^{i}_g (z(\chi))dz/d\chi$ is the normalized radial selection function of lens galaxies for tomographic bin $i$, and $W^{\rm i}_{\kappa}$ is the tomographic lensing efficiency of the source sample
% \begin{linenomath*}
\begin{equation}
    W^{i}_{\rm \kappa} (\chi)= \frac{3\Omega_{\rm m} H_0^2}{2} \int_{\chi}^{\infty} d\chi' n'_{\rm s} [z(\chi')]\frac{\chi}{a(\chi)}\frac{\chi' - \chi}{\chi'}\,,
\end{equation}
% \end{linenomath*}
with $n_{\rm{g/s}}^i(z)$ the normalized redshift distribution of the lens/source galaxies in tomography bin $i$. 
For the galaxy-galaxy lensing observable, we use the Limber approximation \citep{Limber:53, LoVerde:2008re} which simplifies the Eq.~\ref{eq:Cl_exact} to
% \begin{linenomath*}
\begin{equation}\label{eq:Cl_limber}
    C^{ij}_{\rm g\kappa}(\ell)  = \int d\chi \frac{W^{i}_{\rm g}(\chi) W^{j}_{\rm \kappa}(\chi)}{\chi^2} P_{\rm g\kappa}\bigg[k=\frac{l + 1/2}{\chi},z(\chi)\bigg]\,.
\end{equation}
% \end{linenomath*}
In the absence of other modeling ingredients that are described in the next section, we have $C^{ij}_{\rm g\kappa}(\ell) \equiv C^{ij}_{\rm gm}(\ell)$ (similarly $P_{\rm g\kappa} \equiv P_{\rm gm}$). As described in \citet{Fang_nonlimber}, even at the accuracy beyond this analysis, it is sufficient to use  the Limber approximation for the galaxy-galaxy lensing observable, while for galaxy clustering this may cause significant cosmological parameter biases. 

To evaluate galaxy clustering statistics using Eq.~\ref{eq:Cl_exact}, we split the predictions into small and large scales. The non-Limber correction is only significant on large scales where non-linear contributions to the matter power spectra as well as galaxy biasing are sub-dominant. Therefore we use the Limber approximation for the small-scale non-linear corrections and use non-Limber corrections strictly on large scales using linear theory. Schematically, i.e., ignoring contributions from redshift-space distortions and lens magnification \cite[see][for details]{y3-generalmethods}, the galaxy clustering angular power spectrum between tomographic bins $i$ and $j$ is given by:
% \begin{linenomath*}
\begin{widetext}
\begin{align}
    &C_{\rm gg}^{ij} (\ell) \nonumber\\
    &= \int d\chi\, \frac{W_{\rm g}^i(\chi)W_{\rm g}^j(\chi)}{\chi^2} \left[P_{\rm gg}\left(\frac{\ell+0.5}{\chi},\chi\right)- b_1^{i} b_1^{j} P_{\rm lin}\left(\frac{\ell+0.5}{\chi},\chi\right)\right]\nonumber\\
    &+\frac{2}{\pi}\int d\chi_1\,b_{1}^i W_{\rm g}^i(\chi_1) D[z(\chi_1)]\int d\chi_2\,b_{1}^j W_{\rm g}^j(\chi_2)D[z(\chi_2)] \int\frac{dk}{k}k^3 P_{\rm lin}(k,0)j_\ell(k\chi_1)j_\ell(k\chi_2)\,,
\label{eq:Cl-DD_rewrite}
\end{align}
% \end{linenomath*}
\end{widetext}
where $D(z(\chi)$) is the growth factor, and $P_{\rm lin}$ is the linear matter power spectrum. The full model of galaxy clustering, including the contributions from other modeling ingredients like redshift-space distortions and lens magnification that we describe below, is detailed in \citet{Fang_nonlimber} and \citet{y3-generalmethods}. 

The real-space projected statistics of interest can be obtained from these angular correlations via:
% \begin{linenomath*}
\begin{align}\label{eq:2pt_exact}
    w^{ij}(\theta) &= \sum \frac{2\ell + 1}{4\pi} \overline{P_{\ell}}[\cos(\theta)] \ C^{ij}_{\rm gg}(\ell) \\
    \gamma_{\rm t}^{ij}(\theta) &= \sum \frac{2\ell + 1}{4\pi \ell (\ell + 1)} \overline{P_{\ell}^2}[\cos(\theta)] \ C^{ij}_{\rm g\kappa}(\ell)
\end{align}
% \end{linenomath*}
where $\overline{P_{\ell}}$ and $\overline{P_{\ell}^2}$ are bin-averaged Legendre Polynomials (see \citet{y3-covariances} for exact expressions). 

\subsection{The rest of the model}
\label{sec:model_rest}

To describe the statistics measured from data, we have to model various other physical phenomena that contribute to the signal to obtain unbiased inferences. In this section, we describe the leading sources of these modeling systematics. We have also validated in \citet{y3-generalmethods} that higher-order corrections do not bias our results. 

\subsubsection{Intrinsic Alignment} 
Galaxy-galaxy lensing aims to isolate the percent-level coherent shape distortions, or shear, of background source galaxies due to the gravitational potential of foreground lens galaxies. The local environment, however, including the gravitational tidal field, can also impact the intrinsic shapes of source galaxies and contribute to the measured shear signal. This interaction between the source galaxies and their local environment, generally known as ``intrinsic alignments'' (IA) is non-random. When there is a non-zero overlap between the source and lens redshift distributions, IA can have a non-zero contribution to the galaxy-galaxy lensing signal. To account for this effect, we model IAs using the ``tidal alignment and tidal torquing'' (TATT) model \citep{Blazek_2019}. Ignoring higher-order effects, such as lens magnification (see \citep{y3-gglensing,y3-2x2ptmagnification}), IA contributes to the galaxy-shear angular power spectra through the correlation of lens density and the $E$-mode component of intrinsic source shapes: $C^{ij}_{\rm g\kappa}(\ell) \to C^{ij}_{\rm g\kappa}(\ell) + C^{ij}_{\rm gI_{\rm E}}(\ell)$. The $C^{ij}_{\rm gI_{\rm E}}(\ell)$ term is detailed in \citet*{y3-generalmethods}, \citet*{y3-cosmicshear2}, \citet*{y3-gglensing}, and \citet{Blazek_2019}. Within our implementation of the TATT framework, $C^{ij}_{\rm gI_{\rm E}}(\ell)$ for all tomographic bin combinations $i$ and $j$ can be expressed using five IA parameters --- $a_1$ and $a_2$ (normalization of linear and quadratic alignments); $\alpha_1$ and $\alpha_2$ (their respective redshift evolution); and $b_{\rm ta}$ (normalization of a density-weighting term) --- and the linear lens galaxy bias. Therefore this model captures higher order contributions to the intrinsic alignment of source galaxies as compared to the simpler non-linear linear alignment (NLA) model that was used in the DES Y1 analysis \citep{Bridle_King_07,Hirata_IA,Krause2017,Abbott_2018}. In principle, there are also contributions at one-loop order in PT involving the non-linear galaxy bias and non-linear IA terms. However, in this analysis, we neglect these terms as we expect them to be subdominant, and they can be largely captured through the free $b_{\rm ta}$ parameter (see \citet{Blazek_2015} for further discussion of these terms). 

\subsubsection{Magnification}
All the matter between the observed galaxy and the observer acts as a gravitational lens. Hence, the galaxies get magnified, increasing the size of galaxy images (parameterized by the magnification factor, $\mu$) and increasing their total flux. The galaxy magnification decreases the observed number density due to stretching of the local sky, whereas increasing the total flux results in an increase in number density (as intrinsically fainter galaxies, which are more numerous, can be observed). This changes the galaxy-galaxy angular power spectrum to: $C^{ij}_{\rm gg}(\ell) \to C^{ij}_{\rm gg}(\ell) + 2C^{ij}_{\rm \mu g}(\ell) + C^{ij}_{\rm \mu \mu}(\ell) $ and the galaxy-shear angular power spectrum to $C^{ij}_{\rm g\kappa}(\ell) \to C^{ij}_{\rm g\kappa}(\ell) + C^{ij}_{\rm \mu I_{\rm E}}(\ell) + C^{ij}_{\rm \mu \kappa}(\ell)$. The auto and cross-power spectra with magnification are again given by Eq.~\ref{eq:Cl_exact}. For example, $C^{ij}_{\rm \mu g}(\ell) = 2(\mu^i - 1) C^{ij}_{\rm g\kappa}(\ell)$, where, as described below, we fix $\mu^i$ for the five tomographic bins to $[1.31,-0.52,0.34,2.25,1.97]$. We refer the reader to \citet{y3-generalmethods} for the detailed description of the equations for each of the power spectra.

The magnification coefficients are computed with the \balrog\ image simulations \citep{Suchyta_2016,y3-balrog} in a process described in \citet{y3-2x2ptmagnification}. Galaxy profiles are drawn from the DES deep fields \citep{y3-deepfields} and injected into real DES images \citep{Morganson_2018}. The full photometry pipeline \citep{y3-gold} and \redmagic\ sample selection are applied to the new images to produce a simulated \redmagic\ sample with the same selection effects as the real data. To compute the impact of magnification, the process is repeated, this time applying a constant magnification to each injected galaxy. The magnification coefficients are then derived from the fractional increase in number density when magnification is applied. This method captures both the impact of magnification on the galaxy magnitudes and the galaxy sizes, including all numerous sample selection effects. A similar procedure is repeated to estimate the magnification coefficients for the \maglim sample. We refer the reader to \citet*{y3-2x2ptmagnification} for further details about the impact of magnification on our observable and their constraints from data. 

\subsubsection{Non-locality of galaxy-galaxy lensing}  \label{sec:pm_theory}
The configuration-space estimate of the galaxy-galaxy lensing signal is a
non-local statistic. The galaxy-galaxy lensing
signal of source galaxy at redshift $z_{\rm s}$ by the matter around
galaxy at redshift $z_{\rm l}$ at transverse
distance $R$ is related to the mass density of matter around lens
galaxy by:
% \begin{linenomath*}
\begin{equation}\label{eq:gt_theory}
    \gamma_{\rm t}(R;z_{\rm g},z_{\rm s}) = \frac{\Delta \Sigma (R;z_{\rm g})}{\Sigma_{\rm crit} (z_{\rm g},z_{\rm s})},
\end{equation}
% \end{linenomath*}
where,  $\Sigma_{\rm crit}$ is the critical surface mass density given by :
\begin{equation}
\Sigma_{\rm crit} (z_{\rm g}, z_{\rm s}) = \frac{c^2}{4\pi G} \frac{D_{\mathrm{A}}(z_{\mathrm{s}})}{D_{\mathrm{A}}(z_{\rm g})D_{\mathrm{A}}(z_{\mathrm{g}},z_{\mathrm{s}})}\,.
\end{equation}
Here $D_{\mathrm{A}}$ is the angular diameter distance, $z_{\mathrm{l}}$ is the redshift of the lens and $z_{\mathrm{s}}$ is the redshift of the source. In Eq.~\ref{eq:gt_theory}, $\Delta \Sigma(R;z_{\rm g}) = \bar{\Sigma}(0,R; z_{\rm g}) - \Sigma(R;z_{\rm g})$ and $\Sigma(R;z_{\rm g})$ is the surface mass density at a transverse separation $R$ from the lens and $\bar{\Sigma}(0,R)$ is the average surface mass density within a separation $R$ from that lens. Through the $\bar{\Sigma}(0,R)$ term, $\gamma_{\rm t}$  at any scale $R$ is dependent on the mass distribution at all scales less than $R$. This makes $\gamma_{\rm t}$  highly non-local, and any model that is valid only on large scales above some $r_{\rm min}$ will break down more rapidly than for a more local statistic like \wtheta. However, as the dependence on small scales is through the \textit{mean} surface mass density, the impact of the mass distribution inside $r_{\rm min}$ on $\gammat$ can be written as:
% \begin{linenomath*}
\begin{equation}
    \gamma_{\rm t}(R;z_{\rm g},z_{\rm s}) = \frac{1}{\Sigma_{\rm crit}(z_{\rm g},z_{\rm s})} \bigg[\Delta \Sigma_{\rm model}(z_{\rm g}) + \frac{B(z_{\rm g})}{R^2} \bigg],
\end{equation}
% \end{linenomath*}
where $\Delta \Sigma^{\rm model}$ is the prediction from a model (which is given by PT here) that is valid on scales above $r_{\rm min}$ (also see \citep{Baldauf_2010}). Here, $B$ is the effective total residual mass below $r_{\rm min}$ and is known as the point mass (PM) parameter. In this analysis we use the thin redshift bin approximation 
% \gary{[just the lens bin needs to be thin?]}\SP{for this step, i think yes} 
(see Appendix~\ref{app:pm} for details of this validation) and hence the average $\gamma_{\rm t}$ signal between lens bin $i$ and source bin $j$ can be written as:
% \begin{linenomath*}
\begin{equation}
    \gamma^{ij}_{{\rm t}} = \gamma^{ij}_{{\rm t, model}} + G^{ij}/\theta^2,
\end{equation}
% \end{linenomath*}
where,

% \begin{linenomath*}
\begin{equation}\label{eq:pm_Cij}
    G^{ij} = B^i \, \int dz_{\rm g} \ dz_{\rm s} \ n^{i}_{{\rm g}} \ n^{j}_{{\rm s}} \ \Sigma^{-1}_{\rm crit}(z_{\rm g},z_{\rm s}) \ \chi^{-2}(z_{\rm g}) \equiv B^i \, \beta^{ij} \,.
\end{equation}
% \end{linenomath*}
Here $B^i$ is the PM for lens bin $i$, $n^{i}_{{\rm g}}$ is the redshift distribution of lens galaxies for tomographic bin $i$, $n^{j}_{{\rm s}}$ is the redshift distribution of source galaxies for tomographic bin $j$. 

However, instead of directly sampling over the parameters $B^i$ for each tomographic bin, we implement an analytic marginalization scheme as described in \citet{MacCrann:2019ntb}. We modify our inverse-covariance when calculating the likelihood as described in \S\ref{sec:cov_pm}. We note that this scheme of adding and marginalizing over the PM parameters is equivalent to alternative procedures \citep{Park_21, Asgari_2020} for mitigating the impact of unmodeled non-linear small scale physics to the large scales. 

\section{Data description}
\label{sec:data}
\subsection{DES Y3}

The full DES survey was completed in 2019 using the Cerro Tololo Inter-American Observatory (CTIO) 4-m Blanco telescope in Chile and covered approximately 5000 square degrees of the South Galactic Cap. This 570-megapixel Dark Energy Camera \citep{Flaugher15} images the field in five broadband filters, \textit{grizY}, which span the wavelength range from approximately 400nm to 1060nm. The raw images are processed by the DES Data Management team \citep{Sevilla11, Morganson18} and after a detailed object selection criteria on the first three years of imaging data (detailed in \citet{Abbott_2018}), the Y3 \gold data set containing 400 million sources is constructed (single-epoch and coadd images are available\footnote{https://des.ncsa.illinois.edu/releases/dr1} as Data Release 1). We further process this \gold data set to obtain the lens and source catalogs described in the following sub-sections.

\subsubsection{\redmagic lens galaxy sample}
\label{sec:redmagic_def}
The principal lens sample used in this analysis is selected with the \redmagic algorithm \citep{Rozo_2016} run on DES Year 3 data. \redmagic selects Luminous Red Galaxies (LRGs) according to the magnitude-color-redshift relation of red-sequence galaxies, calibrated using an overlapping spectroscopic sample. This procedure is based on selecting galaxies above a threshold luminosity that fit (using $\chi^2_{\rm RM}$ as goodness-of-fit criteria) this \redmagic template of magnitude-color-redshift relation to a threshold better than $\chi^2_{\rm RM} < \chi^2_{\rm max}$. The value of $\chi^2_{\rm max}$ is chosen such that the sample has a constant co-moving space density and is typically less than 3. 
% This sample has a threshold luminosity and constant co-moving density. 
The full \redmagic algorithm is described in \citet{Rozo_2016}, and after application of this algorithm to DES Y3 data, we have approximately 2.6 million galaxies.
% \SP{Modify to include the $\chi^2$ definition and the fiducial value. Also maybe if any modifications in the red galaxy template compared to the \citet{Rozo_2016}?} 

\citet*{y3-galaxyclustering} found that the \redmagic number density fluctuates with several observational properties of the survey, which imprints a non-cosmological bias into the galaxy clustering. To account for this we assign a weight to each galaxy, which corresponds to the inverse of the angular selection function at that galaxy's location. The computation and validation of these weights are described in \citet{y3-galaxyclustering}.

\subsubsection{\maglim lens galaxy sample}

DES cosmological constraints are also derived using a second
lens sample, $\maglim$, selected by applying the criterion  $i < 4z+18$ to the \gold catalog, where $z$ is the photometric redshift estimate given by the Directional Neighbourhood Fitting ($\texttt{DNF}$) algorithm \citep{DNF2016}.
This selection is shown by \citet*{y3-2x2maglimforecast} to be optimal in terms of its 2$\times$2pt cosmological constraints.
We additionally apply a lower magnitude cut, $i>17.5$, to remove contamination from bright objects. The resulting sample has about 10.7 million galaxies. 

Similarly to \texttt{redMaGiC}, we correct the impact of observational systematics on the \maglim galaxy clustering by assigning a weight to each galaxy, as described and validated in \citet{y3-galaxyclustering}. This sample is then used in \citet*{y3-2x2ptaltlensresults} to obtain cosmological constraints from the combination of galaxy clustering and galaxy-galaxy lensing from DES Y3 data. We refer to \citet{y3-2x2ptaltlensresults} for a detailed description of the sample and its validation.

\subsubsection{Source galaxy shape catalog}

To estimate the weak lensing shear of the observed source galaxies, we use the \metacal algorithm \citep{Sheldon_2017, huff2017metacalibration}. This method estimates the response of a shear estimator to artificially sheared galaxy images and incorporates improvements like better PSF estimation \citep{y3-piff}, better astrometric methods \citep{y3-gold} and inclusion of inverse variance weighting. The details of the method applied to our galaxy sample are presented in \citet{y3-shapecatalog}. This methodology does not capture the object-blending effects and shear-dependent detection biases and we use image simulations to calibrate this bias as detailed in \citet{y3-imagesims}. The galaxies that pass the selection cuts designed to reduce systematic biases (as detailed in \citet*{y3-shapecatalog}) are used to make our source sample shape catalog. This catalog consists of approximately 100 million galaxies with effective number density of $n_{\rm eff} = 5.6$ galaxies per ${\rm arcmin}^2$ and an effective shape noise of $\sigma_{\rm e} = 0.26$.

\subsection{$\buzzard$ Simulations}

% \subsubsection{$\buzzard$ sims}
The \buzzard\ simulations are $N$-body lightcone simulations that have been populated with galaxies using the \textsc{Addgals} algorithm \citep{Wechsler2021}, endowing each galaxy with positions, velocities, spectral energy distributions, broad-band photometry, half-light radii and ellipticities. In order to build a lightcone that spans the entire redshift range covered by DES Y3 galaxies, we combine three lightcones constructed from simulations with box sizes of $1.05,\, 2.6 \textrm{ and } 4.0\, (h^{-3}\, \rm Gpc^3)$, mass resolutions of $3.3\times10^{10},\, 1.6\times10^{11},\, 5.9\times10^{11}\, h^{-1}M_{\odot}$, spanning redshift ranges $0.0< z \leq 0.32$, $0.32< z \leq 0.84$ and $0.84< z \leq 2.35$ respectively. Together these produce $10,000$ square degrees of unique lightcone. The lightcones are run with the \textsc{L-Gadget2} $N$-body code, a memory optimized version of \textsc{Gadget2} \citep{Springel_2005}, with initial conditions generated using \textsc{2LPTIC} at $z=50$ \citep{Crocce2012}. From each $10,000$ square degree catalog, we can create two DES Y3 footprints.

The \textsc{Addgals} model uses the relationship, $P(\delta_{R}|M_r)$, between a local density proxy, $\delta_{R}$, and absolute magnitude $M_r$ measured from a high-resolution subhalo abundance matching (SHAM) model in order to populate galaxies into these lightcone simulations. The \textsc{Addgals} model reproduces the absolute--magnitude--dependent clustering of the SHAM. 
Additionally, we employ a conditional abundance matching (CAM) model, assigning redder SEDs to galaxies that are closer to massive dark matter halos, in a manner that allows us to reproduce the color-dependent clustering measured in the Sloan Digital Sky Survey Main Galaxy Sample (SDSS MGS) \citep{Wechsler2021, DeRose2021}. 

These simulations are ray-traced using the spherical-harmonic transform (SHT) configuration of \textsc{Calclens}, where the SHTs are performed on an $N_{\rm side}=8192$ \textsc{HealPix} grid \citep{Becker2013}. The lensing distortion tensor is computed at each galaxy position and is used to deflect the galaxy angular positions, apply shear to galaxy intrinsic ellipticities, including effects of reduced shear, and magnify galaxy shapes and photometry. We have conducted convergence tests of this algorithm and found that resolution effects are negligible on the scales used for this analysis \citep{DeRose2019}.

Once the simulations have been ray-traced, we apply DES Y3-specific masking and photometric errors. To mask the simulations, we employ the Y3 footprint mask but do not apply the bad region mask \citep{y3-gold}, resulting in a footprint with an area of 4143.17 square degrees. Each set of three $N$-body simulations yields two Y3 footprints that contain 520 square degrees of overlap. In total, we use 18 \buzzard realizations in this analysis. 

We apply a photometric error model to simulate wide-field photometric errors in our simulations. To select a lens galaxy sample, we run the \redmagic\ galaxy selection on our simulations using the same configuration as used in the Y3 data, as described in \citet*{y3-galaxyclustering}. A weak lensing source selection is applied to the simulations using PSF-convolved sizes and $i$-band SNR to match the non-tomographic source number density, 5.9 $\textrm{arcmin}^{-2}$, from the \metacal\ source catalog. This matching was performed using a slightly preliminary version of the \metacal\ catalog, so this number density is slightly different from the final \metacal\ catalog that is used in our DES Y3 analyses. We employ the \textit{fiducial} redshift estimation framework (see \S\ref{sec:sourcez}) to our simulations in order to place galaxies into four source redshift bins with number densities of 1.46 $\rm arcmin^{-2}$ each. Once binned, we match the shape noise of the simulations to that measured in the \metacal\ catalog per tomographic bin, yielding shape noise values of $\sigma_{e}=[0.247, 0.266, 0.263, 0.314]$.

Two-point functions are measured in the \buzzard\ simulations using the same pipeline used for the DES Y3 data, where we set \metacal\ responses and inverse variance weights equal to 1 for all galaxies, as these are not assigned in our simulation framework. We have opted to make measurements without shape noise in order to reduce the variance in the simulated analyses using these measurements. Lens galaxy weights are produced in a manner similar to that done in the data and applied to measure our clustering and lensing signals. The clustering and galaxy-galaxy lensing predictions match the DES \redmagic\ measurements to $10-20\%$ accuracy over most scales and tomographic bins, except for the first lens bin, which disagrees by $50\%$ in \wtheta. We refer the reader to Fig. 4 in \citet*{y3-simvalidation} for a more detailed comparison.

\subsection{Tomography and measurements}
In this section we detail the estimation of the photometric redshift distribution of our source galaxy sample and two lens galaxy samples. These three samples are qualitatively different and have different redshift attributes, requiring different redshift calibration methods detailed below.

\subsubsection{\redmagic redshift methodology}
\label{sec:lensz}
We split the \redmagic sample into $N_{\rm z,g} = 5$ tomographic bins, selected on the \redmagic redshift point estimate quantity ZREDMAGIC. The bin edges used are $z=0.15, 0.35, 0.50, 0.65, 0.80, 0.90$. The first three bins use a luminosity threshold of $L_{\min} > 0.5 L_{*}$ and are known as the high-density sample. The last two redshift bins use a luminosity threshold of $L_{\min} > 1.0 L_{*}$ and are known as the high-luminosity sample. The galaxy number densities (in the units of ${\rm arcmin}^{-2}$) for the five tomographic bins are $\langle n_{\rm g} \rangle = 0.022,0.038,0.059,0.03,0.025$.

The redshift distributions are computed by stacking four samples from the PDF of each \redmagic galaxy, allowing for non-Gaussianity of the PDF. We find an average individual redshift uncertainty of $\sigma_z/(1+z) < 0.0126$ in the redshift range used from the variance of these samples. We refer the reader to \citet{Rozo_2016} for more details on the algorithm of redshift assignment for \redmagic galaxies and to \citet{y3-lenswz} for more details on the calibration of redshift distribution of the Y3 \redmagic sample.  

\subsubsection{\maglim redshift methodology}

We use $\texttt{DNF}$ \citep{DNF2016} for splitting the $\maglim$ sample into tomographic bins and estimating the redshift distributions. $\texttt{DNF}$  uses a training set from a spectroscopic database as reference, and then provides an estimate of the redshift of the object through a nearest-neighbors fit in a hyperplane in color and magnitude space.

We split the $\maglim$ sample into $N_{\rm z,g} = 6$ tomographic bins from $z=0.2$ and $z=1.05$, selected using the $\texttt{DNF}$ photometric redshift estimate. The bin edges are $[0.20, 0.40, 0.55, 0.70, 0.85, 0.95, 1.05]$. The galaxy number densities (in the units of ${\rm arcmin}^{-2}$) for the six tomographic bins of this sample are $\langle n_{\rm g} \rangle = 0.15,0.107,0.109,0.146,0.106,0.1$. The redshift distributions in each bin are then computed by stacking the $\texttt{DNF}$ PDF estimates of each $\maglim$ galaxy. See \citet{y3-2x2ptaltlensresults} for a more comprehensive description and validation of this methodology and \citet{y3-2x2ptaltlenssompz} for estimation of redshift distributions of this sample using the same methodology as used for source galaxies that is described below. 

\subsubsection{Source redshift methodology}
\label{sec:sourcez}
The description of the tomographic bins of source samples and the methodology for calibrating their photometric redshift distributions are summarized in \citet*{y3-sompz}. Overall, the redshift calibration methodology involves the use of self-organizing maps \citep{y3-sompz}, clustering redshifts \citep{y3-sourcewz} and shear-ratio \citep{y3-shearratio} information. The Self-Organizing Map Photometric Redshift (SOMPZ) methodology leverages additional photometric bands in the DES deep-field observations \citep{y3-deepfields} and the \balrog\ simulation software \citep{balrog_21} to characterize a mapping between color space and redshifts. This mapping is then used to provide redshift distribution samples in the wide field, after including the uncertainties from sample variance and galaxy flux measurements in a way that is not subject to selection biases. The clustering redshift methodology performs the calibration by analyzing cross-correlations between \redmagic and spectroscopic data from Baryon Acoustic Oscillation Survey (BOSS) and its extension (eBOSS). Candidate $n_{\rm s}(z)$ distributions are drawn from the posterior distribution defined by the combination of SOMPZ and clustering-redshift likelihoods.
These two approaches provide us the mean redshift distribution of source galaxies and uncertainty in this distribution. The shear-ratio calibration uses the ratios of small-scale galaxy-galaxy lensing data, which are largely independent of the cosmological parameters but help calibrate the uncertainties in the redshift distributions. We include it downstream in our analysis pipeline as an external likelihood, as briefly described in \S\ref{sec:shear_ratio} and detailed in \citet*{y3-shearratio}. 

\begin{figure}
\includegraphics[width=0.48\textwidth]{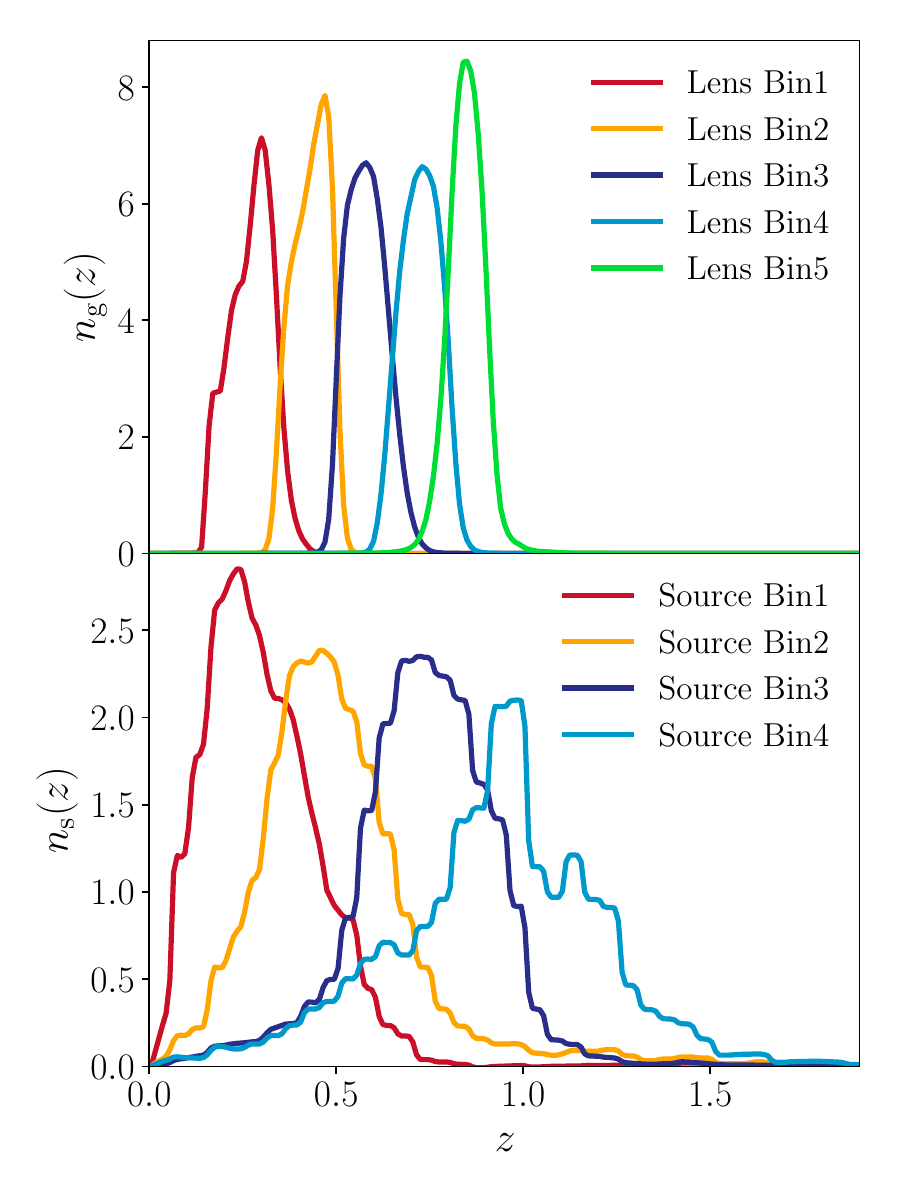}
\caption[]{Comparison of the normalized redshift distributions of various tomographic bins of the source galaxies and \redmagic lens galaxies in the data.}
\label{fig:nz_comp}
\end{figure}

Finally, we split the source catalog into $N_{\rm z,s} = 4$ tomographic bins. The mean redshift distribution of \redmagic lens galaxies and source galaxies are compared in Fig.~\ref{fig:nz_comp}. We refer the reader to \citet*{y3-2x2ptaltlensresults} for \maglim sample redshift distribution. 

\subsubsection{2pt measurements}\label{sec:2pt_data}

For galaxy clustering, we use the Landy-Szalay estimator \citep{Landy_Szalay} given as:
% \begin{linenomath*}
\begin{equation}
    w(\theta) = \frac{DD - 2DR + RR}{RR}
\end{equation}
% \end{linenomath*}
where $DD$, $DR$ and $RR$ are normalized weighted number counts of galaxy-galaxy, galaxy-random and random-random pairs within angular and tomographic bins. For lens tomographic bins, we measure the auto-correlations in $N_{\theta} = 20$ log-spaced angular bins ranging from 2.5 arcmin to 250 arcmin. Each lens galaxy in the catalog ($g_i$) is weighted with its systematic weight $w_{\rm g_i}$. This systematic weight aims to remove the large-scale fluctuations due to changing observing conditions at the telescope and Galactic foregrounds. Our catalog of randoms is 40 times larger than the galaxy catalog. The validation of this estimator and systematic weights of the lens galaxies is presented in \citet{y3-galaxyclustering}. In total we have $N_{\wtheta} = N_{\rm z,g} \times N_{\theta} = 100$ measured $\wtheta$ datapoints.

The galaxy-galaxy lensing estimator used in this analysis is given by:
% \begin{linenomath*}
\begin{equation}
    \gamma_{\rm t}(\theta) = \frac{\sum_k w_{\textrm{r}_k}}{\sum_i w_{\textrm{g}_i}} \frac{\sum_{ij} w_{\textrm{g}_i} w_{\textrm{s}_j} e^{\rm LS}_{{\rm t},ij}}{\sum_{kj} w_{\textrm{r}_k} w_{\textrm{s}_j}} - \frac{\sum_{kj} w_{\textrm{r}_k} w_{\textrm{s}_j} e^{\rm RS}_{{\rm t},kj}}{\sum_{kj} w_{\textrm{r}_k} w_{\textrm{s}_j}}
\end{equation}
% \end{linenomath*}
where $e^{\rm LS}_{{\rm t},ij}$ and $e^{\rm RS}_{{\rm t},kj}$ is the measured tangential ellipticity of source galaxy $j$ around lens galaxy $i$ and random point $k$ respectively. The weight $w_{\textrm{g}_i}$ is the systematic weight of lens galaxy as described above, $w_{\textrm{r}_k}$ is the weight of random point that we fix to 1 and $w_{\textrm{s}_j}$ is the weight of the source galaxy that is computed from inverse variance of the shear response weighted ellipticity of the galaxy (see \citet*{y3-shapecatalog} for details). 
This estimator has been detailed and validated in \citet{Singh_2017} and \citet*{y3-gglensing}. We measure this signal for each pair of lens and source tomographic bins and hence in total we have  $N_{\gammat} = N_{\rm z,g} \times N_{\rm z,s} \times N_{\theta} = 400$ measured $\gammat$ datapoints. 

We analyze both of these measured statistics jointly and hence we have in total $N_{\rm data} = N_{\wtheta} + N_{\gammat} = 500$ datapoints. Our measured signal to noise (SNR)\footnote{The SNR is calculated as $\sqrt{(\vec{\mathbfcal{D}}\, {\mathbfcal{C}}^{-1}\,  \vec{\mathbfcal{D}})}$, where $\vec{\mathbfcal{D}}$ is the data under consideration and $\mathbfcal{C}$ is its covariance.}, using \redmagic lens sample, of $\wtheta$ is 171 \citep{y3-galaxyclustering}, of  $\gammat$ is 121 \citep{y3-gglensing}; giving total joint total SNR of 196. In the \S\ref{sec:param_inf}, we describe and validate different sets of scale cuts for the linear bias model (angular scales corresponding to (8,6)Mpc/$h$ for $w(\theta),\gamma_{\rm t}(\theta)$ respectively) and the non-linear bias model ((4,4)Mpc/$h$). After applying these scale cuts, we obtain the joint SNR, that we analyze for cosmological constraints, as 81 for the linear bias model and 106 for the non-linear bias model.\footnote{Using a more optimal SNR estimator, SNR$= \frac{(\vec{\mathbfcal{D}}^{\rm data}\, {\mathbfcal{C}}^{-1}\,  \vec{\mathbfcal{D}}^{\rm model})}{\sqrt{(\vec{\mathbfcal{D}}^{\rm model}\, {\mathbfcal{C}}^{-1}\,  \vec{\mathbfcal{D}}^{\rm model})}}$, where $\vec{\mathbfcal{D}}^{\rm data}$ is the measured data and $\vec{\mathbfcal{D}}^{\rm model}$ is the bestfit model, we get SNR=79.5 for the linear bias model scale cuts of (8,6)Mpc/$h$.} 

\subsubsection{Shear ratios}\label{sec:shear_ratio}
As will be detailed in \S\ref{sec:scale_cuts}, in this analysis, we remove the small scales' non-linear information from the 2pt measurements that are presented in the above sub-section. However, as presented in \citet*{y3-shearratio}, the ratio of $\gammat$ measurements for the same lens bin but different source bins is well described by our model (see \S\ref{sec:stat_theory}) even on small scales. Therefore we include these ratios (referred to as shear-ratio henceforth) as an additional independent dataset in our likelihood. In this shear-ratio datavector, we use the angular scales above 2Mpc/$h$ and less than our \textit{fiducial} scale cuts for 2pt measurements described in \S\ref{sec:scale_cuts} (we also leave two datapoints between 2pt scale cuts and shear-ratio scale cuts to remove any potential correlations between the two). The details of the analysis choices for shear-ratio measurements and the corresponding covariance matrix are detailed in \citet*{y3-shearratio} and \citet*{y3-3x2ptkp}.

\subsection{Covariance}
\label{sec:cov}

In this analysis, the covariance between the statistic $\wtheta$ and $\gammat$ (${\mathbfcal{C}}$) is modeled as the sum of a Gaussian term ($\mathbfcal{C}_{\rm G}$), trispectrum term ($\mathbfcal{C}_{\rm NG}$) and super-sample covariance term ($\mathbfcal{C}_{\rm SSC}$). The analytic model used to describe ($\mathbfcal{C}_{\rm G}$) is described in \citet{y3-covariances}. The terms $\mathbfcal{C}_{\rm NG}$ and $\mathbfcal{C}_{\rm SSC}$ are modeled using a halo model framework as detailed in \citet{Krause:2016jvl} and \citet{Krause2017}. The covariance calculation has been performed using the CosmoCov package \citep{Fang:2020vhc}, and the robustness of this covariance matrix has been tested and detailed in \citet{y3-covariances}. We also account for two additional sources of uncertainties that are not included in our \textit{fiducial} model using the methodology of analytical marginalization \citep{Bridle_2002} as detailed below.

\subsubsection{Accounting for LSS systematics}

As described in \citet{y3-galaxyclustering}, we modify the $w(\theta)$ covariance to analytically marginalize over two sources of uncertainty in the correction of survey systematics: the choice of correction method, and the bias of the \textit{fiducial} method as measured on simulations. 

These systematics are modelled as 
\begin{equation}
w^{\prime}(\theta) = w(\theta) + A_{1} \Delta w_{\rm method}(\theta) + A_{2} w_{\rm r. \, s. \, bias}(\theta) \,,
\end{equation}

where $\Delta w_{\rm method}(\theta)$ is the difference between two systematics correction methods: Iterative Systematic Decontamination (\isd) and Elastic Net (\enet), and $w_{\rm r. \, s. \, bias}(\theta)$ is the residual systematic bias measured on Log-normal mocks. Both terms are presented in detail in \citet{y3-galaxyclustering}. Also note that here $A_{1}$ and $A_{2}$ are arbitrary amplitudes.  

We analytically marginalise over these terms assuming a unit Gaussian as the prior on the amplitudes $A_{1}$ and $A_{2}$. The measured difference is a $1\sigma$ deviation from the prior center. The final additional covariance term to be added to the \textit{fiducial} covariance is:

\begin{equation}
\Delta \mathbfcal{C} =  {\bf \Delta w_{\rm method}} {\bf \Delta w_{\rm method}}^{T} \ + \ {\bf w_{r. \, s. \, bias}} {\bf w_{r. \, s. \, bias}}^{T} \,.
\end{equation}

The systematic contribution to each tomographic bin is treated as independent so the covariance between lens bins is not modified. However, we verified that both 0\% correlation and 100\% correlation between the tomographic bins (hence bounding the likely effect) on a simulated analyses resulted in negligible differences in the cosmological parameter constraints.

\subsubsection{Point mass analytic marginalization}
\label{sec:cov_pm}
As mentioned in \S\ref{sec:pm_theory}, we modify the inverse covariance to perform analytic marginalization over the PM parameters. As detailed in \citet{MacCrann:2019ntb}, using the generalization of the Sherman-Morrison formula, this procedure changes our \textit{fiducial} inverse-covariance ${\mathbfcal{C}}^{-1}$ to ${\mathbfcal{C}}^{-1}_{\rm wPM}$ as follows:
% \begin{linenomath*}
\begin{equation}
    {\mathbfcal{C}}^{-1}_{\rm wPM} = {\mathbfcal{C}}^{-1} - {\mathbfcal{C}}^{-1} {\mathbfcal{U}} ({\mathbfcal{I}} + {\mathbfcal{U}}^{\rm T} {\mathbfcal{C}}^{-1} {\mathbfcal{U}})^{-1} {\mathbfcal{U}}^{\rm T} {\mathbfcal{C}}^{-1} \,.
\end{equation}
% \end{linenomath*}

Here ${\mathbfcal{C}}^{-1}$ is the inverse of the halo-model covariance as described above, $\mathbfcal{I}$ is the identity matrix and $\mathbfcal{U}$ is a $N_{\rm data} \times N_{\rm z,g}$ matrix where the $i$-th column is given by $\sigma_{B^i} \vec{t}^{i}$. Here $\sigma_{B^i}$ is the standard deviation of the  Gaussian prior on point mass parameter $B^i$ and $\vec{t}^{i}$ is given as:
% \begin{linenomath*}
\begin{equation}
    \bigg(\vec{t}^{i} \bigg)_{a} = \begin{cases}
0 & \parbox{5cm}{if $a$-th element does not correspond to $\gammat$ and if lens-redshift of $a$-th element $\neq i$} \\  
\\
\beta^{ij}\theta_{a}^{-2} &\text{otherwise}
\end{cases}
\end{equation}
% \end{linenomath*}
where the expression for $\beta^{ij}$ is shown in Eq.\ref{eq:pm_Cij}. We evaluate that term at fixed \textit{fiducial} cosmology as given in Table \ref{tab:params_all}. In our analysis we put a wide prior on PM parameters $B^i$ by choosing $\sigma_{B^i} = 10000$ which translates to the effective mass residual prior of $10^{17} M_{\odot}/h$ (see Eq.~\ref{eq:pm_halo}).

\subsection{Blinding and unblinding procedure}
We shield our results from observer bias by randomly shifting our results and datavector at various phases of the analysis \citep{y3-blinding}. This procedure prevents us from knowing the impact of any particular analysis choice on the inferred cosmological constraints from our data until all analysis choices have been made. This procedure, as well as the decision tree used to unblind, is detailed in \citet{y3-3x2ptkp}, which is also employed here. Therefore, all of our cosmology results acquired with fiducial galaxy samples described in this section are achieved using analysis choices that were validated prior to unblinding (see \S~\ref{sec:param_inf}). The results obtained by changing analysis choices (and with a different galaxy sample), after unblinding, are confined to \S~\ref{sec:bchi2_cosmo} and \S~\ref{sec:halomass} of the main article, and in the Appendix~\ref{app:bchi2}. 

\begin{table}[H]
\centering 
% \resizebox{\textwidth}{!}
\tabcolsep=0.11cm
\begin{tabular}{|c| c c c|}
\hline
% \hline
Model & Parameter & Prior & Fiducial  \\ \hline
% & & & \\
& \multicolumn{3}{c|}{\textbf{Cosmology}} \\ 

% & & & \\
\multirow{24}{*}{\shortstack[c]{Common\\ Parameters}} & $\Omega_{\rm m}$ & $\mathcal{U}[0.1, 0.9]$ & 0.3 \\
 & $A_s\times 10^{-9}$ & $\mathcal{U}[0.5, 5]$ & $2.19$\\
 
& $\Omega_{\rm b}$ & $\mathcal{U}[0.03, 0.07]$ & 0.048 \\

& $n_{\rm s}$ & $\mathcal{U}[0.87, 1.06]$ & 0.97\\

& $h$ & $\mathcal{U}[0.55, 0.91]$ &  0.69\\

& $\Omega_{\nu}h^2 \times 10^{-4}$ & $\mathcal{U}[6.0, 64.4]$ & 8.3 \\  
% & & & \\
\cline{2-4}
% & & & \\
& \multicolumn{3}{c|}{\textbf{Intrinsic Alignment}} \\ 

& $a_1$ & $\mathcal{U}[-5.0, 5.0]$ & 0.7\\
& $a_2$ & $\mathcal{U}[-5.0, 5.0]$ & -1.36\\
& $\alpha_1$ & $\mathcal{U}[-5.0, 5.0]$ & -1.7\\
& $\alpha_2$ & $\mathcal{U}[-5.0, 5.0]$ & -2.5\\
& $b_{\rm ta}$ & $\mathcal{U}[0.0, 2.0]$ & 1.0\\ 
% & & & \\
\cline{2-4}
% & & & \\
& \multicolumn{3}{c|}{\textbf{Lens photo-$z$}} \\  
& $\Delta z_{\rm g}^{1}$ & $\mathcal{G}[0.006, 0.004]$ & 0.0  \\ 
& $\Delta z_{\rm g}^{2}$ & $\mathcal{G}[0.001, 0.003]$ & 0.0  \\ 
& $\Delta z_{\rm g}^{3}$ & $\mathcal{G}[0.004, 0.003]$ & 0.0  \\ 
& $\Delta z_{\rm g}^{4}$ & $\mathcal{G}[-0.002, 0.005]$ & 0.0 \\ 
& $\Delta z_{\rm g}^{5}$ & $\mathcal{G}[-0.007, 0.01]$ & 0.0  \\ 
& $\sigma z_{\rm g}^{5}$ & $\mathcal{G}[1.23, 0.054]$ & 1.0  \\ 
% & & & \\
\cline{2-4}
% & & & \\
& \multicolumn{3}{c|}{\textbf{Shear Calibration}} \\ 
& \shortstack[c]{$m^{1}$}   & $\mathcal{G}[-0.0063, 0.0091]$ & 0.0 \\ 
& \shortstack[c]{$m^{2}$}   & $\mathcal{G}[-0.0198, 0.0078]$ & 0.0 \\ 
& \shortstack[c]{$m^{3}$}   & $\mathcal{G}[-0.0241, 0.0076]$ & 0.0 \\ 
& \shortstack[c]{$m^{4}$}   & $\mathcal{G}[-0.0369, 0.0076]$ & 0.0 \\ 
% & & & \\
\cline{2-4}
% & & & \\
& \multicolumn{3}{c|}{\textbf{Source photo-$z$}} \\  
& $\Delta z_{\rm s}^{1}$ & $\mathcal{G}[0.0, 0.018]$ & 0.0  \\ 
& $\Delta z_{\rm s}^{2}$ & $\mathcal{G}[0.0, 0.015]$ & 0.0  \\ 
& $\Delta z_{\rm s}^{3}$ & $\mathcal{G}[0.0, 0.011]$ & 0.0  \\ 
& $\Delta z_{\rm s}^{4}$ & $\mathcal{G}[0.0, 0.017]$ & 0.0 \\ 
\cline{2-4}
& \multicolumn{3}{c|}{\textbf{Point Mass}} \\ 
& \shortstack[c]{$B_i$\\ $i \in [1,5]$}  & $\mathcal{G}[0.0, 10^4]$ & 0.0\\ 
% & {$B_i$\\ $i \in [1,5]$} & $\mathcal{G}[0.0, 10^4]$ & 0.0\\  
% & & & \\
\hline
% & & & \\
& \multicolumn{3}{c|}{\textbf{Cosmology}} \\ 
$w$CDM & $w$ & $\mathcal{U}[-2, -0.33]$ &-1.0\\  
% & & & \\
\hline 
% & & & \\
% & & & \\
& \multicolumn{3}{c|}{\textbf{Galaxy Bias}} \\  
\multirow{2}{*}{\shortstack[c]{\textit{Linear}\\ \textit{Bias}}} &
\shortstack[c]{$b_1^{i}$\\ $i \in [1,3]$}  & $\mathcal{U}[0.8, 3.0]$ & 1.7\\ 
% & & & \\
& \shortstack[c]{$b_1^{i}$\\ $i \in [4,5]$}  & $\mathcal{U}[0.8, 3.0]$ & 2.0\\ 
% & & & \\
\hline
% & & & \\
& \multicolumn{3}{c|}{\textbf{Galaxy Bias}} \\  
\multirow{9}{*}{\shortstack[c]{\textit{Non-linear}\\ \textit{Bias}}} &
\shortstack[c]{$b_1^{i}\sigma_8$\\ $i \in [1,3]$}  & $\mathcal{U}[0.67, 2.52]$ & 1.42\\ 
% & & & \\
& \shortstack[c]{$b_1^{i}\sigma_8$\\ $i \in [4,5]$}  & $\mathcal{U}[0.67, 2.52]$ & 1.68\\ 
% & & & \\

& \shortstack[c]{$b_2^{i}\sigma^2_8$\\ $i \in [1,3]$}  & $\mathcal{U}[-3.5, 3.5]$ & 0.16\\ 
% & & & \\
& \shortstack[c]{$b_2^{i}\sigma^2_8$\\ $i \in [4,5]$}  & $\mathcal{U}[-3.5, 3.5]$ & 0.35\\ 
% & & & \\

% % \cline{2-4}
% & & & \\
% & \shortstack[c]{$b_2^{i}\sigma^2_8$\\ $i \in [1,5]$} & $\mathcal{U}[0.8, 3.0]$ &\shortstack[c]{Lens Galaxy\\ non-linear bias} \\ 
% & & & \\
\hline
\end{tabular}
\caption{The parameters varied in different models, their prior range used ($\mathcal{U}[X, Y] \equiv$ Uniform prior between $X$ and $Y$; $\mathcal{G}[\mu, \sigma] \equiv$ Gaussian prior with mean $\mu$ and standard-deviation $\sigma$) in this analysis and the \textit{fiducial} values used for simulated likelihood tests.}
\label{tab:params_all}
\end{table}

\section{Validation of parameter inference}
\label{sec:param_inf}
We assume the likelihood to be a multivariate Gaussian
% \begin{linenomath*}
\begin{equation}
    \ln \mathcal{L}(\vec{\mathbfcal{D}}|\Theta) = -\frac{1}{2} [\vec{\mathbfcal{D}} - \vec{\mathbfcal{T}}(\Theta)]^{\rm T} \, {\mathbfcal{C}}^{-1}_{\rm wPM} \,  [\vec{\mathbfcal{D}} - \vec{\mathbfcal{T}}(\Theta)] \,.
\end{equation}
% \end{linenomath*}
Here $\vec{\mathbfcal{D}}$ is the measured $\gammat$ and $\wtheta$ datavector of length $N_{\rm data}$ (if we use all the angular and tomograhic bins), $\vec{\mathbfcal{T}}$ is the theoretical prediction for these statistics for the parameter values given by  $\Theta$, and ${\mathbfcal{C}}^{-1}_{\rm wPM}$ is the inverse covariance matrix of shape $N_{\rm data} \times N_{\rm data}$ (including modifications from the PM marginalization term).

For our analysis we use the \textsc{Polychord} sampler with the settings described in \citet{y3-samplers}. The samplers probe the posterior ($\mathcal{P}(\Theta | \vec{\mathbfcal{D}})$) which is given by:
% \begin{linenomath*}
\begin{equation}
    \mathcal{P}(\Theta | \vec{\mathbfcal{D}}) = \frac{\mathcal{L}(\vec{\mathbfcal{D}}|\Theta) {\rm P}(\Theta)}{{\rm P}(\vec{\mathbfcal{D}})}
\end{equation}
% \end{linenomath*}
where ${\rm P}(\Theta)$ are the priors on the parameters of our model, described in \S\ref{sec:prior}, and ${\rm P}(\vec{\mathbfcal{D}})$ is the evidence of data.

To estimate the constraints on the cosmological parameters, we have to marginalize the posterior over all the rest of the multi-dimensional parameter space. We quote the mean and 1$\sigma$ variance of the marginalized posteriors when quoting the constraints. However, note that these marginalized constraints can be biased if the posterior has significant non-Gaussianities, particularly in the case of broad priors assigned to poorly constrained parameters. The maximum-a-posteriori (MAP) point is not affected by such "projection effects"; therefore, we also show the MAP value in our plots. However, we note that in high-dimensional parameter space with a non-trivial structure, it is difficult to converge on a global maximum of the whole posterior (also see \citet{Joachimi_2021} and citations therein).

\subsection{Analysis choices}
\label{sec:analysis_choices}
In this subsection, we detail the galaxy bias models that we use, describe the free parameters of our models, and choose priors on those parameters. 
\subsubsection{PT Models}
\label{sec:pt_models}
In this analysis, we test two different galaxy bias models: 
\begin{enumerate}
    \item \textit{Linear bias} model: The simplest model to describe the overdensity of galaxies, valid at large scales, assumes it to be linearly biased with respect to the dark matter overdensity (see \S\ref{sec:Pk_pred}). In this model, for each lens tomographic bin $j$, the average bias of galaxies is given by a constant free parameter $b^j_1$. 
    \item \textit{Non-linear bias} model: 
    To describe the clustering of galaxies at smaller scales robustly, we also implement a one--loop PT model. As described in \S\ref{sec:Pk_pred}, in general, this model has five free bias parameters for each lens tomographic bin. For each tomographic bin $j$, we fix two of the non-linear parameters to their co-evolution value given by: $b^{j}_{\rm s} = (-4/7) (b^j_1 - 1)$ and $b^{j}_{\rm 3nl} = b^j_1 - 1$ \citep{McDonald2009,Saito2014a}, while set $b^{j}_{\rm k} = 0$ \citep{p2020perturbation}. Therefore, in our implementation, we have two free parameters for each tomographic bin: linear bias $b^{j}_1$ and non-linear bias $b^{j}_2$. This allows us to probe smaller scales with minimal extra degrees of freedom, obtaining tighter constraints on the cosmological parameters while keeping the biases due to projection effects, as described below, in control. 
    
    As we describe below, in order to test the robustness of our model, we analyze the bias in the marginalized constraints on cosmological parameters. However, given asymmetric non-Gaussian degeneracies between the parameters of the model (particularly between cosmological parameters and poorly constrained non-linear bias parameters $b^{j}_2$ and intrinsic alignment parameters), the marginalized constraints show projection effects. We find that imposing priors on the non-linear bias model parameters in combination with $\sigma_8$, as $b^{j}_1 \sigma_8$ and $b^{j}_2 \sigma^2_8$ removes much of the posterior projection effect. 
    As detailed later, these parameters are sampled with flat priors. We emphasize that the flat priors imposed on these non-linear combinations of parameters are non-informative, and our final constraints on $b^{j}_1$ and $b^{j}_2$ are significantly tighter than the projection of priors on these parameters.
    
\end{enumerate}

\subsubsection{Cosmological Models}
\label{sec:cosmo_models}
We report the constraints on two choices of the cosmological model:
\begin{enumerate}
    \item Flat \lcdm\ : We free six cosmological parameters the total matter density $\Omega_{\rm m}$, the baryonic density $\Omega_{\rm b}$,  the spectral index $n_{\rm s}$, the Hubble parameter $h$, the amplitude of scalar perturbations $A_s$ and $\Omega_\nu h^2$ (where $\Omega_\nu$ is the massive neutrino density). We assume a a flat cosmological model, and hence the dark energy density, $\Omega_\Lambda$, is fixed to be $\Omega_\Lambda = 1 - \Omega_{\rm m}$.
    \item Flat \wcdm : In addition to the six parameters listed above, we also free the dark energy equation of state parameter $w$. Note that this parameter is constant in time and $w=-1$ corresponds to \lcdm\ cosmological model. 
\end{enumerate}

\subsubsection{Scale cuts}\label{sec:scale_cuts}

The complex astrophysics of galaxy formation, evolution, and baryonic processes like feedback from active galactic nuclei (AGN), supernova explosions, and cooling make higher-order non-linear contributions that we do not include in our model. The contribution from these poorly understood effects can exceed our statistical uncertainty on the smallest scales; hence we apply scale cuts chosen so that our PT models give unbiased cosmological constraints.

As mentioned earlier, marginalizing over a multi-dimensional parameter space can lead to biased 2D parameter constraints due to projection effects. To calibrate this effect for each of our models, we first perform an analysis using a \textit{baseline} datavector constructed from the \textit{fiducial} values of that model. 
We then run our MCMC chain on the \textit{contaminated} datavector that includes higher-order non-linearities, and we measure the bias between the peak of the marginalized \textit{baseline} contours and the peak of the marginalized \textit{contaminated} contours. 

From a joint analysis of 3D galaxy-galaxy and galaxy-matter correlation functions at fixed cosmology in simulations \citep{p2020perturbation}, we find that the \textit{linear bias} model is a good description above 8Mpc/$h$ while the two-parameter \textit{non-linear bias} model describes the correlations above 4Mpc/$h$. We convert these physical co-moving distances to angular scale cuts for each tomographic bin and treat them as starting guesses. Then for each model, we iterate over scale cuts until we find the minimum scales at which the bias between marginalized \textit{baseline} and \textit{contaminated} contours is less than $0.3\sigma$. For the \lcdm\ model, we impose this criterion on the $\Omega_{\rm m}-S_8$ projected plane, and for the $w$CDM model, we impose this criterion on all three 2D plane combinations constructed out of $\Omega_{\rm m}$, $S_8$ and $w$. Further validation of these cuts is performed using simulations in \ref{sec:sims} and \citet*{y3-simvalidation}.

\subsubsection{Priors and Fiducial values}
\label{sec:prior}

We use locally non-informative priors on the cosmological parameters to ensure statistically independent constraints on them. Although our constraints on cosmological parameters like the Hubble constant $h$, spectral index $n_{\rm s}$ and baryon fraction $\Omega_{\rm b}$ are modest compared to surveys like \textit{Planck}, we have verified that our choice of wide priors does not bias the inference on our cosmological parameters of interest, $\Omega_{\rm m}$ and $S_8$. 

When analyzing the \textit{linear bias} model, we use a wide uniform prior on these linear bias parameters, given by $0.5 < b^{j}_1 < 3$. For the \textit{non-linear bias} model, as mentioned above, we sample the parameters $b^{j}_1 \sigma_8$ and $b^{j}_2 \sigma^2_8$. We use uninformative uniform priors on these parameters for each tomographic bin $j$ given by $0.67 < b^{j}_1 \sigma_8 < 3.0$ and $-4.2 < b^{j}_2 \sigma^2_8 < 4.2$. At each point in the parameter space, we calculate  $\sigma_8$ and retrieve the bias parameters $b^{j}_1$ and $b^{j}_2$ from the sampled parameters to get the prediction from the theory model. The \textit{fiducial} values of the linear bias parameters $b^{j}_1$ used in our simulated likelihood tests are motivated by the  recovered bias values in N-body simulations and are summarized in Table \ref{tab:params_all}. 
For the non-linear bias parameters, the \textit{fiducial} values of $b^{j}_2$ are obtained from the interpolated $b_1-b_2$ relation extracted from 3D tests in \mice simulations (see Fig. 8 of \citet*{p2020perturbation}) for the \textit{fiducial} $b^j_1$ for each tomographic bin.

For the intrinsic alignment parameters, we again choose uniform and uninformative priors. As the IA parameters are directly dependent on the source galaxy population, it is challenging to motivate a reasonable choice of prior from other studies. The \textit{fiducial} values of these parameters required for the simulated test are motivated by the Y1 analysis as detailed in \citet{Samuroff_2019}.

We impose an informative prior for our measurement systematics parameters, lens photo-$z$ shift errors ($\Delta z^j_{\rm g}$), lens photo-$z$ width errors ($\sigma z^j_{\rm g}$), source photo-$z$ shift errors ($\Delta z^j_{\rm s}$) and shear calibration biases ($m^j$) for various tomographic bins $i$. The photo-$z$ shift parameter changes the redshift distributions for lenses (g) or sources (s) for any tomographic bin $j$, used in the theory predictions (see \S\ref{sec:stat_theory}) as $n^j_{\rm g/s}(z) \longrightarrow n^j_{\rm g/s}(z - \Delta z^j_{\rm g/s})$, while the photo-$z$ width results in  $n^j_{\rm g}(z) \longrightarrow n^j_{\rm g}(\sigma z^j_{\rm g}[z - \langle z \rangle^j] + \langle z \rangle^j)$, where $\langle z \rangle^j$ is the mean redshift of the tomographic bin $j$. Lastly, the shear calibration uncertainity modifies the galaxy-galaxy lensing signal prediction between lens bin $i$ and source bin $j$ as $\gamma^{ij}_{\rm t} \longrightarrow (1 + m^j)\gamma^{ij}_{\rm t}$.

For the source photo-$z$, we refer the reader to \citet*{y3-sompz} for the characterization of source redshift distribution, \citet*{y3-sourcewz} for reducing the uncertainity in these redshift distribution using cross-correlations with spectroscopic galaxies and \citet*{y3-hyperrank} for a validation of the shift parameterization using a more complete method based on sampling the discrete distribution realizations. For the shear calibration biases, we refer the reader to \citet{y3-imagesims} which tests the shape measurement pipeline and determine the shear calibration uncertainity while accounting for effects like blending using state-of-art image simulation suite. For the priors on the lens photo-$z$ shift and lens photo-$z$ width errors, we refer the reader to \citet*{y3-lenswz}, which cross-correlated the DES lens samples with spectroscopic galaxy samples from Sloan Digital Sky Survey to calibrate the photometric redshifts of lenses (also see \citet{y3-2x2ptaltlensresults} and \citet{y3-2x2ptaltlenssompz} for further details on \maglim redshift calibration). 

In this paper we fix the magnification coefficients to the best-fit values described in \citet*{y3-2x2ptmagnification, y3-generalmethods}, but we refer the reader to \citet*{y3-2x2ptmagnification} for details on the impact of varying the magnification coefficients on the cosmological constraints. 
Note that in our tests to obtain scale cuts for cosmological analysis using simulated datavectors (described below),  we remain conservative and fix the shear systematics to their \textit{fiducial} parameter values and analyze the datavectors at the mean source redshift distribution $n_{\rm s}(z)$, as shown in Fig.~\ref{fig:nz_comp}. This procedure, after fixing the systematic parameters, results in tighter constraints and ensures that the impact of baryons and non-linear bias on the cosmological inference is over-estimated. Therefore, we expect our recovered scale cuts to be conservative.

\begin{figure*}
\centering
\subfloat{%
       \includegraphics[width=0.49\textwidth]{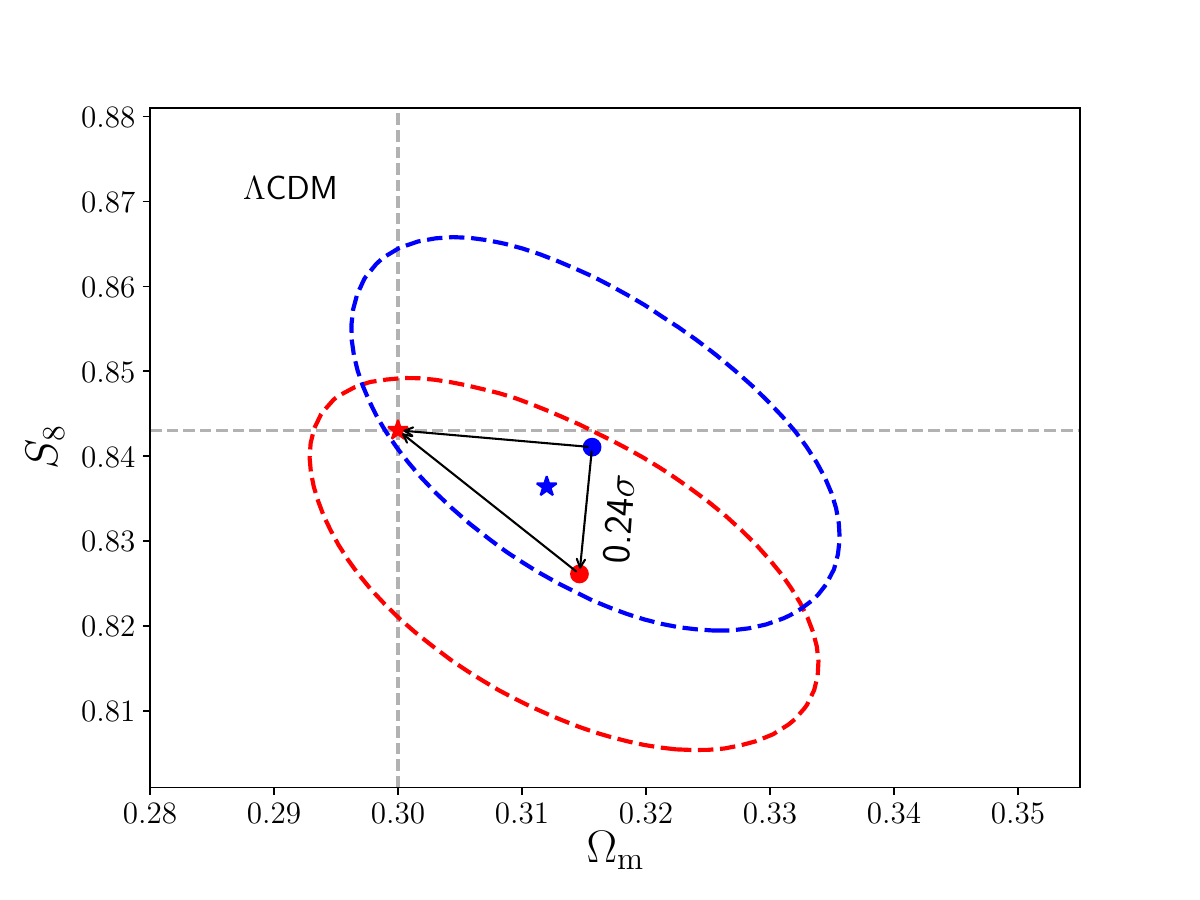}
     }
\hfill
\subfloat{%
      \includegraphics[width=0.49\textwidth]{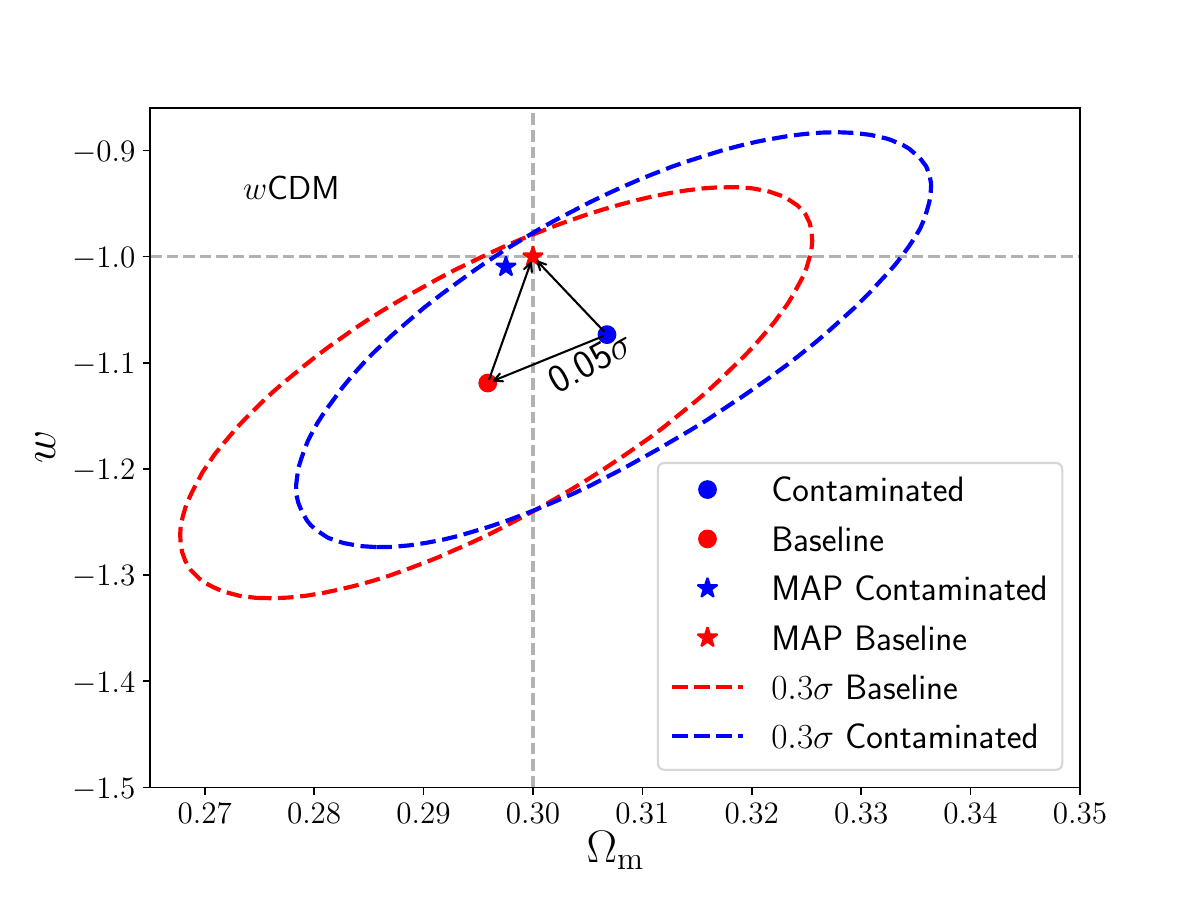}
     }
    \caption[]{Simulated datavector parameter constraints from a datavector contaminated with non-linear bias + baryons but analyzed with a linear bias + \textsc{Halofit} model. Dashed grey lines mark the truth values for the simulated datavector. The left panel shows contours for \lcdm, and the right panel shows \wcdm. The scale cuts are (8,6) Mpc/$h$ for $w(\theta)$ and $\gamma_{\rm t}$ respectively. In both panels, we compare the peak of the marginalized constraints in the 2D  parameter plane for the contaminated datavector (blue circle) and the baseline datavector  (red square). The peaks of the marginalized baseline contours are within 0.3$\sigma$ of the peaks of the marginalized contaminated contours, which is our criterion for acceptable scale cuts. We also show the corresponding maximum posterior value obtained for all the contours (with a star symbol), obtained using the methodology described in the main text.}
\label{fig:sim_lin}    
\end{figure*}

\subsection{Simulated Likelihood tests}\label{sec:simlike_analysis}

We perform simulated likelihood tests to validate our choices of scale cuts, galaxy bias model and the cosmological model (including priors and external datasets when relevant). In this analysis we focus on determining and validating the scale cuts using \redmagic lens galaxy sample and we refer the reader to \citet{y3-2x2ptaltlensresults} for validation using the \maglim lens galaxy sample. We require that the choices adopted return unbiased cosmological parameters. This first step based on the tests on noiseless datavectors in the validation is followed by tests on cosmological simulations. 

\subsubsection{Scale cuts for the linear bias model}
\label{sec:sc_linbias}

Our baseline case assumes linear galaxy bias and no baryonic impact on the matter-matter power spectrum. We use the linear bias values for the five lens bins (in order of increasing redshift) $b_1 = 1.7, 1.7, 1.7, 2.0$, and  $2.0$. We compare the cosmology constraints from the baseline datavector with a simulated datavector contaminated with contributions from non-linear bias and baryonic physics. For baryons, the non-linear matter power spectra ($P^{\rm cont}_{\rm mm}$) used in generating the contaminated datavector is estimated using following prescription: 
\begin{equation}
    P^{\rm cont}_{\rm mm} = \bigg(\frac{P^{\rm hydro-sim}_{\rm mm}}{P^{\rm DM-only}_{\rm mm}}\bigg) P^{\textsc{Halofit}}_{\rm mm},
\end{equation}
where, $P^{\rm hydro-sim}_{\rm mm}$ and $P^{\rm DM-only}_{\rm mm}$ are the matter power spectra measured from a full hydrodynamical simulation and dark matter only simulation respectively. We use the measurements from the OWLS-AGN simulations, which is based on hydrodynamical simulations that include the effects of supernovae and AGN feedback, metal-dependent radiative cooling, stellar evolution, and kinematic stellar feedback \citep{Le_Brun_2014}
To capture the effect of non-linear bias, we use the \textit{fiducial} $b^j_2$ values as described in the previous section and fix the bias parameters $b^j_s$ and $b^j_{\rm 3nl}$ to their co-evolution values. 
% To capture the effect of baryons, we use the OWLS-AGN datavector, which is based on hydrodynamical simulations that include the effects of supernovae and AGN feedback, metal-dependent radiative cooling, stellar evolution, and kinematic stellar feedback \citep{Le_Brun_2014}. 

\begin{figure*}
\includegraphics[width=\columnwidth]{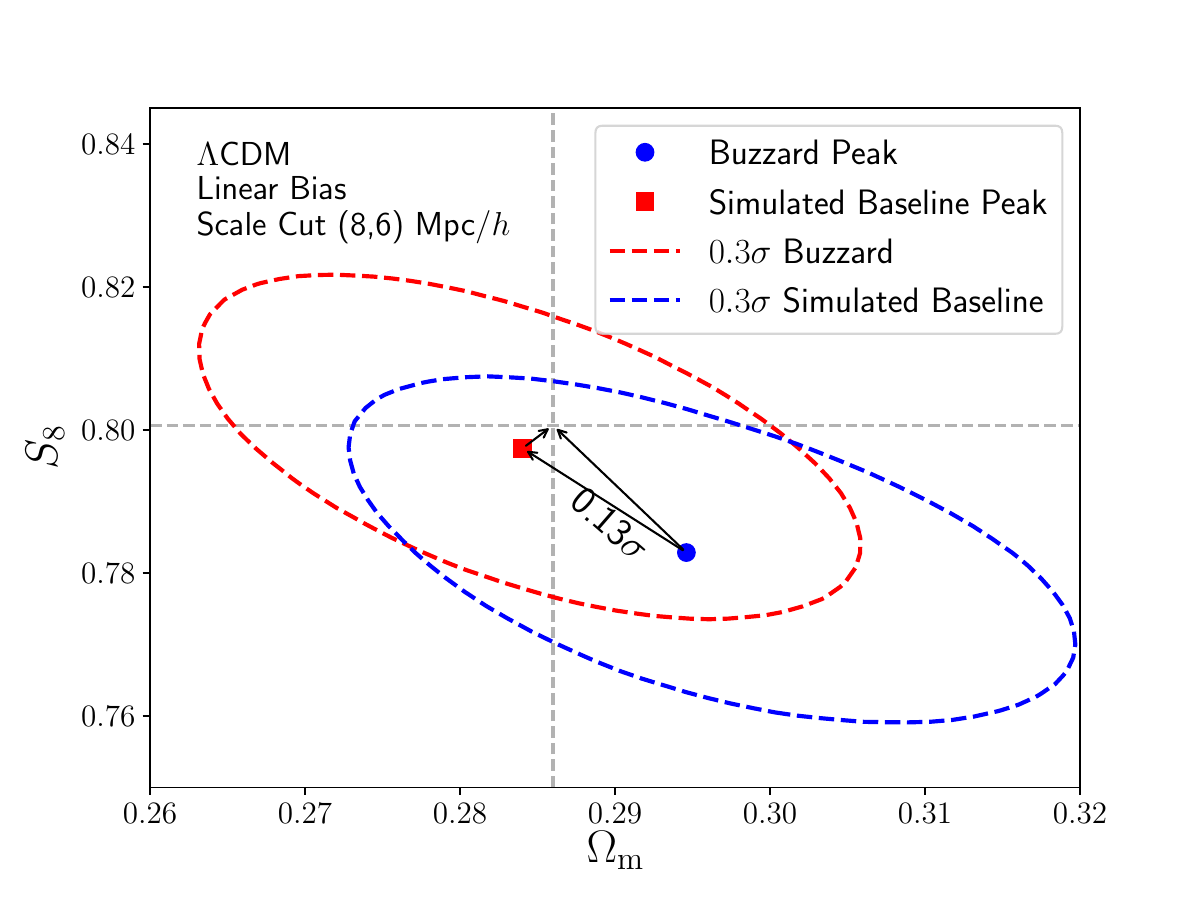}
\includegraphics[width=\columnwidth]{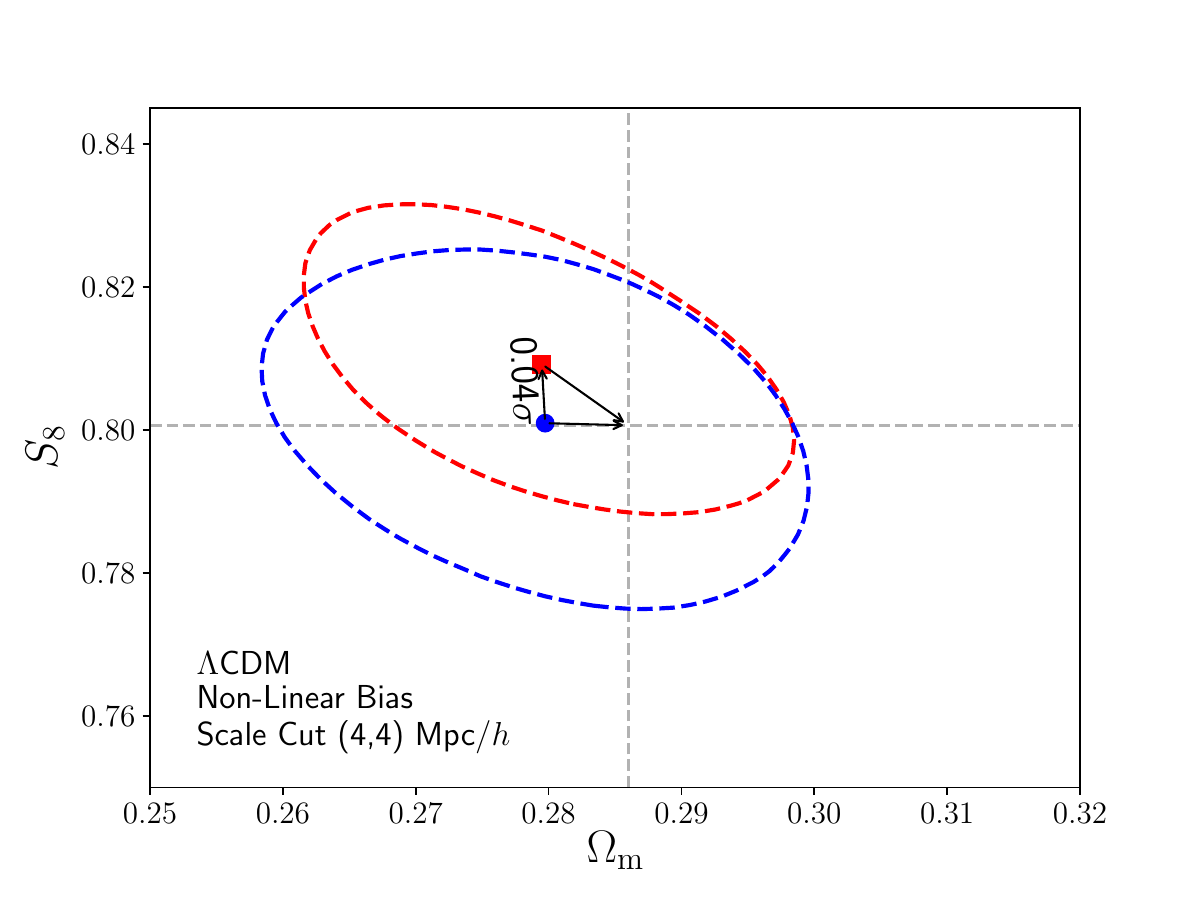}
\caption[]{The blue contours show constraints from \buzzard simulations (blue contours) compared with  \buzzard-like theory datavector (red contours) in the $\Lambda$CDM cosmological model.
The left (right) panel shows the constraints for linear (non-linear) bias models with the scale cuts given in the legend. The linear and non-linear bias values are extracted from fits to the 3D correlation functions ($\xigg$ and $\xigm$). We see that both the scale-cut choices satisfy our validation criterion. 
}
\label{fig:bcc_des_lcdm}
\end{figure*}

\begin{figure*}
\includegraphics[width=\columnwidth]{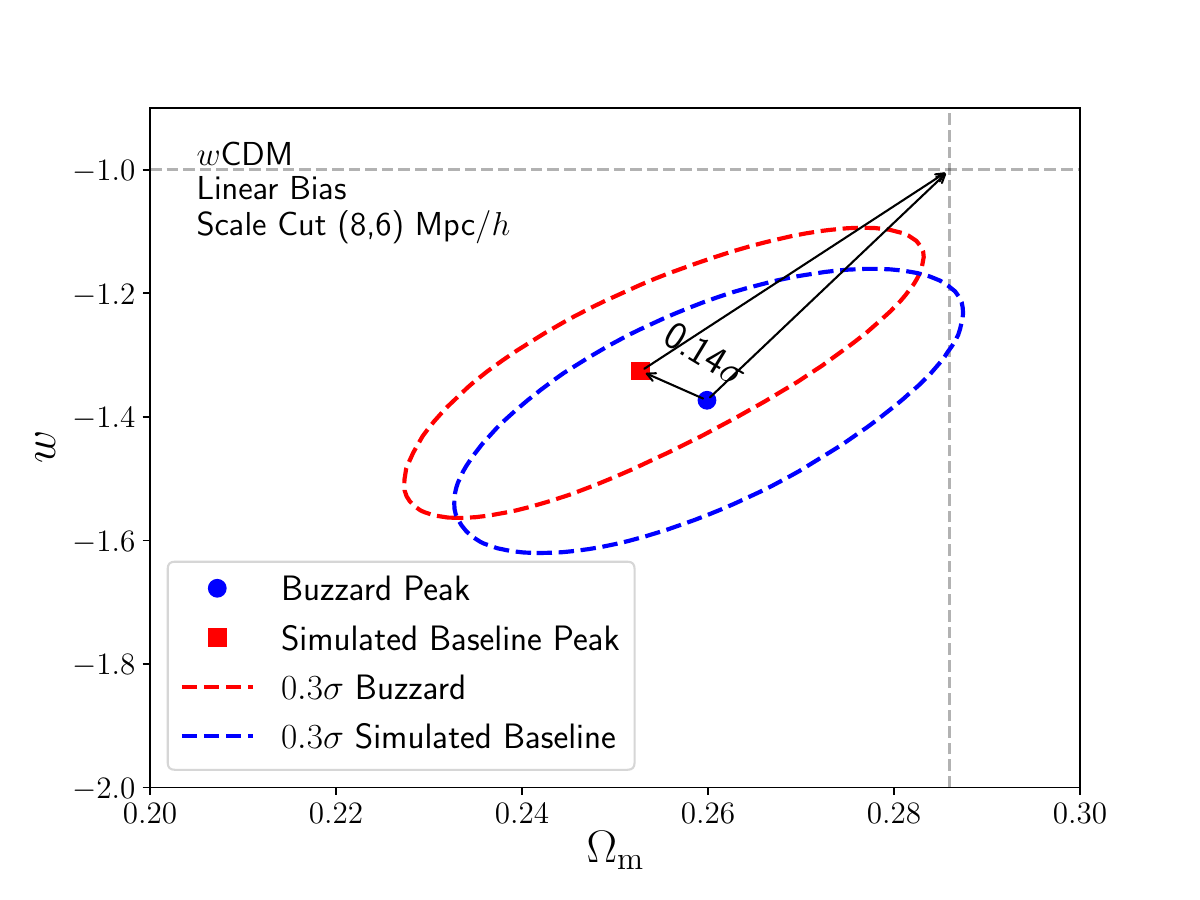}
\includegraphics[width=\columnwidth]{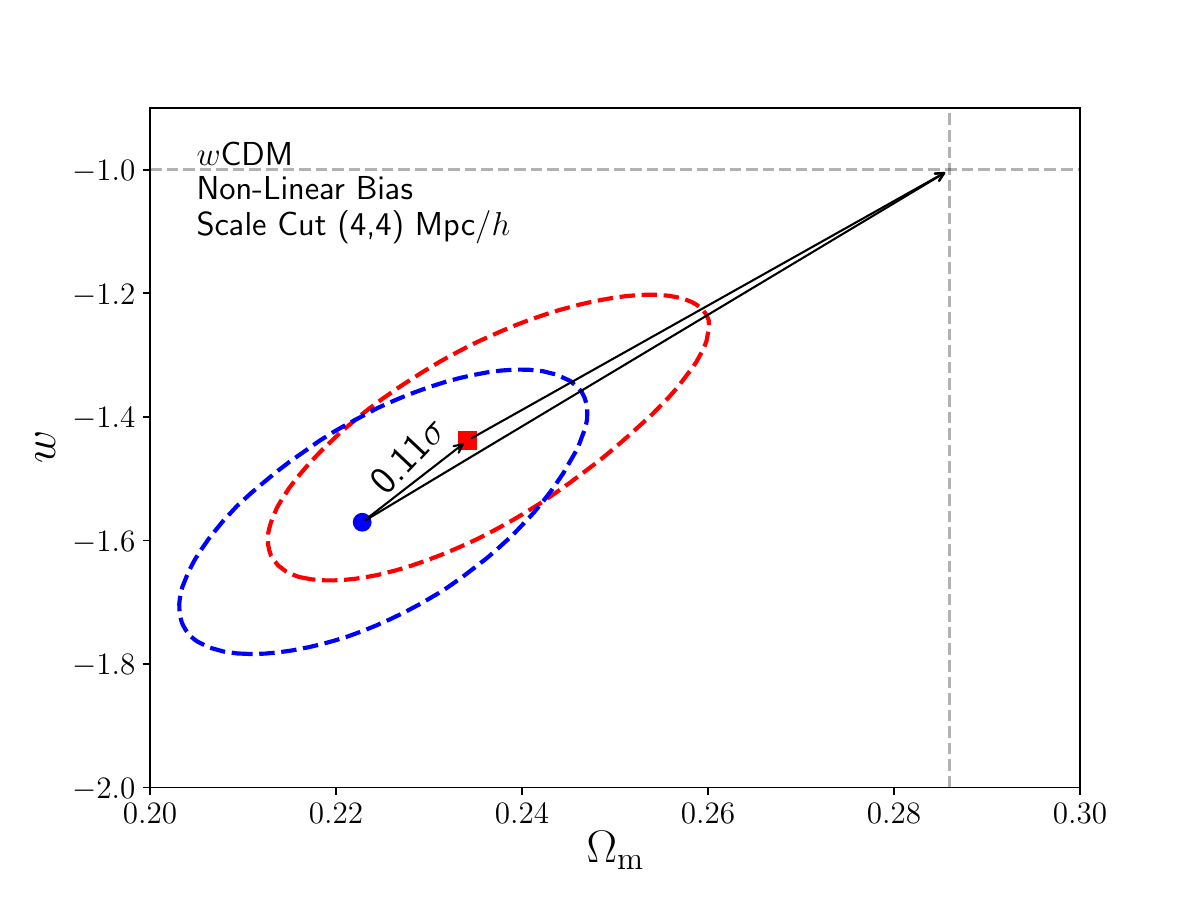}
\caption[]{Same as Fig.~\ref{fig:bcc_des_lcdm} but for $w$CDM cosmology.
}
\label{fig:bcc_des_wcdm}
\end{figure*}

Fig.~\ref{fig:sim_lin} shows the 0.3$\sigma$ contours when implementing the angular cuts corresponding to (8,6) Mpc/$h$ for $w(\theta)$ and $\gamma_{\rm t}$. The left panel is for $\Lambda$CDM, and the right panel for $w$CDM (only the $w- \Omega_{\rm m}$ plane is shown, but we also verified that the criterion is satisfied in the $\Omega_{\rm m}-S_8$ and $S_8-w$  planes). The figure shows the peaks of marginalized contaminated and baseline posteriors in 2D planes with blue and red markers respectively. We find that a 0.24$\sigma$ marginalized contaminated contour intersects the peak of baseline marginalized posterior in $\Lambda$CDM model, while same is true for a 0.05$\sigma$ contour in $w$CDM model. We find that for the linear bias model, (8,6) Mpc/$h$ scale cuts pass the above-mentioned criteria that the distance between the peaks of baseline and contaminated contours is less than 0.3$\sigma$. In Fig.~\ref{fig:sim_lin}, we also show the MAP parameter values for each run using a star symbol. In order to obtain the MAP value, we use the Nelder-Mead algorithm \citep{Nelder_Mead_65} to minimize the posterior value after starting the optimization from the highest posterior point of the converged parameter inference chain. We find that the MAP point also lies within 0.3$\sigma$ of the true cosmology, further validating the inferred scale cuts (although, note the caveats about MAP mentioned in \S\ref{sec:param_inf}). We note that we analyze smaller scales of $\gamma_{\rm t}$ compared to $w(\theta)$ statistic because it is a less significant measurement, hence can tolerate greater modeling uncertainty. Moreover, we use the point mass marginalization scheme (see \S~\ref{sec:pm_theory} and Appendix~\ref{app:pm}) that effectively makes $\gamma_{\rm t}$ a local statistic (c.f. \cite{Abbott_2018}).

\subsection{\buzzard simulation tests}
\label{sec:sims}
Finally, we validate our model with mock catalogs from cosmological simulations for analysis choice combinations that pass the simulated likelihood tests. These tests, and tests of cosmic shear and $3\times2$-point analyses, are presented in full in \citet*{y3-simvalidation}, and we summarize the details relevant for $2\times2$-point analyses here. We use the suite of Y3 \buzzard simulations described above. We again require that our analysis choices return unbiased cosmological parameters. In order to reduce the sample variance, we analyze the mean datavector constructed from 18 \buzzard realizations.

\subsubsection{Validation of linear bias model}
We have run simulated $2\times 2$-point analyses on the mean of the measurements from all 18 \buzzard simulations. We compare our model for $w(\theta)$ and $\gamma_{\rm t}(\theta)$ to our measurements at the true \buzzard cosmology, leaving only linear bias and lens magnification coefficients free. In this case, we have ten free parameters in total, and we find a chi-squared value of 13.6 for 285 data points using our \textit{fiducial} scale cuts and assuming the covariance of a single simulation, as appropriate for application to the data. Note that the datavector is mean of multiple realization, so we expect a low chi-squared value for the bestfit curve and a principled accounting of error is presented in \citet*{y3-simvalidation}. This analysis assumes true source redshift distributions, and we fix the source redshift uncertainties to zero as a conservative choice. This results in cosmological constraints where the mean two-dimensional parameter biases are $0.23\sigma$ in the $S_8-\Omega_{\rm m}$ plane and $0.18\sigma$ in the $w-\Omega_{\rm m}$ plane. These biases are consistent with noise, as they have an approximately $1/\sqrt{18}\sigma$ error associated with them (assuming 1$\sigma$ error from a single realization). We perform a similar analysis using calibrated photometric redshift distributions where we use \redmagic lens redshift distributions, and use the SOMPZ redshift distribution estimates of source galaxies. These are weighted by the likelihood of those samples given the cross-correlation of our source galaxies with redMaGiC and spectroscopic galaxies (we refer the reader to Appendix F of \citet{y3-simvalidation} for detailed procedure). This procedure results in the mean two-dimensional parameter biases of $0.07\sigma$ in the $S_8-\Omega_{\rm m}$ plane and $0.05\sigma$ in the $w-\Omega_{\rm m}$ plane.

The left panels of Fig.~\ref{fig:bcc_des_lcdm} and Fig.~\ref{fig:bcc_des_wcdm} show the 0.3$\sigma$ constraints obtained from analyzing linear galaxy bias models in $\Lambda$CDM and $w$CDM cosmologies on the \buzzard datavector in blue colored contours. Since we expect the marginalized posteriors to be affected by the projection effects, we compare these contours to a simulated noiseless baseline datavector obtained at the input cosmology of \buzzard (denoted by gray dashed lines in Fig.~\ref{fig:bcc_des_lcdm} and Fig.~\ref{fig:bcc_des_wcdm}, also see \cite{DeRose2019}). We find that similar to results obtained with simulated datavectors in previous section, our parameter biases are less than the threshold of 0.3$\sigma$ for the fiducial scale cuts.  For a more detailed discussion how these shift compare with probability to exceed (PTE) values of exceeding a $0.3\sigma$ bias, see Section V of \citet*{y3-simvalidation}.

Also note that as changing the input truth values of the parameters impacts the shape of the multi-dimensional posterior, we find that the effective magnitude and direction of the projection effects of the baseline contours (comparison of red contours in  Fig.~\ref{fig:sim_lin} with Fig.~\ref{fig:bcc_des_lcdm} and Fig.~\ref{fig:bcc_des_wcdm}) are different. 

\subsubsection{Scale cuts for non-linear bias model}
Likewise, we have run simulated $2\times 2$-point analyses including our non-linear bias model on the mean of the measurements from all 18 simulations. Similar to the procedure used to determine the linear bias scale cuts in \S\ref{sec:sc_linbias}, we iterate over scale cuts for each tomographic bin defined from varying physical scale cuts. 

We compare our model for $w(\theta)$ and $\gamma_{\rm t}(\theta)$ to our measurements at the true \textsc{Buzzard} cosmology, leaving our bias model parameters and magnification coefficients free, which adds 15 free parameters. We find a $\chi^2$ value of 15.6 for 340 data points using our non-linear bias scale cuts and assuming the covariance of a single simulation. Simulated analyses using true redshift distributions result in cosmological constraints where the associated mean two-dimensional parameter biases for these analyses are $0.04\sigma$ in the $S_8-\Omega_{\rm m}$ plane and $0.11\sigma$ in the $w-\Omega_{m}$ plane. This is again consistent with noise due to finite number of realizations.

In the right panel of \fig{fig:bcc_des_lcdm} we show the constraints on $\Omega_{\rm m}$ and $S_8$ from the mean Buzzard $2\times2$pt measurements for \lcdm\ cosmological model. The results for non-linear bias models are shown, where we find, the criterion for unbiased cosmology is satisfied for the choice of scale cuts of (4,4)Mpc/$h$ for $(w(\theta),\gamma_{\rm t}(\theta))$ respectively. Again for a more detailed discussion how these shift compare with PTE values of exceeding a $0.3\sigma$ bias, see \citet*{y3-simvalidation}. The \fig{fig:bcc_des_wcdm} shows the same analysis for \wcdm\ cosmological model in the $\Omega_{\rm m}$ and $w$ plane, where we find similar results. We therefore use (4,4)Mpc/$h$ as our validated scale cuts when analyzing data with non-linear bias model.

\section{Results}
\label{sec:results}
\begin{figure}
\includegraphics[width=\columnwidth]{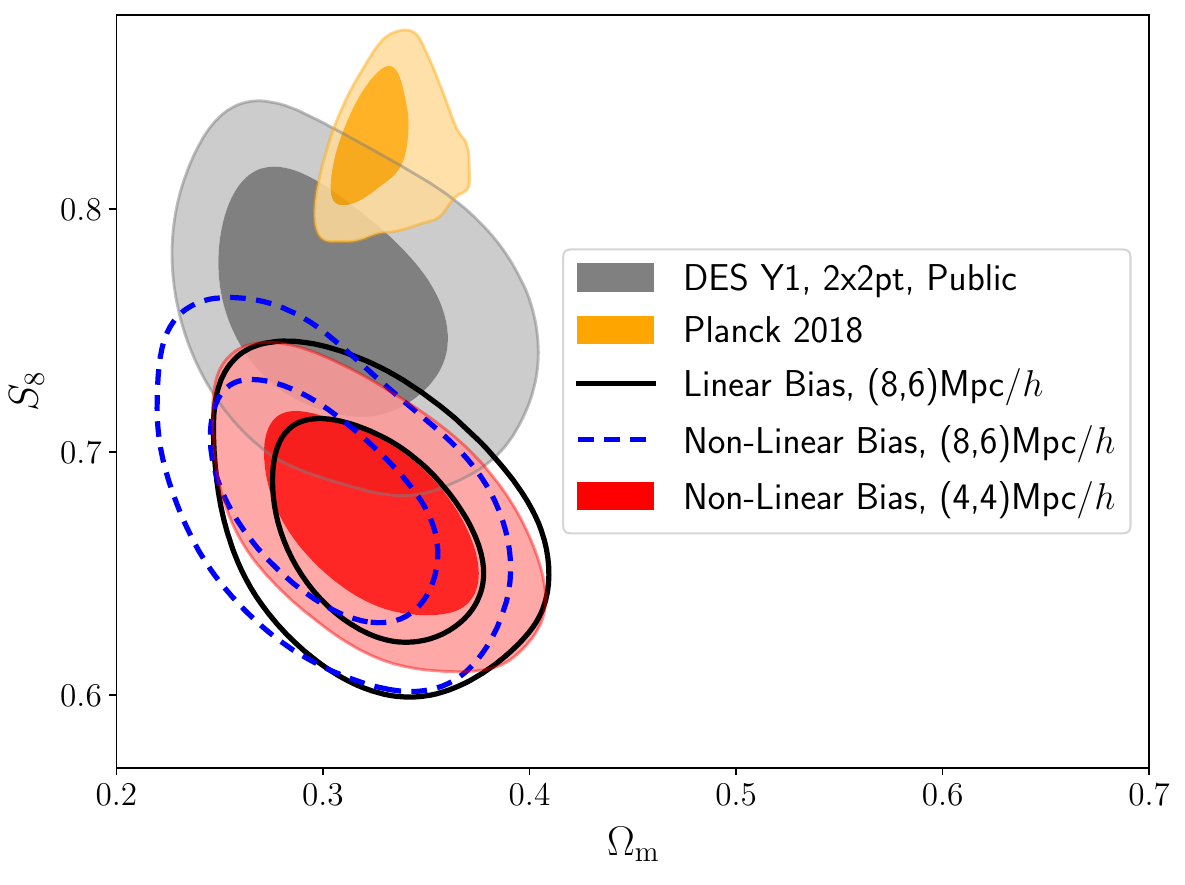}
\caption[]{Comparison of the $2\times2$pt $\Lambda$CDM constraints, using \redmagic lens galaxy sample, for both linear bias and non-linear bias models at their respectively defined scale cuts given in the legend. We find a preference for a low value of $S_8$, compared to DES Y1 $2\times2$pt public result \citep{Abbott_2018} and Planck 2018 public result \citep{Planck_2018_cosmo}, with both models of galaxy bias which we investigate in \S\ref{sec:internal_consistency}. We also show that analyzing smaller scales using the non-linear galaxy bias model leads to 17\% better constraints in the $\Omega_{\rm m} - S_8$ plane. }\label{fig:des_comp}
\end{figure}

In this section we present the $2\times2$pt cosmology results using the DES Y3 \redmagic lens galaxy sample and study the implications of our constraints on  galaxy bias. 
\subsection{\redmagic cosmology constraints}
\label{sec:fid_cosmo_res}
In Fig.~\ref{fig:des_comp}, we compare the constraints on the cosmological parameters obtained from jointly analyzing $\wtheta$ and $\gammat$ with both linear and non-linear bias models. 
% The 2x2pt constraints on $\om$ are at the 10\% level with DES Y3. 
We find $\om = 0.325^{+0.033}_{-0.034}$ from the linear bias model (a 10\% constraint) at the \textit{fiducial} scale cuts of (8,6) Mpc/$h$ (for $(w(\theta),\gamma_{\rm t}(\theta))$ respectively), while using the non-linear bias model at same scale cuts gives completely consistent constraints. 
% us $\om = 0.30^{+0.035}_{-0.037}$. 
We also show the results for the scale cuts of (4,4) Mpc/$h$ using the non-linear bias model where we find $\om=0.323^{+0.034}_{-0.035}$. These marginalized constraints on $\om$ are completely consistent with the public DES-Y1  $2\times 2$pt results \citep{Abbott_2018} and \Planck results (including all  three correlations between temperature and E-mode polarization, see \citet{Planck_2018_cosmo} for details). 

With the analysis of linear bias model with (8,6) Mpc/$h$ scale cuts (referred to as \textit{fiducial}  model in following text), we find $S_8 = 0.668^{+0.026}_{-0.033}$. As is evident from the contour plot in Fig.~\ref{fig:des_comp}, our constraints prefer lower $S_8$ compared to previous analyses.  We use the Monte-Carlo parameter difference distribution methodology (as detailed in \citet*{y3-tensions}) to assess the tension between our \textit{fiducial} constraints and \Planck results. Using this criterion, we find a tension of 4.1$\sigma$, largely driven by the differences in the $S_8$ parameter. We find similar constraints on $S_8$ from the non-linear bias as well for both the scale cuts. We investigate the cause of this low $S_8$ value in the following sub-sections. We note that the shift to slightly lower $\om$ with the non-linear bias model compared to the linear bias model at (8,6) Mpc/$h$ scale cuts can arise in a noisy datavector. This is in contrast to the analysis done in \S~\ref{sec:param_inf} with noiseless datavectors to validate the scale cuts. 

Note that the non-linear bias model at (4,4) Mpc/$h$ scale cuts results in tighter constraints in the $\om-S_8$ plane. 
We estimate the total constraining power in this $\om-S_8$ plane by estimating 2D figure-of-merit (FoM), which is defined as ${\rm FoM}_{p_1,p_2} = 1/\sqrt{[{\rm det \, Cov}(p_1,p_2)]}$, for any two parameters $p_1$ and $p_2$ \citep{Huterer_2001, Wang_2008}. This statistic here is proportional to the inverse of the confidence region area in the 2D parameter plane of $\om-S_8$.
We find that the non-linear bias model at  (4,4) Mpc/$h$ results in a $17$\% increase in constraining power compared to the linear bias model at (8,6) Mpc/$h$.

\subsection{Comparison with \maglim results}

\begin{figure}
\includegraphics[width=\columnwidth]{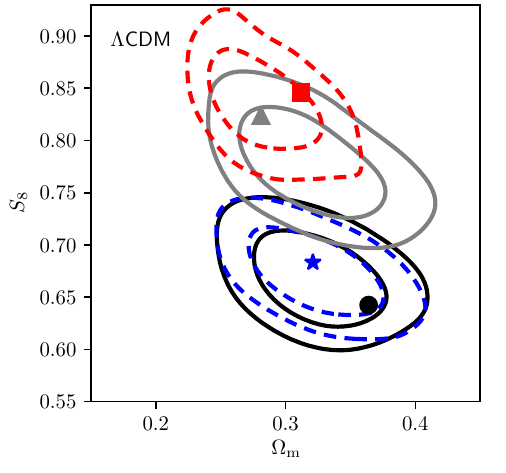}
\includegraphics[width=\columnwidth]{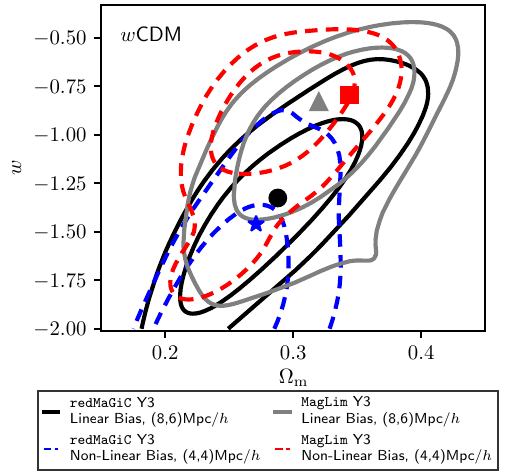}
\caption[]{Comparing the constraints from $2\times2$pt between the \redmagic and Maglim samples. The black dot and blue star denote the MAP point estimate for \redmagic linear and non-linear bias model respectively, while the gray triangle and red square show the same for the \maglim sample. 
}\label{fig:maglim_comp}
\end{figure}

In Fig.~\ref{fig:maglim_comp}, we show the comparison of the cosmology constraints obtained from $2\times 2$pt analysis using the \maglim sample (see \citet*{y3-2x2ptaltlensresults}) with the results obtained here with the \redmagic lens galaxy sample. The top panel compares the $\Omega_{\rm m}- S_8$ contours assuming $\Lambda$CDM cosmology while the bottom panel compares the $\Omega_{\rm m}- w$ contours assuming $w$CDM cosmology. We compare both the linear bias and the non-linear bias model at the (8,6) Mpc/$h$ and (4,4) Mpc/$h$ scale cuts respectively. We again find that the $S_8$ constraints obtained with the \redmagic sample are low compared to the \maglim sample for both linear and non-linear bias models. As the source galaxy sample, the measurement pipeline and the modeling methodology used are the same for the two $2\times 2$pt analysis, this suggests that the preference for low $S_8$ in our \textit{fiducial} results is driven by the Y3 \redmagic lens galaxy sample, which we investigate in the following sub-sections. 

We also show the maximum a posteriori (MAP) estimate in the $\Omega_{\rm m}- S_8$ and the $\Omega_{\rm m}- w$ planes, in order to estimate the projection  effects arising from marginalizing over the large multi-dimensional space to these two dimensional contours (see Fig.~\ref{fig:sim_lin} and Fig.~\ref{fig:bcc_des_wcdm}). We find that the non-linear bias model suffers from mild projection effects within the $w$CDM model (although note the caveats about the MAP estimator mentioned in \S\ref{sec:param_inf}). We also emphasize that using the non-linear galaxy bias model with smaller scale cuts gives similar improvement in the figure-of-merit of the cosmology contours shown in Fig.~\ref{fig:maglim_comp}, using both \redmagic and \maglim lens galaxy samples. 

\begin{figure*}
\includegraphics[width=\textwidth]{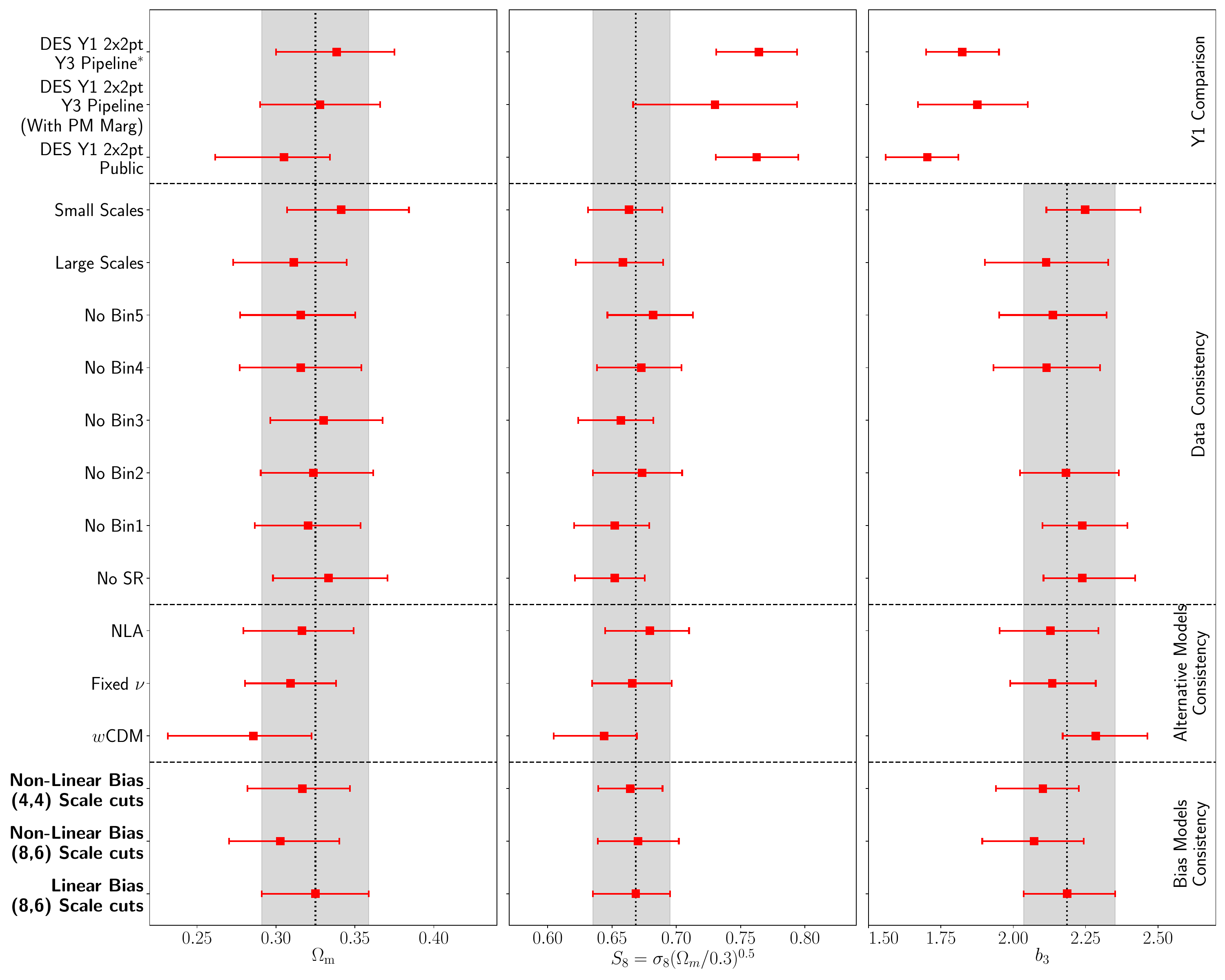}
\caption[]{The consistency of the \redmagic $2\times2$pt cosmology and galaxy bias constraints when changing the analysis choices (see \S\ref{sec:internal_consistency} for details). We also compare our constraints to the DES Y1 public $2\times2$pt results as well as its re-analysis with the current analysis pipeline ($*$ -- we fix the point mass parameters when re-analyzing the DES Y1 data due to the large degeneracy between point mass parameters and cosmology at the scale cuts described and validated in \citet{Abbott_2018}).}
\label{fig:2x2pt_consistency}
\end{figure*}

\subsection{Internal consistency of the \redmagic results}
\label{sec:internal_consistency}

To investigate the low $S_8$ constraints in the \textit{fiducial} analysis of the \redmagic galaxy sample, we first check various aspects of the modeling pipeline. In Fig.~\ref{fig:2x2pt_consistency}, we show the constraints on $\om$, $S_8$ and galaxy bias for the third tomographic bin $b_3$, for various robustness tests. We choose to show the third tomographic bin for the galaxy bias constraints as this bin has the highest signal-to-noise ratio. We divide the figure into three parts, separated by horizontal black lines. The bottom panel shows the marginalized constraints from the results described in the previous subsection (see Fig.~\ref{fig:des_comp}). As mentioned previously, we obtain completely consistent constraints from both linear and non-linear bias models. To check the robustness and keep the interpretation simple, we use the linear bias model using the scale cuts of (8,6) Mpc/$h$ in the following variations.  

In the next part of the Figure, moving upwards from the bottom, we test the robustness of the model. In particular, we check the robustness of the \textit{fiducial} intrinsic alignment model by using the NLA model. We also run the analysis by fixing the neutrino masses to $\Omega_{\nu}h^2 = 0.00083$. This choice of $\Omega_{\nu}h^2$ parameter corresponds to the sum of neutrino masses, $\sum m_{\nu}=0.06$eV at the \textit{fiducial} cosmology described in Table~\ref{tab:params_all} (which is the baseline value used in the \textit{Planck} 2018 cosmology results as well \citep{Planck_2018_cosmo}). Lastly, we test the impact of varying the dark energy parameter using the $w$CDM model. We find entirely consistent constraints for all of the above variations. 

In the next part of the figure, we test the internal consistency of the datavector. Firstly we remove the contribution of shear-ratio information to the total likelihood, resulting in entirely consistent constraints. Also, note that the size of constraints on the cosmological parameters do not change in this case compared to the \textit{fiducial} results. This demonstrates that the majority of constraints on the cosmological and bias parameters are obtained from the $\wtheta$ and $\gammat$ themselves. We also test the impact of removing one tomographic bin at a time from the datavector. We find consistent constraints in all five cases. 
Lastly, we also split the datavector into large and small scales. The small-scales run uses the datavector between  angular scales corresponding to (8,6) Mpc/$h$ and (30,30) Mpc/$h$. The large-scales run uses the datavector between angular scales corresponding to (30,30) Mpc/$h$ and 250 arcmins. When analyzing the large scales, we fix the point-mass parameters to their \textit{fiducial} values (see Table~\ref{tab:params_all}), because of the large degradation of constraining power at these larger-scale cuts due to the degeneracy between point-mass parameters, galaxy bias and cosmological parameter $\sigma_8$ (see Appendix~\ref{app:pm} and \citet{MacCrann:2019ntb}). In both of these cases, we find similar constraints on all parameters, demonstrating that the low $S_8$ does not originate from either large or small scales.

As an additional test of the robustness of the modeling pipeline, we analyze the $\wtheta$ and $\gammat$ measurements as measured from DES Y1 data \citep{Abbott_2018}. Note that in this analysis, we keep the same scale cuts as described and validated in \citet{Abbott_2018}, which are $8$ Mpc/$h$ for $\wtheta$ and $12$ Mpc/$h$ for $\gammat$. To analyze this datavector, while we use the model described in this paper, we fix the point-mass parameters again to zero due to similar reasons as described above in the analysis of large scales. The constraints we obtain are consistent with  the public results  described in \citet{Abbott_2018}. We attribute an approximately $1\sigma$ shift in the marginalized $\Omega_{\rm m}$ posterior to the improvements made in the current model, compared to the model used for the public Y1 results \citep{Krause2017}. In particular, we use the full non-limber calculation, including the effects of redshift-space distortions, for galaxy clustering (also see \citet{Fang_nonlimber}).

\begin{figure}
\includegraphics[width=\columnwidth]{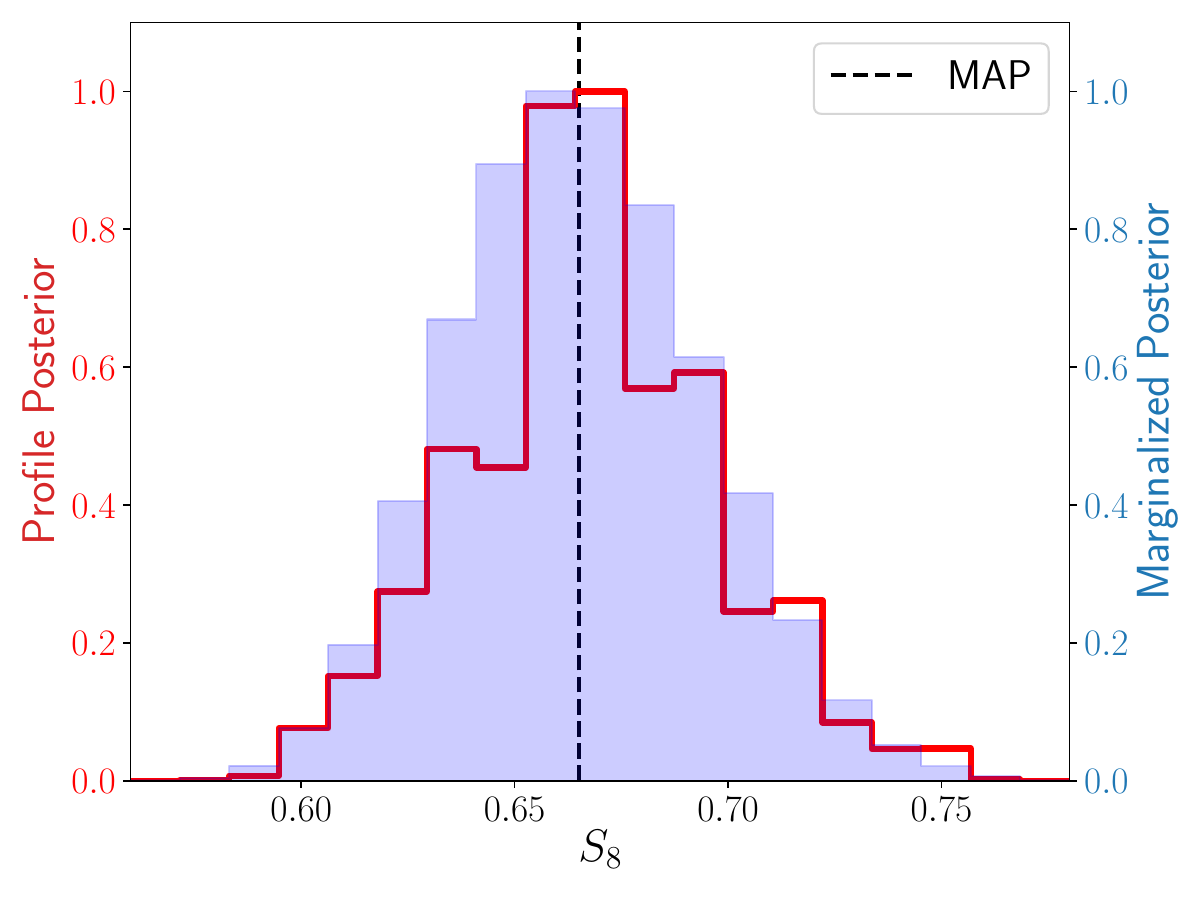}
\caption[]{Comparison of the profile posterior and marginalized posterior on the $S_8$ parameter from the $2\times2$pt \redmagic LCDM chain.}
\label{fig:prof_like}
\end{figure}

Lastly, to assess the impact of projection effects on the $S_8$ parameter, we compare the profile posterior to the marginalized posterior. The profile posterior in Fig.~\ref{fig:prof_like} is obtained by dividing the samples into 20 bins of $S_8$ values and calculating the maximum posterior value for each bin. Therefore, unlike the marginalized posterior, the profile posterior does not involve the projection of a high dimensional posterior to a single $S_8$ parameter. Hence the histogram of the profile posterior is not impacted by projection effects. We compare the marginalized posterior and profile posterior in Fig.~\ref{fig:prof_like}, showing that projection effects have a sub-dominant impact on the marginalized $S_8$ constraints. This demonstrates that projection effects do not explain the preference for low $S_8$ with the \redmagic sample. 
 
In summary,  the results presented in this sub-section demonstrate that our modeling methodology is entirely robust, and hence we believe our data are driving the low $S_8$ constraints with the \redmagic  sample. Moreover, as described above, no individual part of the data drives a low value of $S_8$; therefore, we perform global checks of the datavector in the following sub-sections.

\subsection{Galaxy bias from individual probes}

\begin{figure}
\includegraphics[width=\columnwidth]{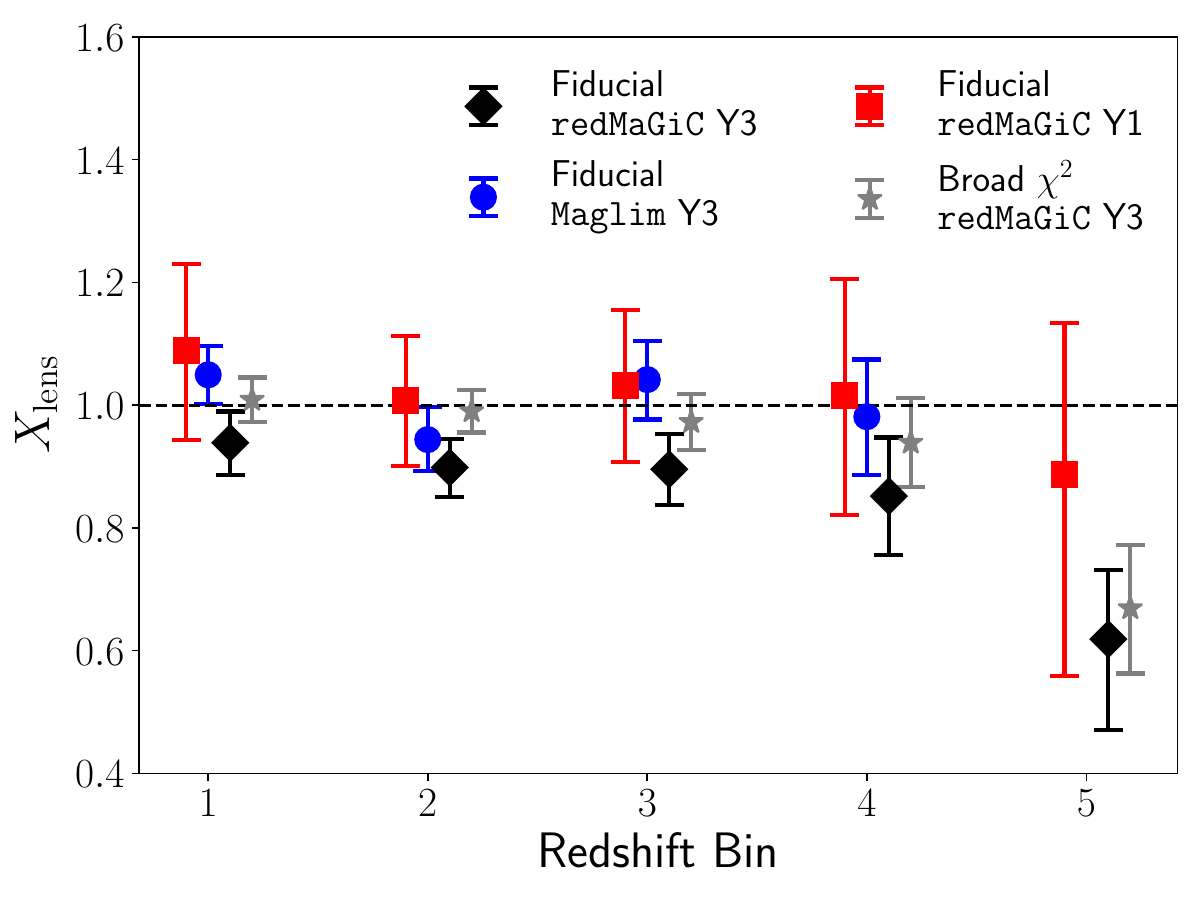}
\caption[]{Constraints on the phenomenological de-correlation parameter, $X_{\rm lens}$, for each tomographic bin obtained from $2\times 2$pt analysis using Y1 \redmagic,  Y3 fiducial \redmagic sample, Y3 broad-$\chi^2$ \redmagic sample (see \S~\ref{sec:bchi2_cosmo}) and Y3 \maglim as the lens galaxies (the cosmological parameters are fixed to the DES Y1 best-fit values \citep{Abbott_2018}).}
\label{fig:Xlens_5X_y1y3}
\end{figure}

In this sub-section, we test the compatibility of the $\wtheta$ and $\gammat$ parts of the datavector. As we will lose the power of complementarity when analyzing them individually, we fix the cosmological parameters to the maximum posterior obtained from the DES Y1 $3\times 2$pt analysis \citep{Abbott_2018}. We find that the best-fit bias values from the $\wtheta$ part of the datavector are systematically higher than $\gammat$ for each tomographic bin. We parameterize this difference in bias values with a phenomenological parameter $X$ for each tomographic bin $i$ as:
\begin{equation}
    X^{i}_{\rm lens} = b^i_{\gammat}/b^i_{\wtheta}
\end{equation}
% $X^{i}_{\rm lens} = b^i_{\gammat}/b^i_{\wtheta}$. 
We refer to $X$ as a "de-correlation parameter" because its effect on the data is very similar to assuming that the mass and galaxy density functions have less than 100\% correlation.  A value of $X=1$ is expected from local biasing models.
The constraints on the parameter $X^{i}_{\rm lens}$ are shown in Fig.~\ref{fig:Xlens_5X_y1y3}. We also compare the constraints of these $X^{i}_{\rm lens}$ parameters obtained from Y1 \redmagic $2 \times 2$pt (see \citet{Abbott_2018} and \citet{gglpaper} for details) and the $2 \times 2$pt datavector using Y3 \maglim lens galaxy sample. For the Y1 \redmagic datavector,  we fix the scale cuts and priors on the calibration of photometric redshifts of lens and source galaxies as described in the \citet{Abbott_2018} and for analysis of Y3 \maglim datavector we follow the analysis choices detailed in \citet*{y3-2x2ptaltlensresults}. In this analysis of all the datavectors, we use the linear bias galaxy bias model while keeping the rest of the model the same as described in \S\ref{sec:model_rest}. We find that the Y1 \redmagic as well as Y3 \maglim $2\times 2$pt data are consistent with $X^{i}_{\rm lens} = 1$ for all the tomographic bins, while \redmagic Y3 $2\times 2$pt data have a persistent preference for $X^{i}_{\rm lens} < 1$ for all the tomographic bins.

Noticeably, we find that for the DES Y1 best-fit cosmological parameters, the Y3 \redmagic datavector prefers a value of $X^{i}_{\rm lens} \sim 0.9$ for each tomographic bin. Therefore, in order to keep the interpretation simple, we use a single parameter $X_{\rm lens}$ to describe the ratio of galaxy bias $b^i_{\gammat}/b^i_{\wtheta}$ for all tomographic bins $ i \in [1,5]$. We constrain this single redshift-independent parameter to be $X_{\rm lens} = 0.9^{+0.03}_{-0.03}$ for Y3 \texttt{redMaGiC}, a 3.5$\sigma$ deviation from $X_{\rm lens} = 1$. Within general relativity, even when including the impact of non-linear astrophysics, we do not expect a de-correlation between galaxy clustering and galaxy-galaxy lensing to be present at more than a few percent level \citep{Desjacques_2018}. We comment on the impact of this de-correlation on the \redmagic cosmology constraints in \S\ref{sec:X_cosmo_impact}. 

Note that the inferred value of $X_{\rm lens}$ depends on the cosmological parameters, because the large-scale amplitudes of galaxy clustering and galaxy-galaxy lensing involve different combinations of galaxy bias, $\sigma_8$ and $\Omega_{\rm m}$. Therefore, a self-consistent inference of $X_{\rm lens}$ requires the full $3\times2$pt datavector and is presented in \citet*{y3-3x2ptkp}. However, the DES Y1 $3\times2$pt best-fit cosmological parameters are fairly close to the DES Y3 $3\times2$pt best-fit parameters, therefore we expect  the results presented here to be good approximations to those obtained with the Y3 $3\times2$pt datavector.

\subsection{Area split of the de-correlation parameter}
In order to further study the properties of this de-correlation parameter $X_{\rm lens}$, we estimate it independently in 10 approximately equal area patches  of the DES Y3 footprint.  We measure the datavectors, $w(\theta)$ and $\gammat$ in each of these 10 patches, using the same methodology presented in \S\ref{sec:2pt_data}. In order to obtain the covariance for each patch, we rescale the \textit{fiducial} covariance (see \S\ref{sec:cov}) of the full footprint to the area of each patch. We then estimate  $X_{\rm lens}$ from each patch while keeping all the other analysis choices the same.

\begin{figure}
\includegraphics[width=\columnwidth]{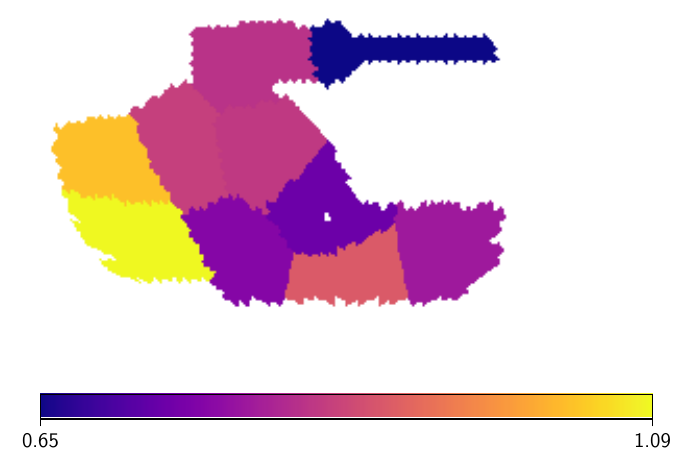}
\caption[]{The DES footprint is split into 10 regions. The color of each area corresponds to the mean value of the constraints on $X_{\rm lens}$ from that particular area, inferred at fixed DES-Y1 cosmology and using the \redmagic lens sample. This plot shows that no special region in the sky (for example, near the galactic plane) drives the preference for low $X_{\rm lens}$. While a variation over the sky in the inferred $X_{\rm lens}$ is expected from analyzing only the $2\times 2$pt data due to the variations in the photometric redshift distribution of source galaxies, we find that the preferred mean value of $X_{\rm lens}$ from the \redmagic sample is significantly lower than the expected value of 1 (see Fig.~\ref{fig:Xrm_mag_scatter}).}
\label{fig:Xlens_area_split}
\end{figure}

In Fig.~\ref{fig:Xlens_area_split} we show the DES footprint split into 10 regions. In this figure, each region is colored in proportion to the mean value of the $X_{\rm lens}$ parameter we obtain using \redmagic as the lens galaxy sample. We run a similar analysis when using \maglim as the lens sample.

\begin{figure}
\includegraphics[width=\columnwidth]{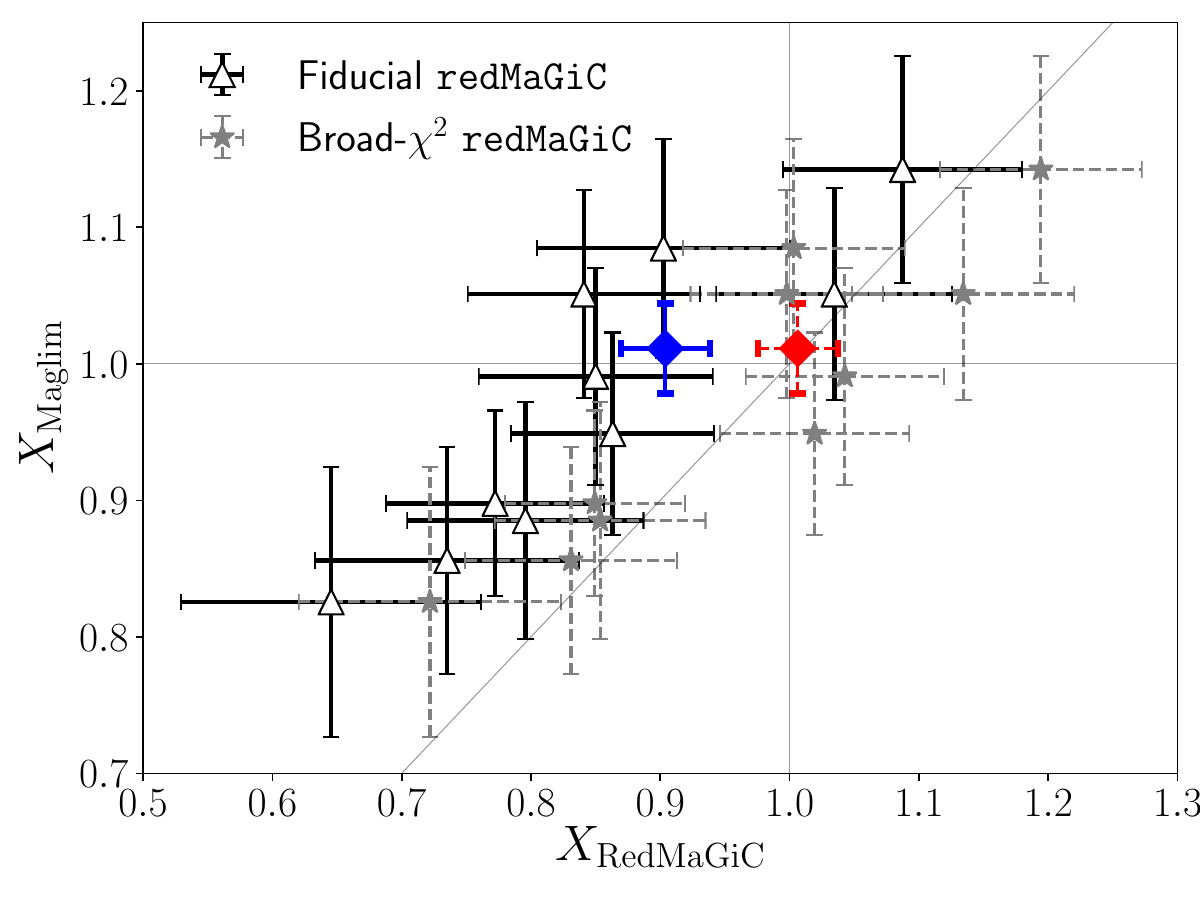}
\caption[]{Each errorbar corresponds to the 68\% credible interval constraints on  $X_{\rm lens}$ from one of the 10 regions (see Fig.~\ref{fig:Xlens_area_split}), using either the \redmagic sample or the \maglim lens galaxy sample. The blue errorbar corresponds to the constraints on $X_{\rm lens}$ from the entire Y3 area using the \maglim sample and the fiducial \redmagic sample, whereas the red errorbar uses the broad-$\chi^2$ galaxy sample (see \S~\ref{sec:bchi2_cosmo}). We find a tight correlation between $X_{\rm Redmagic}$ and $X_{\rm Maglim}$, due to common sources of statistical noise (e.g., photometric redshifts of the source galaxies). For example, if the photometric redshift distribution of the source galaxies fluctuates from the imposed prior for each patch in a different direction, it would shift the inferred $X_{\rm lens}$ similarly for each lens sample. We find that, while the inferred $X_{\rm lens}$ from 10 regions using the \maglim and the broad-$\chi^2$ \redmagic sample fluctuates around its mean value that is close to 1, the inference from the fiducial \redmagic sample fluctuates around a mean value that is significantly lower than 1. This figure shows that the fiducial \redmagic sample prefers $X_{\rm lens}$  to be lower than 1, independent of the sky-area. }
\label{fig:Xrm_mag_scatter}
\end{figure}

In Fig.~\ref{fig:Xrm_mag_scatter} we show a scatter plot between the value of $X_{\rm lens}$ recovered from each of 10 regions using \redmagic and \maglim as lens samples. We find a tight correlation between the value of $X_{\rm lens}$ from the two lens samples, as would be expected if they share noise from sample variance and fluctuations in the source galaxy population. Note that the scatter in the inferred $X_{\rm lens}$ for both the \maglim and the \redmagic samples corresponding to each sky patch (red points) around the mean of full sample (the blue point), is consistent with the expectation. 
This shows that, compared with \maglim, the \redmagic lens sample has a preference for $X_{\rm lens} < 1$ in the whole DES footprint. 
% Fig.~\ref{fig:Xrm_mag_scatter} also suggests a large variation  in the inferred $X_{\rm lens}$ over the footprint, when compared to \maglim. 
This correlation and area independence of the ratio $X_{\rm Redmagic}/X_{\rm Maglim}$ is remarkable and suggests that the potential systematic in the \redmagic sample has a more global origin.

% \begin{figure}
% \includegraphics[width=\columnwidth]{X_rm_mag_area_scatter.pdf}
% \caption[]{Each red errorbar corresponds to the 68\% credible interval constraints on  $X_{\rm lens}$ from each of the 10 regions (see Fig.~\ref{fig:Xlens_area_split}), using either \redmagic or \maglim lens galaxy sample. The blue errorbar corresponds to the constraints on $X_{\rm lens}$ from the full Y3 area. We find a tight correlation between $X_{\rm Redmagic}$ and $X_{\rm Maglim}$, due to common sources of statistical noise (e.g., photometric redshift of the source galaxies). We find that, while $X_{\rm Maglim}$ fluctuates around its mean value that is close to 1, $X_{\rm Redmagic}$ fluctuates around a mean value that is significantly lower than 1. This figure shows that $X_{\rm Redmagic}$ prefers to be lower than 1, independent of the sky-area.}
% \label{fig:Xrm_mag_scatter}
% \end{figure}

\begin{figure}
\includegraphics[width=\columnwidth]{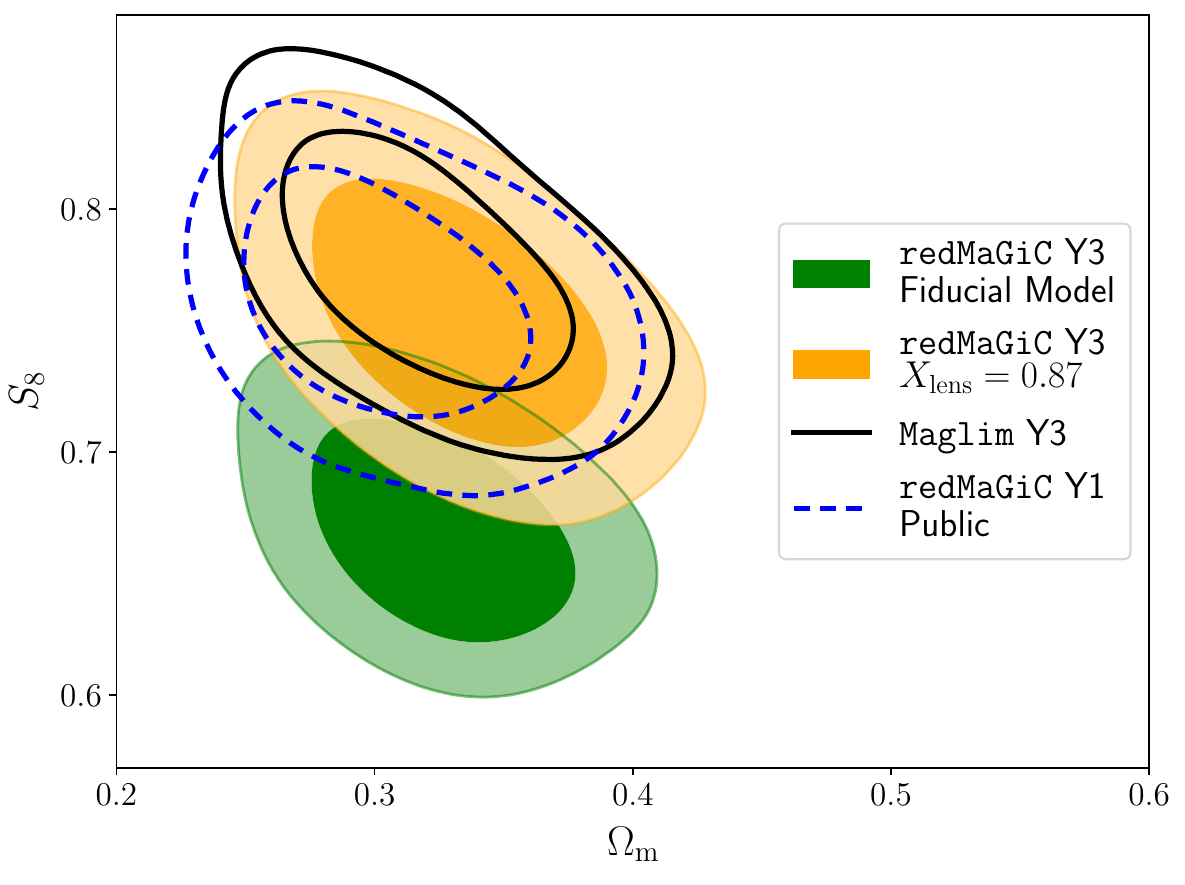}
% \vfil
% \includegraphics[width=\columnwidth]{data_wcdm_comp.pdf}
\caption[]{Comparison of the constraints from $2\times2$pt analysis when using the mean value of $X_{\rm lens}$ parameter for \redmagic lens sample analysis, as estimated and described in \citet*{y3-3x2ptkp}. We find a shift in $S_8$ parameter compared to our \textit{fiducial} results in \S\ref{sec:fid_cosmo_res}, but $\om$ constraints are fully consistent.   }\label{fig:X_comp_main}
\end{figure}

\subsection{Impact of de-correlation on $2\times 2$pt cosmology}
\label{sec:X_cosmo_impact}

To summarize, assuming a standard cosmological model, we have identified that the galaxy-clustering and galaxy-galaxy lensing signal measured using the Y3 \redmagic lens galaxy sample are incompatible with each other (at the set of cosmological parameters preferred by previous studies). We have further identified that this incompatibility is well-captured by a redshift-, scale- and area-independent phenomenological parameter $X_{\rm lens}$. Using Y3 \redmagic lens sample, we detect $X_{\rm lens} \sim 0.9$, at the 3.5$\sigma$ confidence level away from the expected value of $X_{\rm lens} = 1$. This $2\times2$pt analysis is done when the cosmological parameters are fixed to their DES Y1 best-fit values; a self-consistent $X_{\rm lens}$ inference analysis with free cosmological parameters requires the full $3\times2$pt datavector. This is presented in \citet*{y3-3x2ptkp}, where the inferred constraints on this de-correlation parameter are $X_{\rm lens} = 0.87^{+0.02}_{-0.02}$. 

In Fig.~\ref{fig:X_comp_main}, we fix  $X_{\rm lens} = 0.87$ in our model and re-run the Y3 \redmagic $2\times2$pt analysis. We find, as expected, that this has a significant impact on the marginalized $S_8$ values and results in the marginalized constraints $S_8 = 0.76_{-0.037}^{+0.034}$, completely consistent with $2\times2$pt Y1 \redmagic public results as well as Y3 \maglim results. Also note that the marginalized constraints on $\Omega_{\rm m}$ for $X_{\rm lens} = 0.87$ model are $\om = 0.331^{+0.037}_{-0.037}$, which remains consistent with the \textit{fiducial} result. 

\subsection{Broad-$\chi^2$ \redmagic sample}
\label{sec:bchi2_cosmo}
% \begin{figure}
% % \includegraphics[width=\textwidth]{b1_nbar_Mh_sim_marg_wcov.pdf}
% \includegraphics[width=\columnwidth]{xlens_comp_bchi2.pdf}
% \caption[]{
% We show the constraints on $X_{\rm lens}$ for each tomographic bins as well as for all bins combined from both the fiducial and braod-$\chi^2$ \redmagic lens samples. 
% }
% \label{fig:xlens_bchi2}
% \end{figure}

% We show the results with the broad-$\chi^2$ sample in this sub-section. 
In order to further investigate the source of the de-correlation, we modify the $\chi_{\rm RM}^2$ threshold for a galaxy to be classified as a \redmagic galaxy when fitting to the \redmagic template using the procedure as described in \citet{Rozo_2016}. As described in \S~\ref{sec:redmagic_def}, the fiducial \redmagic catalog is generated by implementing the $\chi_{\rm RM}^2$ threshold of 3. This low-$\chi_{\rm RM}^2$ threshold only selects the galaxies that closely match the template. In case there are any residual variations in the \redmagic catalog number densities caused by variations in the colors that are not already corrected using the fiducial weighting scheme (as described in \citet{y3-galaxyclustering}), it would contribute a spurious galaxy clustering signal. This would contribute towards $X_{\rm lens} < 1$, as we found above. In order to test this hypothesis, we increase the threshold criteria and generate another catalog with $\chi_{\rm RM}^2 = 8$ and denote this new sample as the ``broad-$\chi^2$" sample. 
% and hence generating a sample selection that 

\begin{figure}
\includegraphics[width=\columnwidth]{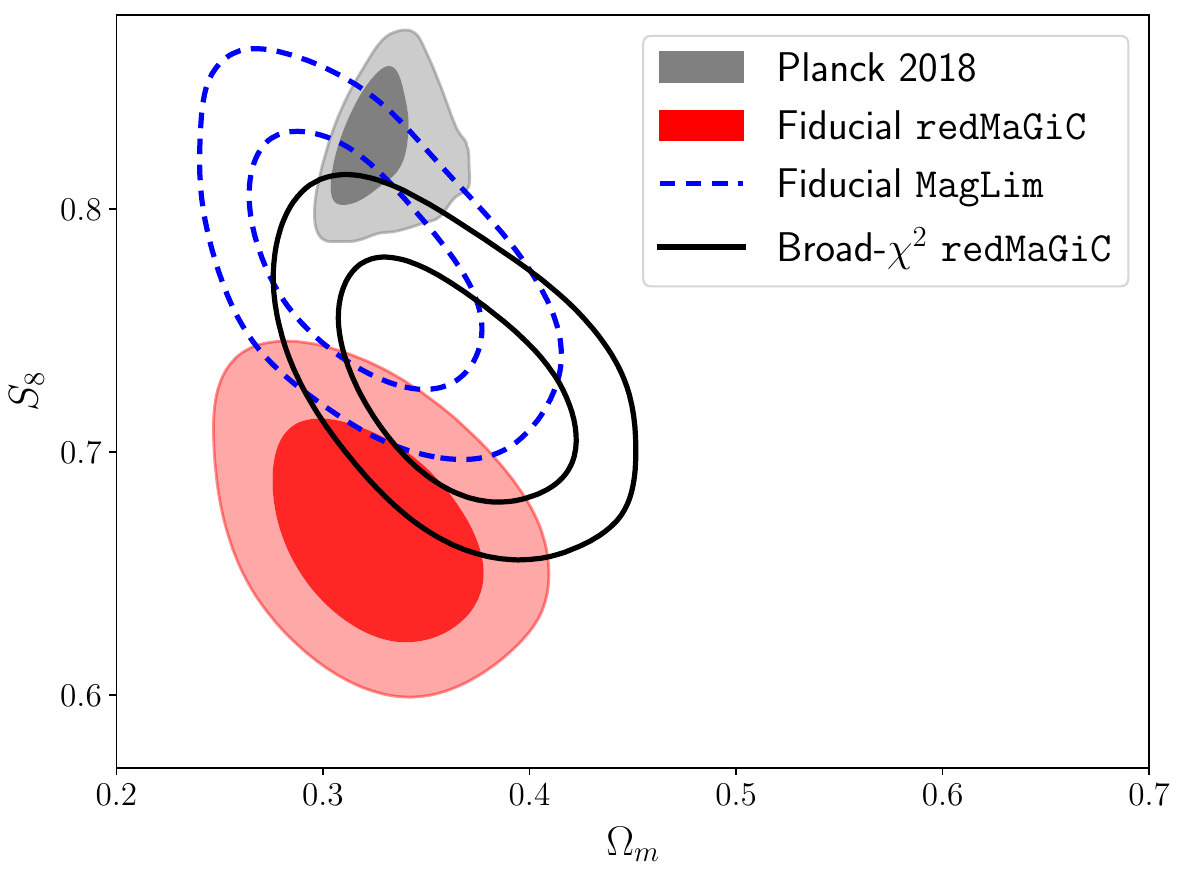}
\includegraphics[width=\columnwidth]{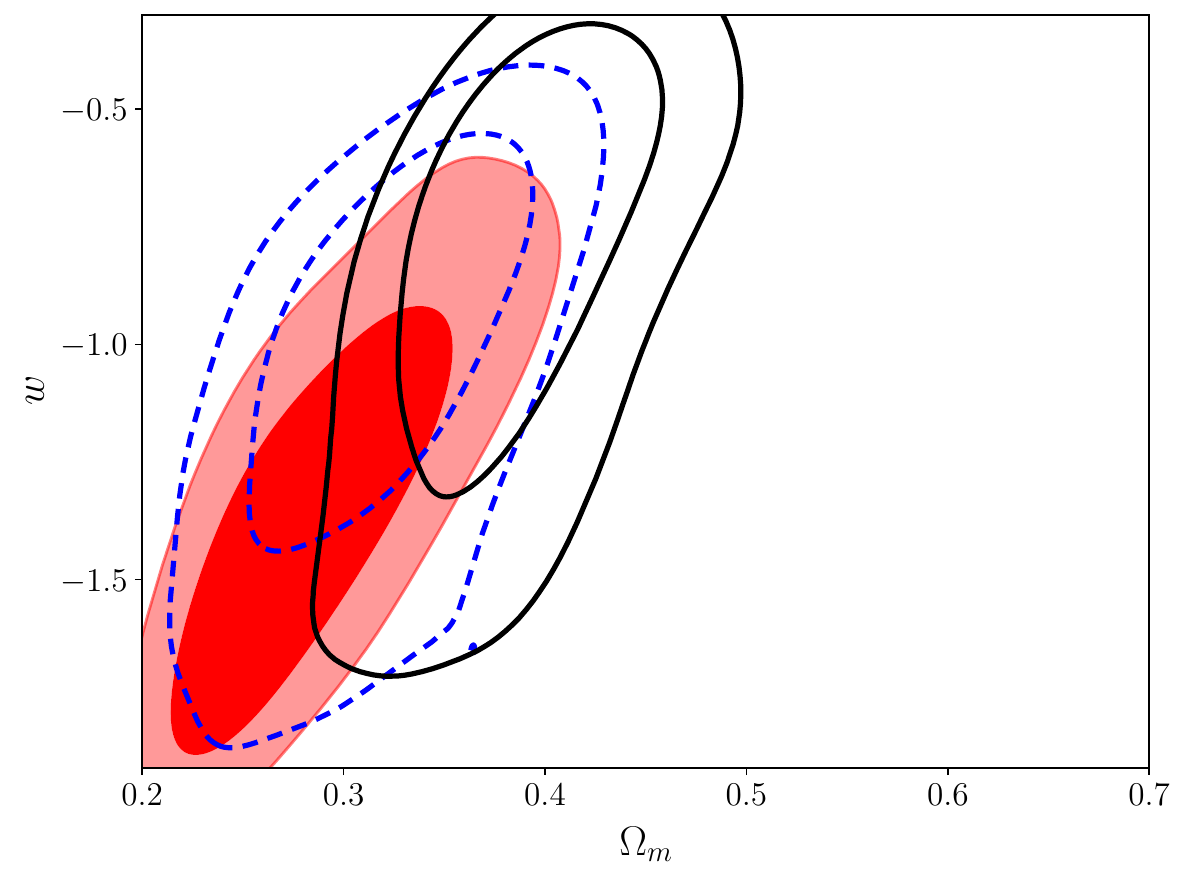}
\caption[]{
Constraints on the cosmological parameters using the linear bias model with the broad-$\chi^2$ \redmagic sample.  The top figure corresponds to the $\Lambda$CDM cosmological model and the bottom figure corresponds to the $w$CDM cosmological model. We also compare the constraints from the fiducial \redmagic and the fiducial \maglim lens galaxy samples. 
}
\label{fig:cosmo_bchi2}
\end{figure}

We show the result for $X^{i}_{\rm lens}$ for all the five tomographic bins in Fig.~\ref{fig:Xlens_5X_y1y3}. 
% the new sample as well as combined X again inferred at Y1-MAP cosmology in Fig.~\ref{fig:xlens_bchi2}. 
We find that with the broad-$\chi^2$ sample, $X^i_{\rm lens}$ is consistent with 1 for the first four tomographic bins. While we still find that for the fifth tomographic bin, $X^5_{\rm lens}<1$, this bin has low constraining power. We also show the inferred $X_{\rm lens}$ from 10 independent regions over the DES footprint in the Fig.~\ref{fig:Xlens_area_split}. We find that with the new sample, the scatter in the inferred $X_{\rm lens}$ is consistent with expected value of 1. Moreover, as shown with a red symbol in the Fig.~\ref{fig:Xlens_area_split}, we find the redshift and area independent $X_{\rm lens}$ to be entirely consistent with 1 using the broad-$\chi^2$ sample. This validates our hypothesis and points towards an uncorrected systematic that might be related to a color-dependent photometric issue in the DES data. Since the shear catalog, as well as the \maglim galaxy catalog, do not select galaxies based on a red-galaxy template, we do not expect this systematic to have an effect on those catalogs. 

In Appendix~\ref{app:bchi2}, we further describe details of this new sample and compare it with our fiducial \redmagic sample. 
With this new sample, we use conservative analysis choices and implement the following approximations: 
\begin{itemize}
    \item We downsample the broad-$\chi^2$ catalog to roughly match the number densities of the fiducial \redmagic sample. This ensures that the validations of analysis choices performed for the redmagic sample, including the covariance, scale cuts, and methodology, remain true for the broad-$\chi^2$ sample as well.
    \item We use a two-parameter model (shift and stretch parameterization) to account for the uncertainty in the lens redshift distribution for each tomographic bin \cite{y3-lenswz}. We implement this model to reduce the impact of the outliers in the assigned galaxy redshifts for this new sample. The Gaussian priors on the shift and stretch parameters are tabulated in Table~\ref{tab:params_nz_bchi2}. 
    % \item We use the non-linear bias model of galaxy biasing to infer the cosmology constraints. Lastly, we 
\end{itemize}
% we use the non-linear bias model to infer the cosmological constraints. 
We show the cosmological constraints from the broad-$\chi^2$ sample in Fig.~\ref{fig:cosmo_bchi2} and find that they are consistent with the results from \maglim sample in both $\Lambda$CDM and $w$CDM cosmological models. Using the $\Lambda$CDM model, we constrain $\om = 0.363^{+0.0375}_{-0.0388}$ and $S_8 = 0.73^{+0.035}_{-0.029}$, and using the $w$CDM model, we constrain $w = -0.821^{+0.1908}_{-0.4341}$.  
% . We find that our constraints on the cosmological parameters are now fully consistent with \maglim sample. 

We note that this analysis is showing the constraints on the cosmological parameters under the approximation that we neglect the contribution to the LSS covariance systematic term. We use the ISD method to get the weights for this sample. Moreover, we assume that the same scale cuts work with this sample as we obtained for the fiducial \redmagic sample. Lastly, we have used the same value of lens magnification as for the fiducial \redmagic sample. We do not expect these choices to have any major effects on the cosmological constraints described above. However, we leave a detailed study optimizing the $\chi_{\rm RM}^2$ value, validating the analysis choices, and obtaining final constraints with \redmagic sample to \citet{DES_Y3_Xf}.

    % \item We do not use the LSS covarinace systematic term.
    % \item We assume the scale cuts hold between the new and the 
    % \item 
% \end{itemize}

\subsection{\redmagic host halo mass inference}
\label{sec:halomass}

\begin{figure}
\includegraphics[width=\columnwidth]{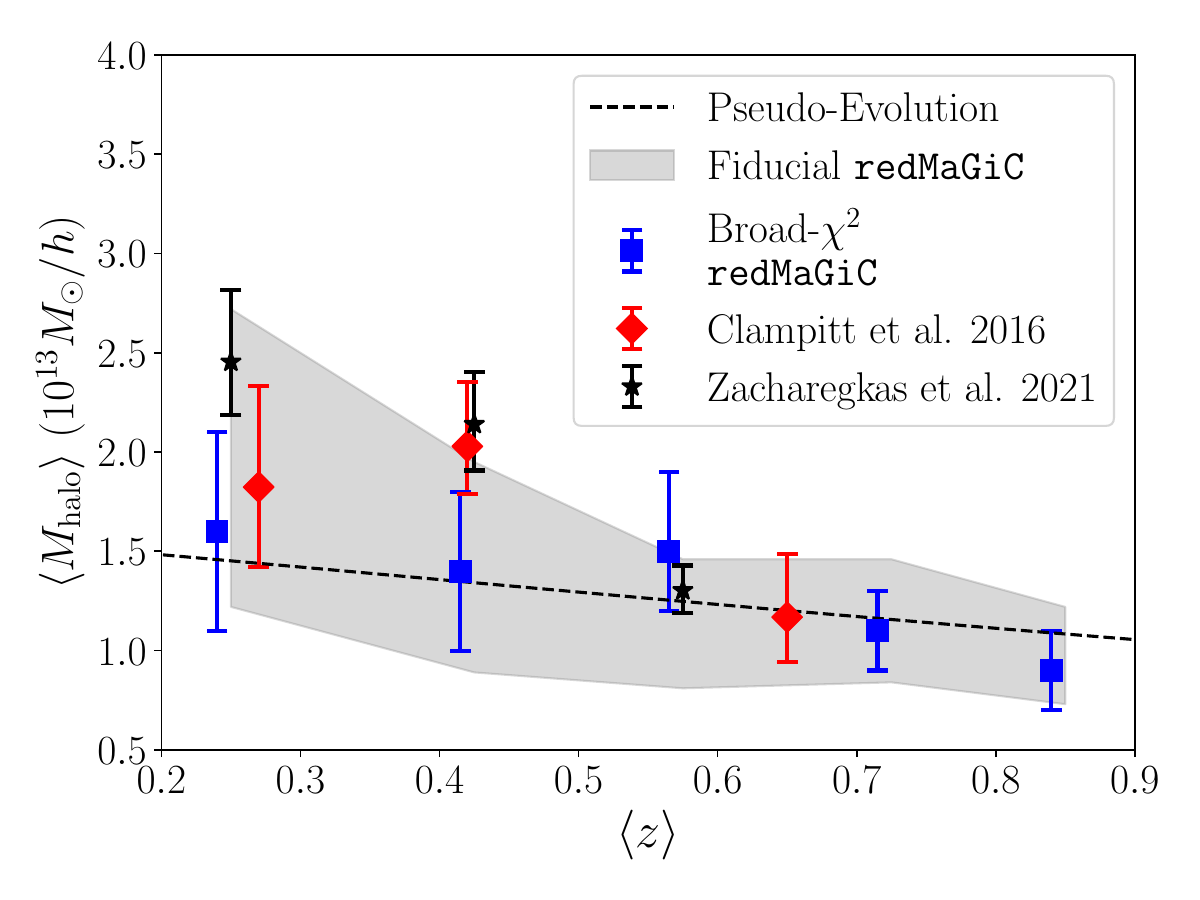}
\caption[]{
This figure shows the inferred constraints on mean host halo masses of \redmagic galaxies for five tomographic bins. We use the HOD framework to make this inference as detailed in the Appendix \ref{app:halo_mass} and use the linear bias constraints obtained using the broad-$\chi^2$ \redmagic sample. We infer the mean host halo masses from the linear bias constraints for all the five tomographic bins. We compare our results to \citet{Clampitt_2016} and \citet{zacharegkas_hod}, and also show the expected pseudo-evolution of a halo having $M_{\rm halo} = 1.6\times 10^{13} M_{\odot}/h$ at $z=0$. 
}
\label{fig:bias_mass_nbar}
\end{figure}
% \SP{Change the text to use the broad-$\chi^2$ sample details}
In the halo model framework (see \citet{COORAY_2002} for a review), the value of the linear bias of a tracer of dark matter can be related to the host halo mass of that tracer. The standard halo occupation distribution (HOD) approach parameterizes the distribution of  galaxies inside  halos, and hence the observed number density as well as the large scale bias values of any galaxy sample can be expressed in terms of its HOD parameters \citep{Berlind_2002, Zheng_2005, Zehavi_2011}. The same HOD parameters can also be used to infer the mean host halo mass of the galaxy sample. We use the constraints on  linear galaxy bias parameters and the co-moving number density to infer the mean host halo mass of the broad-$\chi^2$ \redmagic galaxy sample by marginalizing over HOD parameters.

The details of the halo model framework used here are given in Appendix \ref{app:halo_mass}. Note that we have neglected the effects of assembly bias and the correlation between number density and bias constraints in this analysis. With these caveats in mind, in Fig.~\ref{fig:bias_mass_nbar} we show approximately 25\% constraints on mean host halo mass of broad-$\chi^2$ \redmagic galaxies and the constraints for different tomographic bins show its evolution with redshift. This redshift evolution trend is broadly consistent with the pseudo-evolution of halo masses due to changing background reference density with redshift (see \citep{Diemer_2013} for more details). Therefore we find that the broad-$\chi^2$ \redmagic sample lives in halos of mass of approximately $1.6 \times 10^{13} M_{\odot}/h$, which remains broadly constant with redshift.

We also bracket the uncertainty in the host-halo masses of the fiducial \redmagic sample with a gray band in Fig.~\ref{fig:bias_mass_nbar}. In order to estimate the band, we use the linear bias constraints from the $2\times2$pt analysis with fiducial \redmagic sample, after fixing $X_{\rm lens}=0.87$. This de-correlation parameter results in $w(\theta)$ and $\gamma_{\rm t}$ preferring different linear bias parameters, related by $b^{i}[w(\theta)]/b^{i}[\gamma_{\rm t}(\theta)] = X_{\rm lens} = 0.87$, for all tomographic bins $i$. Therefore, we infer the host halo masses using both linear bias parameter values. The band is estimated by using the lower limit of masses inferred by $b^{i}[\gamma_{\rm t}(\theta)]$ and upper limit of masses inferred by $b^{i}[w(\theta)]$ for all tomographic bins $i$. We find that the broad-$\chi^2$ sample has a slight preference for lower halo masses, but it is consistent with constraints for the fiducial sample. 

We find that our results are also broadly consistent with the analysis of \citep{Clampitt_2016}, which used the \redmagic galaxies of DES Science-Verification dataset and estimated the mean halo masses by studying galaxy-galaxy lensing signal in a broad range of scales (including high signal-to-noise small scales that we remove here) using HOD model.\footnote{Note that we use $M_{\rm 200c}$ as our halo mass definition, which denotes the total mass within a sphere enclosing a mean density which is 200 times the \textit{critical} density of the universe. \citet{Clampitt_2016} work with $M_{\rm 200m}$ as their mass definition, denoting the total mass within a sphere enclosing a mean density which is 200 times the \textit{mean} density of the universe, therefore we convert their constraints to $M_{\rm 200c}$ in the above figure.} We also find broad agreement with a similar study presented in \citet{zacharegkas_hod}, analyzing DES Y3 using the galaxy-galaxy lensing data estimated from the fiducial \redmagic sample and on a wide range of scales with an improved halo model.

% \begin{figure}
% % \includegraphics[width=\textwidth]{b1_nbar_Mh_sim_marg_wcov.pdf}
% \includegraphics[width=\columnwidth]{M_const.pdf}
% \caption[]{
% This figure shows the inferred constraints on mean host halo masses of \redmagic galaxies for five tomographic bins. We use the HOD framework to make this inference as detailed in Appendix \ref{app:halo_mass} and use the bias constraints from a linear bias model $2\times2$pt chain, fixing $X_{\rm lens} = 0.87$. We infer the mean host halo masses from the linear bias constraints, $b^{i}(w(\theta))$ and $b^{i}(\gamma_{\rm t}(\theta))$, where they are related as $b^{i}(w(\theta)/b^{i}(\gamma_{\rm t}(\theta) = X_{\rm lens} = 0.87$ for all the five tomographic bins $i$. We compare our results to \citet{Clampitt_2016} and also show the expected pseudo-evolution of a halo having $M_{\rm halo} = 1.6\times 10^{13} M_{\odot}/h$ at $z=0$. 
% }
% \label{fig:bias_mass_nbar}
% \end{figure}

\section{Conclusions}
\label{sec:conclusions}
This paper has presented the cosmological analysis of the $2\times2$pt datavector of the DES Year 3 dataset using \redmagic lens sample. We refer the reader to \citet{y3-2x2ptaltlensresults} for similar results using \maglim lens sample and \citet*{y3-2x2ptmagnification} for details on the impact of lens magnification on the $2\times2$pt datavector. The $2\times2$pt datavector comprises the 2-point correlations of galaxy clustering and galaxy lensing using five redshift bins for the lens galaxies and four bins for source galaxies. It provides independent constraints on two primary parameters of interest, the mass density $\Omega_{\rm m}$ and amplitude of fluctuations $S_8$. As shown in Fig.~\ref{fig:all2pt_comp}, these constraints are complementary to those from cosmic shear. The combination of $2\times2$pt with cosmic shear is thus better able to constrain $\Omega_{\rm m}, S_8$ as well as the dark energy equation of state parameter $w$. Perhaps more importantly, this provides a robustness check on the results from either approach. 

The estimation and marginalization of galaxy bias parameters is one of the central tasks in extracting cosmology from the $2\times2$pt datavector. We have developed and validated the methodology for this based on perturbation theory. We use a five-parameter description of galaxy bias per redshift bin, with three of the parameters fixed based on theoretical considerations. We validated these choices using mock catalogs built on N-body simulations as detailed in our earlier study \citep{p2020perturbation} and Section \S\ref{sec:sims}.  
We carry out two analyses: the first using linear bias with more conservative scale cuts, and the second using the full PT bias model going down to smaller scales. Other  elements of our model include intrinsic alignments, magnification and ``point mass marginalization'' (see \S\ref{sec:model_rest}). The validation of the analysis choice and scale cuts with simulated datavectors (both idealized and from mock catalogs) are presented in \S\ref{sec:analysis_choices}.

Our cosmological results are presented in Figs. \ref{fig:des_comp}, \ref{fig:maglim_comp} and \ref{fig:2x2pt_consistency}, which show preference for low value of $S_8$ parameter when compared with previous results. We refer the reader to \citet{y3-3x2ptkp}, where, after unblinding the cosmological parameter constraints, we find similar inconsistency in the $S_8$ parameter constraints between Y3 $2\times2$pt \redmagic analysis and Y3 cosmic shear analysis, as well as a high $\chi^2$ using the $\Lambda$CDM model. As detailed in \citet{y3-3x2ptkp}, we discovered that the reason for the high $\chi^2$ of the $3\times2$pt analysis with the \textit{fiducial} model was due to inconsistencies in the galaxy-galaxy lensing and galaxy clustering signals. The source of this inconsistency is still undetermined, however we found that a single parameter $X_{\rm lens}$, representing the ratio of the bias inferred from $w(\theta)$ and $\gamma_{\rm t}$, substantially improves the goodness of fit. This ratio is cosmology-dependent and can only be inferred consistently (along with the other model parameters)  when using the full $3\times2$pt analysis, presented in \citet*{y3-3x2ptkp}.

This ratio is expected to be unity in the absence of galaxy stochasticity, an effect that is expected to be only at the percent level on scales above $\sim 10$ Mpc \citep{Desjacques_2018}. Several previous analyses with similar datasets have also found this ratio to be consistent with unity \citep{Mandelbaum_2013, Cacciato_2012, gglpaper}. However, we detect a value of $X_{\rm lens}=0.87$, below 1 at the 5-$\sigma$ level. This purely phenomenological model assumes no scale or redshift dependence, and we found consistent values of $X_{\rm lens}$ when fitting to different scales (see Fig.~\ref{fig:2x2pt_consistency}) and when fitting separate values for each \textit{lens} redshift bin (see Fig.~\ref{fig:Xlens_5X_y1y3}). 
Considering the constraints from the cosmic shear measurements, no known cosmological effect can produce such a large and coherent deviation in clustering and galaxy lensing. We therefore pursued possible systematic errors that could lead to this unusual result. 
% We were unable to verify any such systematics yet, and this is a work in progress. 
Note that this kind of behavior can arise with potential systematics, for example unaccounted-for impact of photometric uncertainty or background subtraction for large or faint objects on the galaxy selection. This can introduce extra fluctuation of the number density of the lens galaxies across the footprint which will not be captured by the set of survey property maps used in the LSS weights estimation pipeline. 

Fig.~\ref{fig:X_comp_main} shows the $2\times2$pt \redmagic cosmology constraints after fixing $X_{\rm lens}=0.87$, the best fit value from \citet{y3-3x2ptkp}. There is a significant shift in $S_8$, while $\Omega_{\rm m}$ remains stable. Interestingly the resulting contours are fully consistent with the Y1 analysis as well as the $2\times2$pt analysis using the \maglim lens galaxy sample \citep{y3-2x2ptaltlensresults}. We track down the source of this de-correlation to an aggressive threshold on the colors of galaxies to match the red-galaxy template. We find that using a sample with a relaxed threshold, which we call the broad-$\chi^2$ sample, results in cosmological constraints that are consistent with the expectations from \maglim sample. This points towards the existence of a potential color-dependent systematic in the galaxy catalog, and we leave a detailed exploration and mitigation of this systematic to a future study \citep{DES_Y3_Xf}.

We note that although recent analyses using BOSS galaxies have found similar inconsistencies in the galaxy clustering and galaxy-galaxy lensing using \Planck-preferred cosmological parameters (see \citep{Leauthaud_2017, Lange_2021} and references there-in); there are some important differences. In this analysis as well as in \citet{y3-3x2ptkp}, unlike in \citet{Leauthaud_2017}, we do not use any small scale information for galaxy clustering and galaxy-galaxy lensing measurements. Therefore, we are significantly less prone to the impacts of poorly understood small scale non-linear physics, like baryon feedback and galaxy assembly biases \citep{Yuan_2020, Amodeo_2021, zu2020lensing}. Moreover, in \citet{y3-3x2ptkp}, by leveraging all the three two-point functions used in $3\times2$pt, the analysis of the consistency between galaxy-lensing and galaxy-clustering can be carried out while freeing the relevant cosmological parameters. The analysis in this paper fixes the cosmological parameters close to the best-fit cosmology from \citet{y3-3x2ptkp}, hence our results are a good approximation to the analysis using the full $3\times2$pt datavector. Similarly, a few recent studies jointly analyzing galaxy clustering auto-correlations and galaxy-CMB lensing cross-correlations have also reported preference for lower galaxy bias value for the cross-correlation compared to the auto correlations \citep{Hang_2020, Kitanidis_2020}. However similar to above analysis with BOSS galaxies, these studies also fix their cosmological parameters to the best-fit cosmology from \Planck results \citep{Planck_2018_cosmo}, which is different from this study (see \citep{krolewski2021cosmological} for related discussion).

To access the information in the measurements on smaller scales, we use higher-order perturbation theory. We use a hybrid 1-loop perturbation theory model for galaxy bias, capturing the non-linear contributions to the overdensity field till third order. We have tested and validated our model using 3-dimensional correlation functions from DES-mock catalogs in \citet{p2020perturbation} as well as with projected statistics in \citet*{y3-simvalidation}; in this study, we validate the bias model with mocks for the $2\times2$pt \redmagic datavector at scales above 4Mpc/$h$. This validation presented here, along with results in \citet{p2020perturbation}, are then also directly used to validate non-linear bias model for \maglim datavector.  We apply it to the data and find that the non-linear bias model results in a gain in constraining power of approximately 17\% in the $\Omega_{\rm m} - S_8$ parameter plane.

A different approach, the halo occupation distribution in the halo model, enables a connection between the masses of halos in which galaxies live and their large-scale bias. We use our constraints on linear bias parameters (along with the galaxy number density) and estimate the host halo masses of \redmagic galaxies. We marginalize over the halo occupation distribution parameters and obtain 25\% constraints on the mean mass of host halos. We show these constraints, including its evolution with redshift in Fig.~\ref{fig:bias_mass_nbar}, finding halo mass of approximately $1.5 \times 10^{13} M_{\odot}/h$ and its evolution with redshift consistent with the expected pseudo-evolution due to changing background density.

The $2\times2$pt combination of probes plays a crucial role in extracting the most cosmological information from LSS surveys, especially in constraining the matter content of universe ($\Omega_{\rm m}$) and the dark energy equation of state ($w$). In this analysis we measure the combination of galaxy clustering and galaxy-galaxy lensing  at approximately 200$\sigma$; this significance  is expected to dramatically increase with imminent large scale surveys like the Euclid Space Telescope,\footnote{https://www.euclid-ec.org} the Dark Energy Spectroscopic Instrument,\footnote{https://www.desi.lbl.gov} the Nancy G. Roman Space Telescope,\footnote{https://roman.gsfc.nasa.gov} and the Vera C. Rubin Observatory Legacy Survey of Space and Time.\footnote{https://www.lsst.org} In order to optimally analyze these high precision measurements, especially at non-linear small scales, we need better models and ensure their proper validation before applying them to measurements. We have shown that the  hybrid perturbation theory galaxy bias model can be validated with simulations to sufficient accuracy for the present analysis. By relaxing the priors on all five parameters (per redshift bin), the model's accuracy can be improved though the increase in model complexity poses other challenges in parameter estimation. Finally, and perhaps most importantly, we have highlighted how understanding   potential sources of systematic uncertainty is of paramount importance for extracting  unbiased cosmological information in this era of precision cosmology.

\section*{Acknowledgements}
EK is supported by the Department of Energy grant DE-SC0020247 and the David \& Lucile Packard Foundation.
SP and BJ are supported in part by the US Department of Energy Grant No. DE-SC0007901 and NASA ATP Grant No. NNH17ZDA001N. 

Funding for the DES Projects has been provided by the U.S. Department of Energy, the U.S. National Science Foundation, the Ministry of Science and Education of Spain, 
the Science and Technology Facilities Council of the United Kingdom, the Higher Education Funding Council for England, the National Center for Supercomputing 
Applications at the University of Illinois at Urbana-Champaign, the Kavli Institute of Cosmological Physics at the University of Chicago, 
the Center for Cosmology and Astro-Particle Physics at the Ohio State University,
the Mitchell Institute for Fundamental Physics and Astronomy at Texas A\&M University, Financiadora de Estudos e Projetos, 
Funda{\c c}{\~a}o Carlos Chagas Filho de Amparo {\`a} Pesquisa do Estado do Rio de Janeiro, Conselho Nacional de Desenvolvimento Cient{\'i}fico e Tecnol{\'o}gico and 
the Minist{\'e}rio da Ci{\^e}ncia, Tecnologia e Inova{\c c}{\~a}o, the Deutsche Forschungsgemeinschaft and the Collaborating Institutions in the Dark Energy Survey. 

The Collaborating Institutions are Argonne National Laboratory, the University of California at Santa Cruz, the University of Cambridge, Centro de Investigaciones Energ{\'e}ticas, 
Medioambientales y Tecnol{\'o}gicas-Madrid, the University of Chicago, University College London, the DES-Brazil Consortium, the University of Edinburgh, 
the Eidgen{\"o}ssische Technische Hochschule (ETH) Z{\"u}rich, 
Fermi National Accelerator Laboratory, the University of Illinois at Urbana-Champaign, the Institut de Ci{\`e}ncies de l'Espai (IEEC/CSIC), 
the Institut de F{\'i}sica d'Altes Energies, Lawrence Berkeley National Laboratory, the Ludwig-Maximilians Universit{\"a}t M{\"u}nchen and the associated Excellence Cluster Universe, 
the University of Michigan, the National Optical Astronomy Observatory, the University of Nottingham, The Ohio State University, the University of Pennsylvania, the University of Portsmouth, 
SLAC National Accelerator Laboratory, Stanford University, the University of Sussex, Texas A\&M University, and the OzDES Membership Consortium.

The DES data management system is supported by the National Science Foundation under Grant Numbers AST-1138766 and AST-1536171.
The DES participants from Spanish institutions are partially supported by MINECO under grants AYA2015-71825, ESP2015-88861, FPA2015-68048, SEV-2012-0234, SEV-2016-0597, and MDM-2015-0509, 
some of which include ERDF funds from the European Union. IFAE is partially funded by the CERCA program of the Generalitat de Catalunya.
Research leading to these results has received funding from the European Research
Council under the European Union's Seventh Framework Program (FP7/2007-2013) including ERC grant agreements 240672, 291329, and 306478.
We  acknowledge support from the Australian Research Council Centre of Excellence for All-sky Astrophysics (CAASTRO), through project number CE110001020.

This manuscript has been authored by Fermi Research Alliance, LLC under Contract No. DE-AC02-07CH11359 with the U.S. Department of Energy, Office of Science, Office of High Energy Physics. The United States Government retains and the publisher, by accepting the article for publication, acknowledges that the United States Government retains a non-exclusive, paid-up, irrevocable, world-wide license to publish or reproduce the published form of this manuscript, or allow others to do so, for United States Government purposes.

Based in part on observations at Cerro Tololo Inter-American Observatory, 
National Optical Astronomy Observatory, which is operated by the Association of 
Universities for Research in Astronomy (AURA) under a cooperative agreement with the National 
Science Foundation.

The analysis made use of the software tools {\sc SciPy}~\cite{2020SciPy-NMeth}, {\sc NumPy}~\cite{2020NumPy-Array},  {\sc Matplotlib}~\cite{Hunter:2007}, {\sc CAMB}~\cite{Lewis:1999bs,Lewis:2002ah,Howlett:2012mh}, {\sc GetDist}~\cite{Lewis:2019xzd}, {\sc Multinest}~\cite{Feroz_2008,Feroz_2009,Feroz_2019},  {\sc Polychord}~\cite{Handley_2015}, {\sc CosmoSIS}~\cite{Zuntz_2015}, {\sc Cosmolike}~\cite{Krause_2017} and {\sc TreeCorr}~\cite{Jarvis_2004}.

%%%%%%%%%%%%%%%%%%%%%%%%%%%%%%%%%%%%%%%%%%%%%%%%%%

%%%%%%%%%%%%%%%%%%%% REFERENCES %%%%%%%%%%%%%%%%%%

% The best way to enter references is to use BibTeX:

% \bibliographystyle{mnras_2author}
\bibliography{ref} % if your bibtex file is called example.bib

%%%%%%%%%%%%%%%%%%%%%%%%%%%%%%%%%%%%%%%%%%%%%%%%%%

%%%%%%%%%%%%%%%%% APPENDICES %%%%%%%%%%%%%%%%%%%%%

\appendix

\section{Point mass marginalization}
\label{app:pm}
The point mass parameter ($B$) can also be expressed as residual mass bias, $B = \delta M/\pi$ where $\delta M$ is approximately related to the difference between the model and true estimate of halo mass below the scales of our model validity ($r_{\rm min}$). More accurately, $\delta M_{\rm halo}$ can be expressed in terms of galaxy-matter correlation as:
% \begin{linenomath*}
\begin{equation}
\label{eq:pm_halo}
    \delta M = \int_{0}^{r_{\rm min}} dr_p (2\pi r_p) \int_{-\infty}^{\infty} d\Pi \, \Delta \xi_{\rm gm}\bigg(\sqrt{r_p^2 + \Pi^2}, z \bigg), 
\end{equation}
% \end{linenomath*}
where $\Delta \xi_{\rm gm} = \xi^{\rm true}_{\rm gm} - \xi^{\rm model}_{\rm gm}$.

\begin{figure}
\includegraphics[width=\columnwidth]{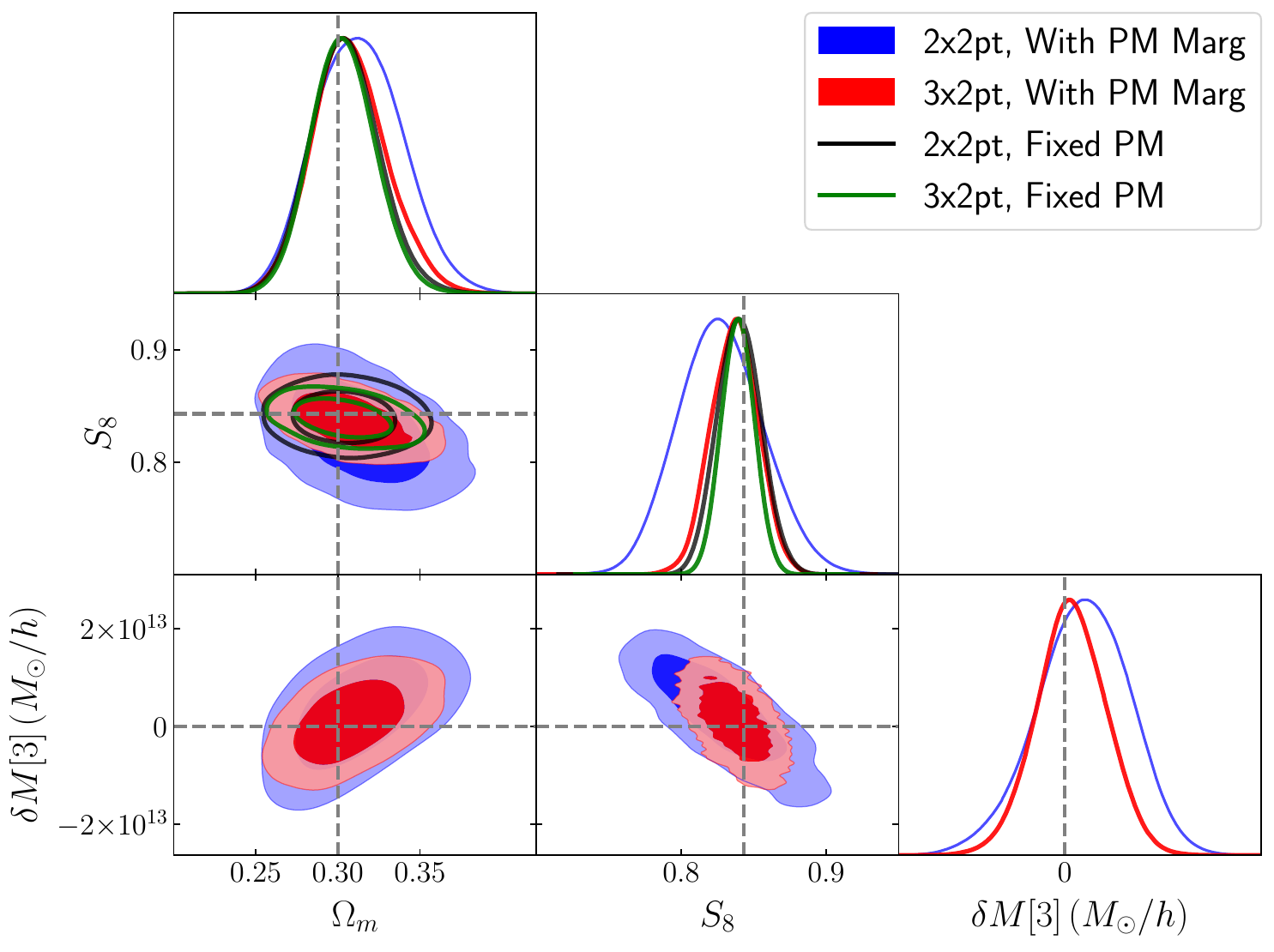}
\caption[]{Effect of point mass marginalization on the constraining power of $2\times2$pt and $3\times2$pt. We see that the constraining power of $2\times2$pt degrades significantly with point mass marginalization, while for $3\times2$pt the change is minimal. Including the shear-shear correlation  breaks the degeneracy between point-mass (we show PM for third bin, $M_{\rm halo}[3]$) and $S_8$, leading to smaller sensitivity of cosmology constraints on point mass constraints. }
\label{fig:pm_effect}
\end{figure}

In Fig.~\ref{fig:pm_effect} we compare the constraining power of $2\times2$pt and $3\times2$pt \textit{simulated} analysis at our \textit{fiducial} scale cuts for different point mass parameter settings. We generate a noiseless theory baseline datavector using the linear bias model and the \textit{fiducial} parameter values given in Table ~\ref{tab:params_all}. In the blue and red filled contours, instead of analytically marginalizing over the point mass parameters, we explicitly sample them when analyzing $2\times 2$pt and $3\times 2$pt datavectors respectively. To test the impact of point mass marginalization on the constraining power, we also show the constraints obtained after fixing the PM parameters to their fiducial value of zero using unfilled contours. The black and green unfilled contours show the constraints using $2\times 2$pt and $3\times 2$pt datavectors respectively. We see that although point mass marginalization has a significant impact on the constraining power of the $2\times2$pt analysis, it has a small impact on the $3\times2$pt analysis. The main reason is that, due to extra constraints from cosmic shear, we break the degeneracy between PM parameters and cosmological parameters, and hence uncertainty in PM parameters do not dilute our cosmology constraints.

\begin{figure}
\includegraphics[width=\columnwidth]{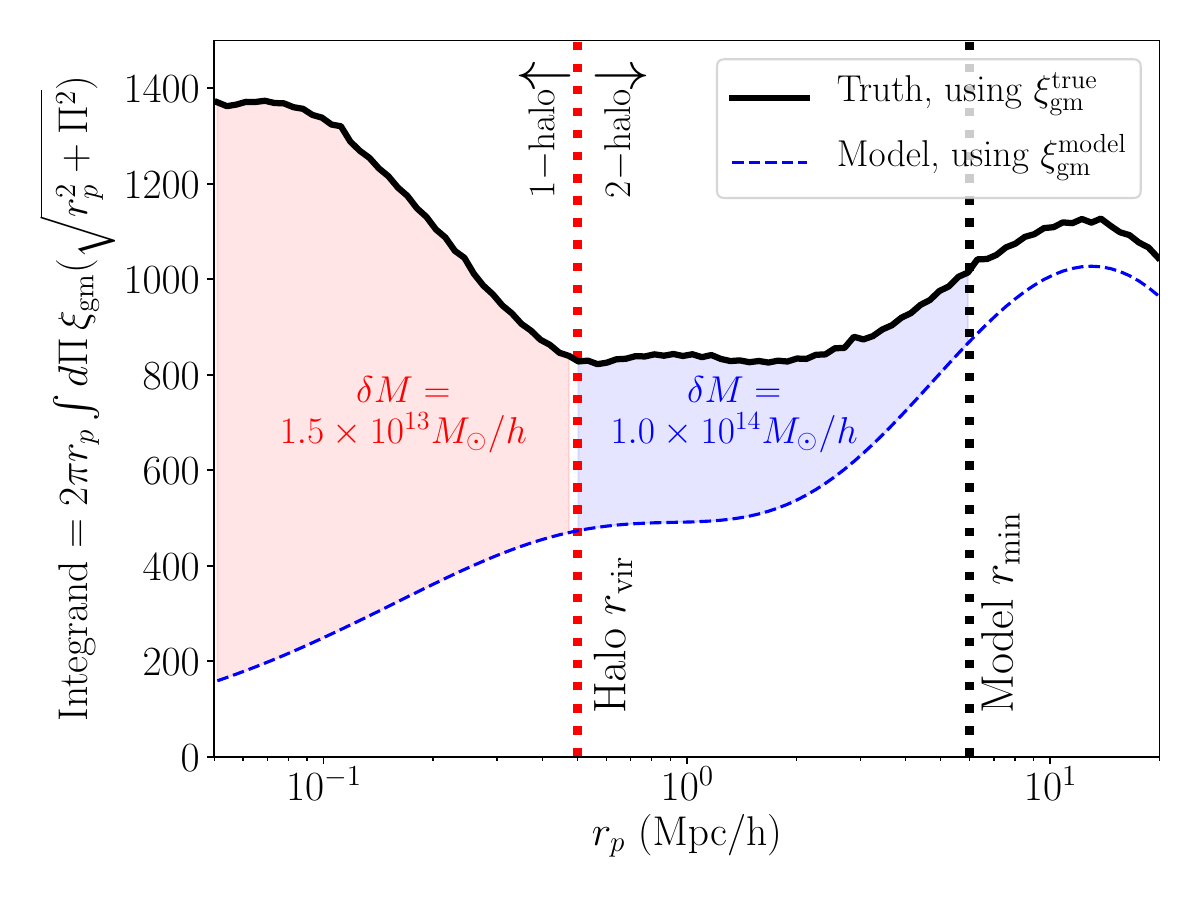}
\caption[]{We show the contribution to the residual mass shown in Eq.~\ref{eq:pm_halo} from different radial regimes. We find a significant contribution from 2-halo regime and therefore we cannot motivate an astrophysical informative prior on the PM parameters, without putting an informative prior on cosmology as well.  
}
\label{fig:pm_prior}
\end{figure}

As PM marginalization degrades the constraining power of $2\times2$pt significantly, it might be desirable to implement an informative prior on the PM parameters. However, motivating an astrophysical prior on the PM parameters is not possible for our scale cuts as the majority of residual mass constraints are contributed from the 2-halo regime, as shown in Fig.~\ref{fig:pm_prior}. For simplicity, we assume all our galaxies occupy the center of $2.5 \times 10^{13} M_{\odot}/h$ mass halos. The input ``truth" curve in black solid line uses $\xi_{\rm gm}$ that is generated using the Navarro-Frenk-White profile \citep{Navarro_1996} in the 1-halo regime ($r < 0.5$ Mpc/$h$) and one--loop PT in the 2-halo regime ($r > 0.5$ Mpc/$h$). Given this input halo mass, the halo model framework predicts the effective large scale linear bias value \citep{COORAY_2002}. The dashed blue curve is generated using a linear bias model, using a linear bias value that is 1$\sigma$ lower from this predicted value. Here $\sigma$ is the uncertainity obtained from $2\times2$pt marginalized constraints on the galaxy bias for first tomographic bin. The area between the two curves below some scale is equal to total $\delta M$ as calculated using Eq.~\ref{eq:pm_halo}.

We show the contribution to $\delta M$ separately for the 1-halo region (below the red dashed line) and 2-halo regimes (up to the scales of 6Mpc/$h$, which are our scale cuts for $\gammat$). We find that the 2-halo regime contributes significantly more than the 1-halo region and the resulting $\delta M$ value is significantly more than the input halo mass of $2.5 \times 10^{13} M_{\odot}/h$. An informative prior would amount to understanding the galaxy-matter correlation and its dependence on cosmology and galaxy bias model from all scales below our scale cuts. Therefore we choose an uninformative wide prior on the point mass parameters.

\begin{figure}
\includegraphics[width=\columnwidth]{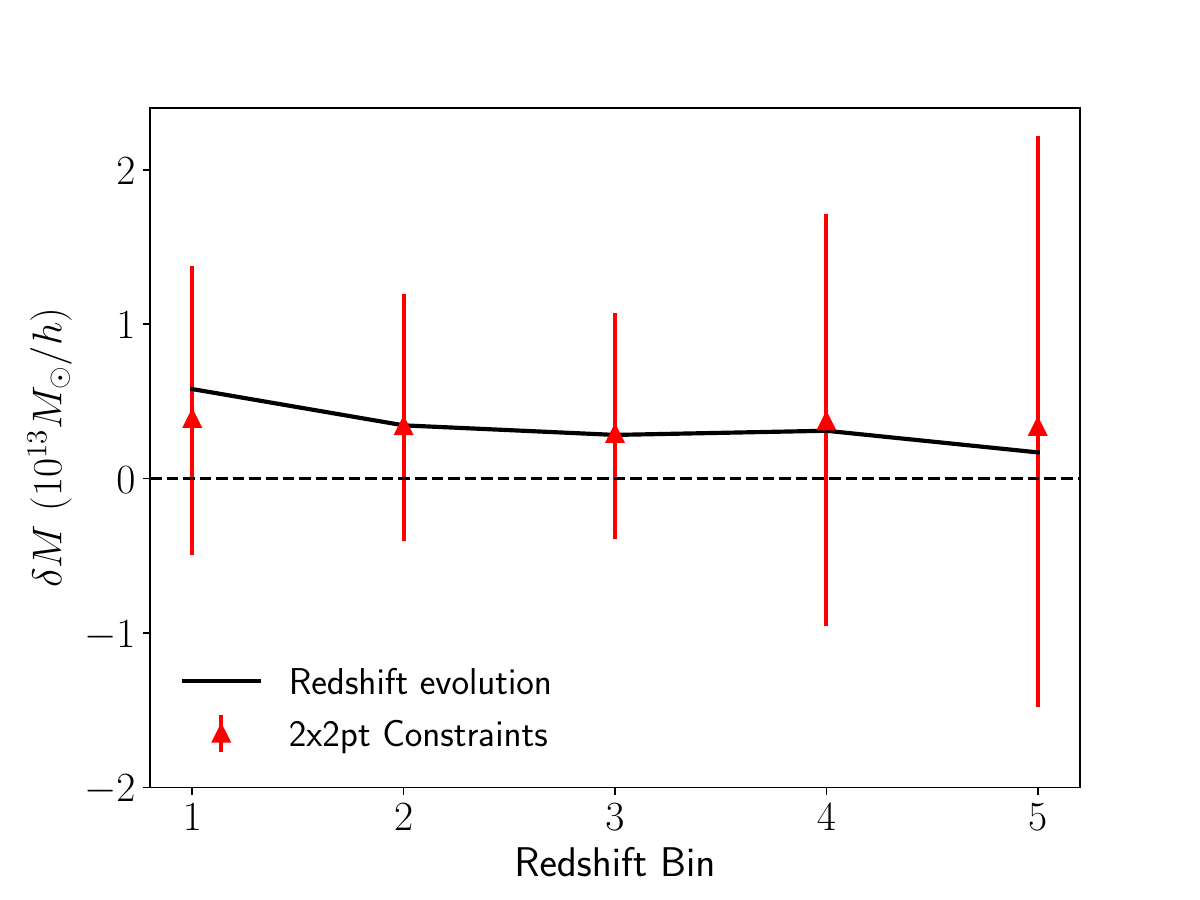}
\caption[]{We show the effect of the evolution of galaxy matter correlation functions on the PM parameters for each tomographic bin in the black line. The red errorbars show the expected errorbars on PM parameters for $2\times2$pt as shown in Fig.~\ref{fig:pm_effect}. The blue errorbars are the constraints from $3\times2$pt.
}
\label{fig:pm_evolve}
\end{figure}

The baseline model parameterization assumes the point mass parameter to be constant within each tomographic bin. We test this assumption implicitly in the suite of \buzzard simulations. The datavector measured in N-body \buzzard simulation will capture the effects of evolving point-mass parameters due to the evolution of the galaxy-matter correlation within a lens tomographic bin. As we have validated that our scale cuts pass our threshold criteria of bias in cosmological parameters being less than 0.3$\sigma$, we can conclude that the effect of point mass parameter evolution is small. Here we also test this effect explicitly by generating a simulated galaxy matter correlation function using the halo model. We assume a constant HOD of the \redmagic galaxies but include the evolution of halo mass function and halo bias to predict the evolution of the galaxy-matter correlation function. The contribution to the PM parameter due to this evolution in each tomographic bin is given by Eq.~\ref{eq:pm_halo}. In Fig.~\ref{fig:pm_evolve}, we show this contribution to each redshift bin by the black solid line. We compare this bias with the expected level of uncertainty in the PM parameters by plotting the marginalized constraints on these parameters as shown in Fig.~\ref{fig:pm_effect} for $2\times2$pt analyses. We see that the uncertainty in PM parameters is significantly greater than the expected bias.

\begin{figure*}
\includegraphics[width=\textwidth]{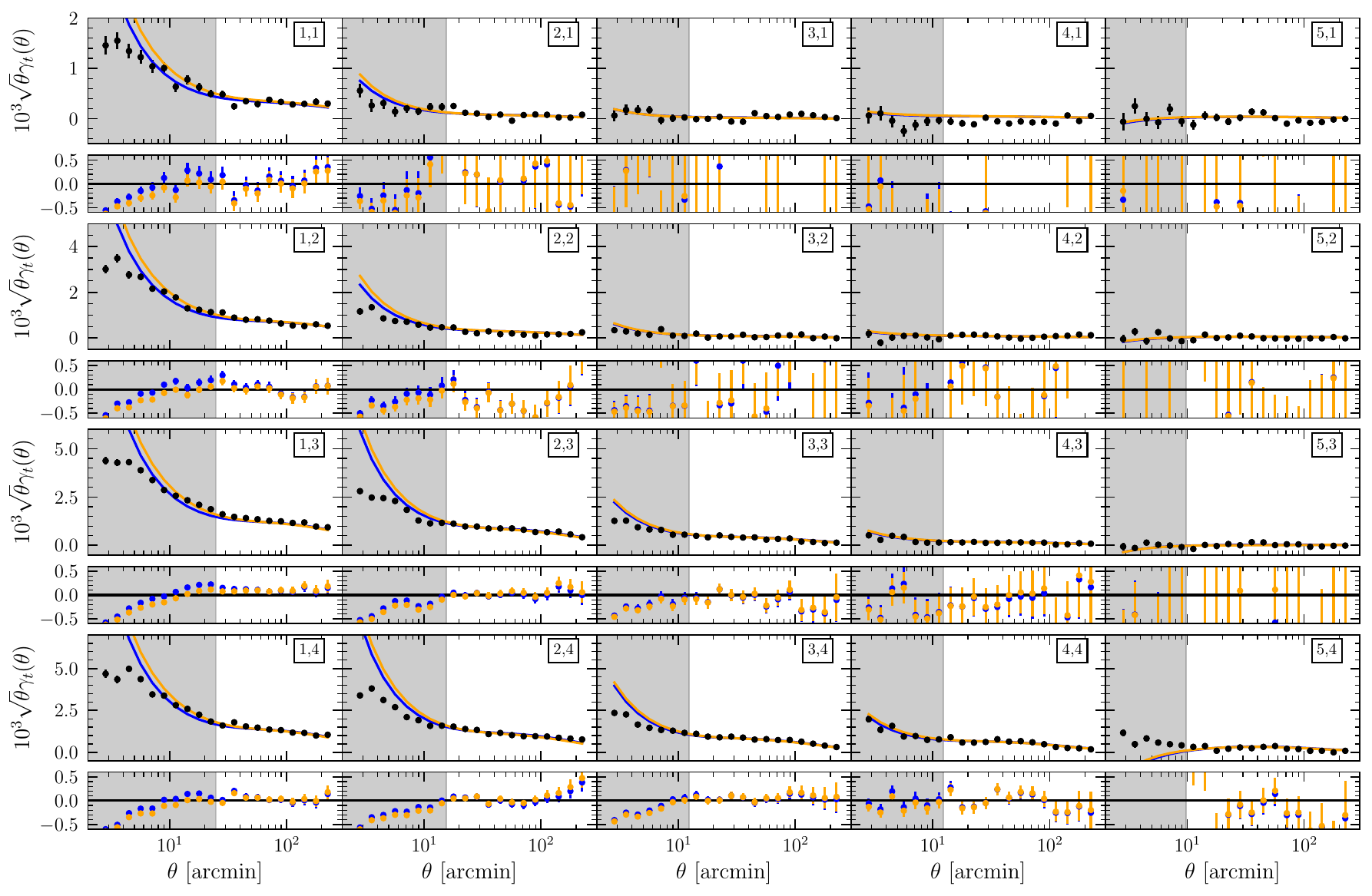}
\includegraphics[width=\textwidth]{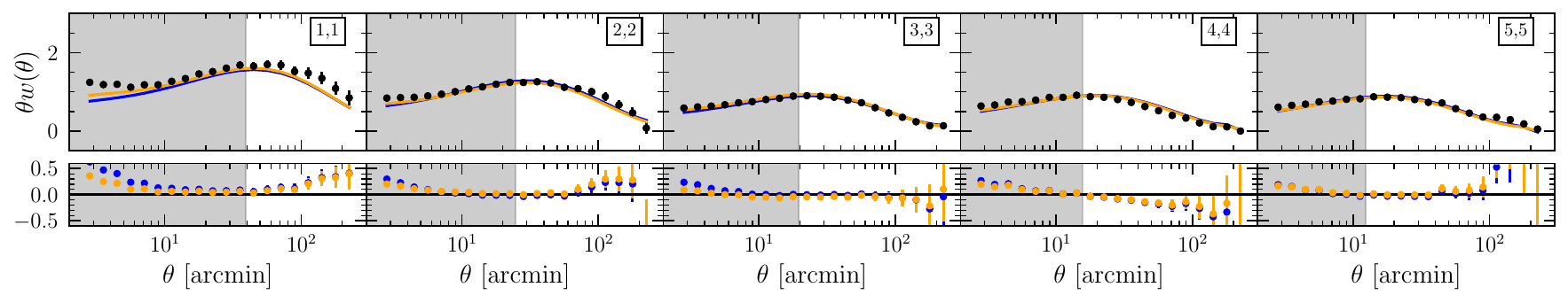}
\caption[]{The measurements of $\wtheta$ and $\gammat$ with \redmagic sample are shown with black dots. We show the best fit using the \textit{fiducial} \textit{Linear bias} model in blue and model with $X_{\rm lens}=0.87$ in orange.  }
\label{fig:data_2pt}
\end{figure*}

\section{Datavector residuals}
We show the comparison between our measurements and best-fit theory datavector in Fig.~\ref{fig:data_2pt}. We show the residuals between data and best-fit theory model from both the \textit{fiducial} model as well as with $X_{\rm lens}=0.87$ model . Using the \textit{fiducial} linear bias model scale cuts of (8,6)~Mpc/$h$ (that leaves 302 datapoints in total), we find a minimum $\chi^2$ of 347.2 and 351.1 for the \textit{fiducial} model and $X_{\rm lens}=0.87$ model respectively.

\section{Broad-$\chi^2$ sample}
\label{app:bchi2}

\begin{figure}
\includegraphics[width=\columnwidth]{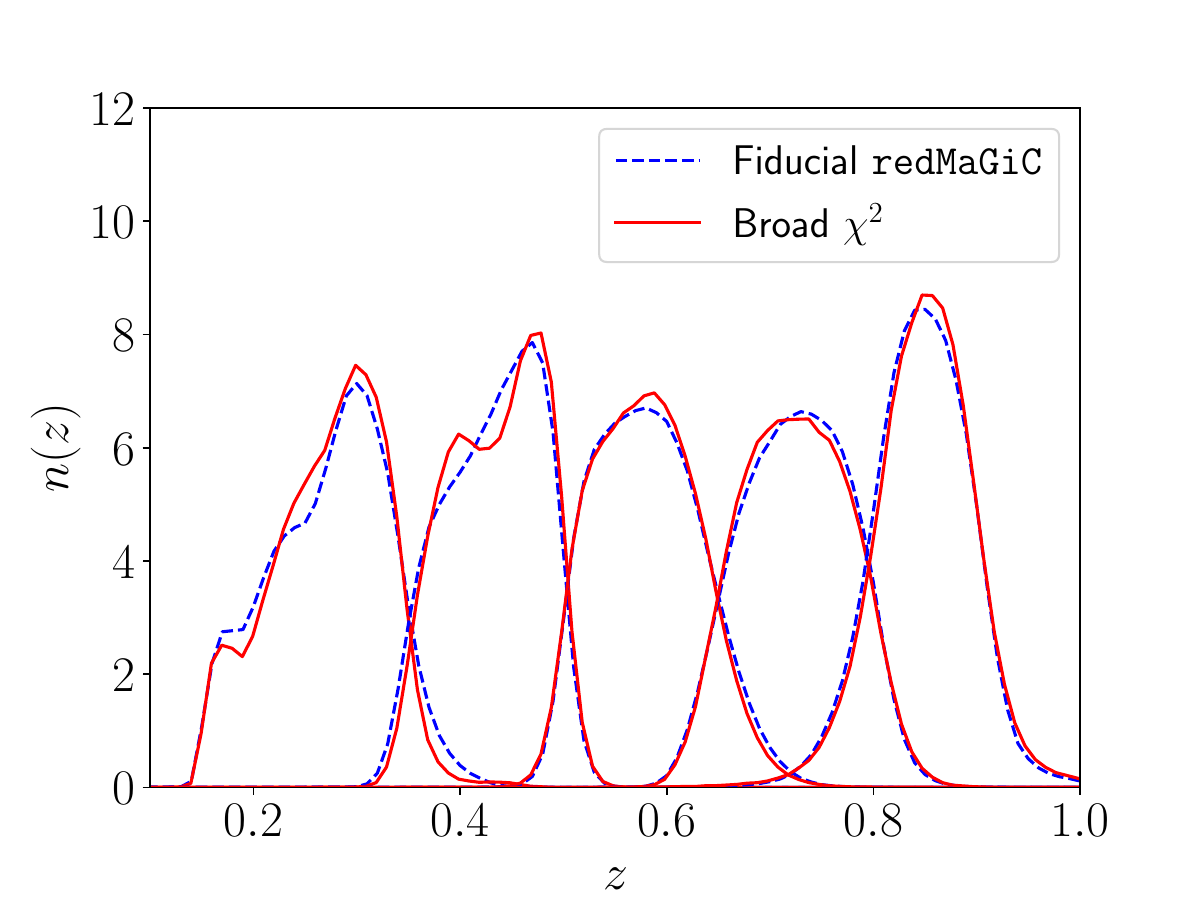}
\caption[]{We compare the redshift distribution of the fiducial \redmagic sample with the broad-$\chi^2$ \redmagic sample. 
% The redshift distribution have been normalized to integrate out to the effective number density for each tomographic bin. 
}
\label{fig:nz_comp_bchi2}
\end{figure}

As detailed in the main text, we generate a new galaxy sample by relaxing the selection criteria and selecting galaxies with goodness-of-fit $\chi^2_{\rm RM}=8$ to the \redmagic template. This new sample is constructed to reduce the sensitivity of any color-dependent photometric issue that might be present in the fiducial \redmagic sample and causing $X_{\rm lens} < 1$. After analyzing the $2\times 2$pt datavector, we do find that this sample prefers $X_{\rm lens} = 1$ and results in $S_8$ consistent with the \maglim galaxy sample. 

In Fig.~\ref{fig:nz_comp_bchi2}, we show the comparison of the lens number densities for the five tomographic bins. We perform the calibration of these redshift distributions using cross-correlations with BOSS/eBOSS data using the same methods described in \citet{y3-lenswz}. The lens photo-z prior that we use are shown in Table.~\ref{tab:params_nz_bchi2}. After downsampling the full catalog by a factor of 2, the number density (in the units of $\textrm{arcmin}^{-2}$) for this sample are $\langle n_{\rm g} \rangle = 0.027,0.04,0.07,0.03,0.03$ for the five tomographic bins. We generate a non-Gaussian covariance corresponding to these number densities. To mitigate the bias caused by wrong parameter values input to theory covariance calculations, we recalculate the covariance matrix using the best-fit parameters of an initial $2\times 2$pt  analysis and show the cosmological constraints corresponding to this new covariance. 

\begin{figure*}
\includegraphics[width=\textwidth]{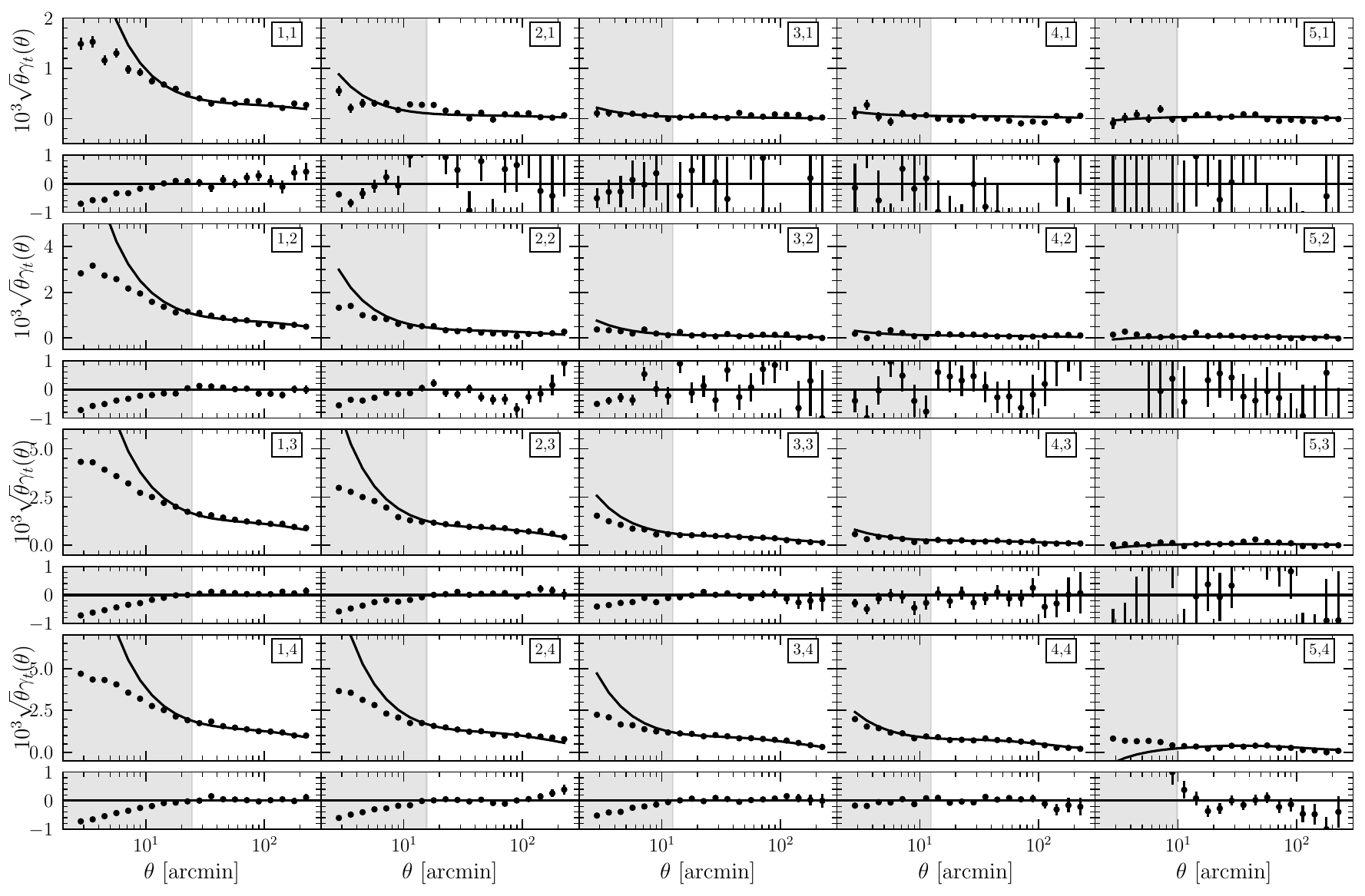}
\includegraphics[width=\textwidth]{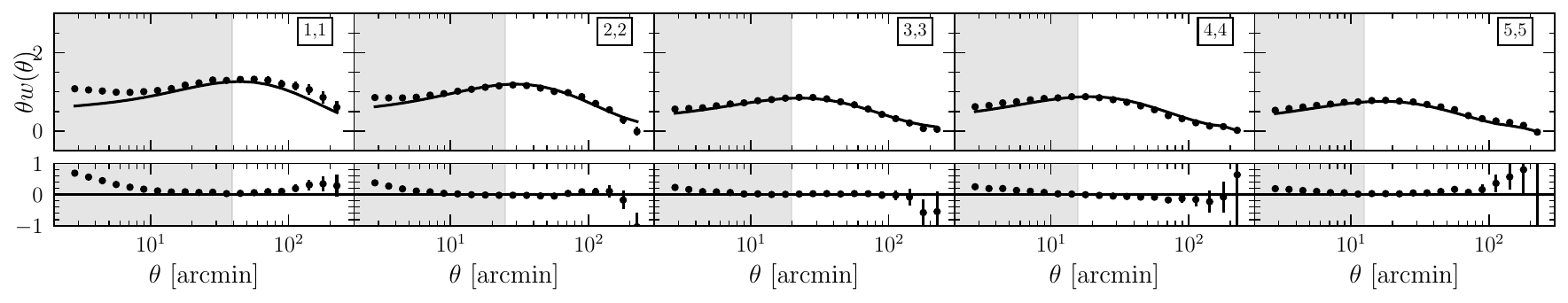}
\caption[]{The measurements of $\wtheta$ and $\gammat$ with the broad-$\chi^2$ \redmagic sample are shown with black dots. We show the best fit model in black.}
\label{fig:bchi2_resid_2pt}
\end{figure*}

\begin{figure*}
\includegraphics[width=\textwidth]{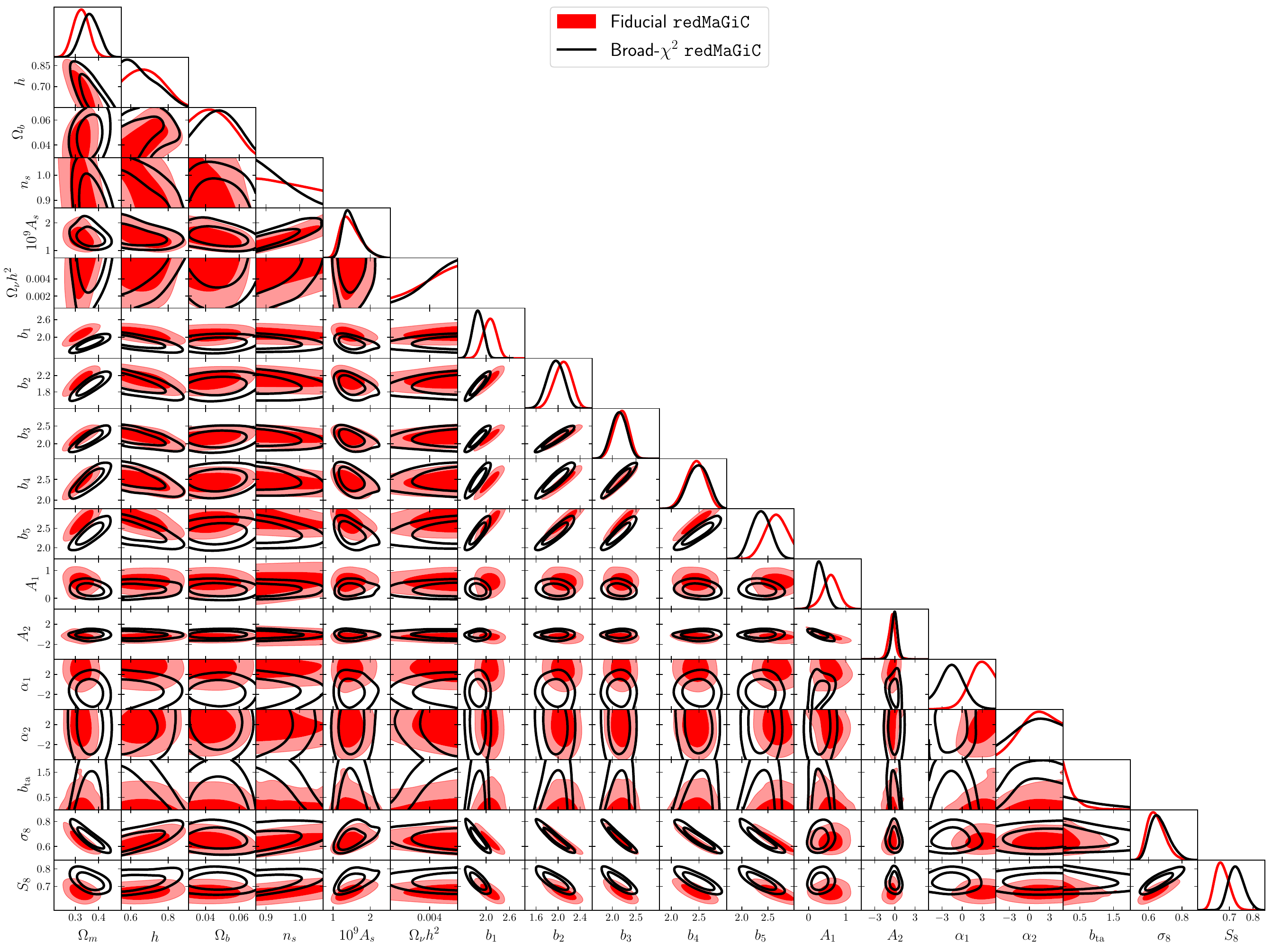}
\caption[]{Comparison of parameter constraints using fiducial and broad-$\chi^2$ \redmagic for all the parameters sampled in the analysis without a tight gaussian prior. We also show the derived parameters $\sigma_8$ and $S_8$.}
\label{fig:data_2pt_allcomp_bchi2}
\end{figure*}

Using the best-fit parameter values obtained with the linear bias model, we show the residuals in Fig.~\ref{fig:bchi2_resid_2pt}. We find a best-fit $\chi^2$ of 353 for 302 datapoints, and both the $w(\theta)$ and $\gamma_{\rm t}$ measurements are fit well with a linear bias, $\Lambda$CDM model. In Fig.~\ref{fig:data_2pt_allcomp_bchi2} we show the parameters constraints and compare them to the ones obtained with the fiducial \redmagic sample. 

% Show the n(z)'s
% Show the datavector comparison
% Also put bestfit in here? 

\begin{table}[H]
\centering 
% \resizebox{\textwidth}{!}
\tabcolsep=0.11cm
\begin{tabular}{|c c|}
\hline
% \hline
Parameter & Prior \\ \hline
% & & & \\
% & \multicolumn{3}{c|}{\textbf{Cosmology}} \\ 

% & & & \\
% \multirow{24}{*}{\shortstack[c]{Common\\ Parameters}} 
% % & & & \\
% & \multicolumn{3}{c|}{\textbf{Lens photo-$z$}} \\  
$\Delta z_{\rm g}^{1}$ & $\mathcal{G}[0.0088, 0.0029]$ \\ 
$\sigma z_{\rm g}^{1}$ & $\mathcal{G}[1.015, 0.035]$  \\ 
$\Delta z_{\rm g}^{2}$ & $\mathcal{G}[-0.0033, 0.0022]$  \\
$\sigma z_{\rm g}^{2}$ & $\mathcal{G}[0.991, 0.028]$  \\ 
$\Delta z_{\rm g}^{3}$ & $\mathcal{G}[0.0076, 0.0029]$  \\
$\sigma z_{\rm g}^{3}$ & $\mathcal{G}[1.096, 0.029]$  \\ 
$\Delta z_{\rm g}^{4}$ & $\mathcal{G}[0.0015, 0.0042]$ \\
$\sigma z_{\rm g}^{4}$ & $\mathcal{G}[1.104, 0.045]$  \\ 
$\Delta z_{\rm g}^{5}$ & $\mathcal{G}[-0.0058, 0.0061]$  \\ 
$\sigma z_{\rm g}^{5}$ & $\mathcal{G}[1.193, 0.056]$  \\ 
% & & & \\
% \cline{2-4}
\hline
% & & & \\
\end{tabular}
\caption{The lens photo-$z$ shift and stretch parameters varied in the analysis using the broad-$\chi^2$ sample and their prior range used ($\mathcal{G}[\mu, \sigma] \equiv$ Gaussian prior with mean $\mu$ and standard-deviation $\sigma$).}
\label{tab:params_nz_bchi2}
\end{table}

\section{Halo mass inference}
\label{app:halo_mass}

\begin{figure*}
\includegraphics[width=\textwidth]{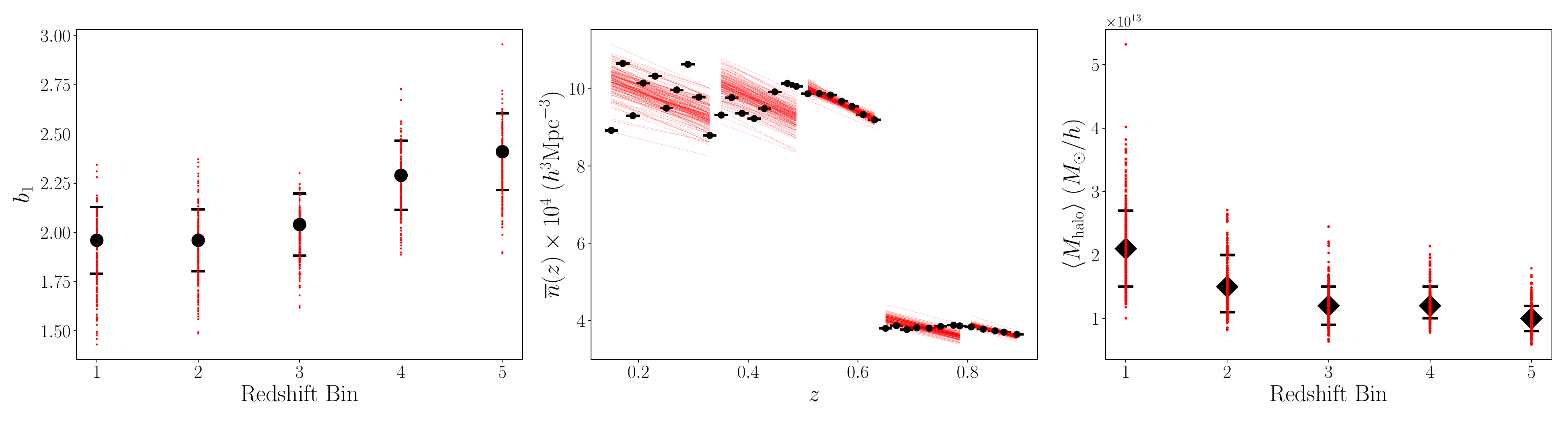}
\caption[]{This figure shows the marginalized constraints on the large-scale bias of \redmagic sample for the five tomographic bins on the left panel. The black dots denote the mean, and the error bars correspond to 68\% credible interval. Using these constraints and co-moving number density (middle panel), we infer the constraints on mean halo mass, as shown in the right panel for five tomographic bins. The red line and dots correspond to MCMC samples. We use the \textit{Linear bias} model with $X_{\rm lens}=0.87$.  }
\label{fig:b1_nbar_Mh}
\end{figure*}

In this section we detail the methodology to infer the host halo mass of our \redmagic lens galaxy sample from the constraints on galaxy bias parameters and number density. We use the halo model framework to make this prediction and parameterize the number of galaxies in a halo of mass $M$ in tomographic bin $j$ as $N^{j}_{\rm g}(M) = N^{j}_{\rm{cen}}(M) + N^{j}_{\rm{sat}}(M)$ where $N^{j}_{\rm{cen}}$ is the number of central galaxies and $N^{j}_{\rm{sat}}$ is the number of satellite galaxies. We parameterize these two components as:
% \begin{linenomath*}
\begin{align}
    N^{j}_{\rm cen} =  \frac{f^{j}_{\rm cen}}{2} \bigg[1 + {\rm erf}\bigg(\frac{ \log{M} - (\log{M_{\rm min}})^j}{(\sigma_{\log{M}})^j} \bigg) \bigg]  \\
    N^j_{\rm sat} = \frac{1}{2} \bigg[1 + {\rm erf}\bigg(\frac{ \log{M} - (\log{M_{\rm min}})^j}{(\sigma_{\log{M}})^j} \bigg) \bigg] \times \bigg( \frac{M_{\rm h}}{M^j_1} \bigg)^{\alpha^j}.
\end{align}
% \end{linenomath*}
Here we have five free parameters, $f^j_{\rm cen}$, $(\log{M_{\rm min}})^j$, $(\sigma_{\log{M}})^j$, $M^j_1$ and $\alpha^j$, that we marginalize over. We can predict the comoving number density ($\overline{n}(z)^j$) and galaxy bias for a given tomographic bin $j$, $b^j_1$, from galaxy HOD as follows:

% \begin{equation}\label{eq:ng}
% ,
% \end{equation}
% \begin{linenomath*}
\begin{align}\label{eq:nbar_b1b2}
\nonumber    \overline{n}^j(z) = \int_{0}^{\infty} dM \frac{dn}{dM} N^{j}_{\rm g}(M) \\
\nonumber    b^{j}_1 = \int dz \frac{n^j_g(z)}{\overline{n}^j(z)} \int_{0}^{\infty} dM \frac{dn}{dM} N^{j}_{\rm g}(M) b^{\rm halo}_{1}(M,z)\\
\end{align}
% \end{linenomath*}
We use the \citet{Tinker_2008} halo mass function ($dn/dM$) and the \citet{Tinker_2010} relation for linear halo bias ($b^{\rm halo}_{1}(M,z)$). 

Therefore, Eqs.~\ref{eq:nbar_b1b2} allow us to predict the number density and galaxy bias values. We then sample these HOD parameters to fit the datavector $\vec{\mathbfcal{D}_H} = \overline{n}^j(z_1)...\overline{n}^j(z_{n}),b^j_1,b^j_2]$ of length $d$ where $\overline{n}^j(z_1)...\overline{n}^j(z_{n})$ are the $n=d-2$ observed comoving number density of \redmagic galaxies as shown in middle panel of Fig.\ref{fig:b1_nbar_Mh} and $b^j_1$ and $b^j_2$ are the marginalized mean bias values obtained at our \textit{fiducial} scale cut. For a given set of HOD parameters ($\Theta_{\rm H}$), the theoretical prediction is given by $\mathbfcal{T}_{\rm H}$ and we write our log-likelihood as:
% \begin{linenomath*}
\begin{multline}
    \ln \mathcal{L}(\vec{\mathbfcal{D}_H}|\Theta) = -\frac{1}{2} \bigg[ (\vec{\mathbfcal{D}_H} - \vec{\mathbfcal{T}_H}(\Theta_{\rm H})) \, {\mathbfcal{C}_H}^{-1} \,  (\vec{\mathbfcal{D}_H} - \vec{\mathbfcal{T}_H}(\Theta_{\rm H}))^{\rm T} \\ -  \ln(|\mathbfcal{C}_H|) \bigg]
\end{multline}
% \end{linenomath*}
In order to account for variation of HOD within a tomographic bin that contributes to the variation on $\overline{n}^j(z)$ within each tomographic bin as seen in Fig.\ref{fig:b1_nbar_Mh}, we implement an analytical marginalization scheme. We change the covariance of our datavector $\mathbfcal{C}_H$ as :
% \begin{linenomath*}
\begin{equation}
    \mathbfcal{C}_H \to \mathbfcal{C}_H + \alpha_{c} \mathbfcal{I}_D
\end{equation}
% \end{linenomath*}
where $\mathbfcal{I}_D$ is a diagonal matrix of dimension $d\times d$ whose diagonal elements equal to 1 from index 1 to d-1, and equal to 0 otherwise. We sample over the parameter $\alpha_{c}$, treating it as a free parameter.

%%%%%%%%%%%%%%%%%%%%%%%%%%%%%%%%%%%%%%%%%%%%%%%%%%

% Don't change these lines
% \bsp	% typesetting comment
\label{lastpage}

% \bibliographystyle{apsrev4-1}
% \bibliography{ref} 

\end{document}